\documentclass[useAMS,usenatbib]{mn2e/mn2e}
\usepackage{graphicx}
\usepackage{txfonts}
\usepackage{url}
\usepackage{color}
\usepackage{enumitem}

\title[Carbon dust in C-stars]{Constraining dust properties in Circumstellar Envelopes of C-stars in the Small Magellanic Cloud: optical constants and grain size of Carbon dust}
\author[Nanni et al.]{Ambra Nanni$^1$,
Paola Marigo$^1$,
Martin A.T. Groenewegen$^2$,
Bernhard Aringer$^1$,
L\'eo Girardi$^3$, \newauthor
Giada Pastorelli$^1$,
Alessandro Bressan$^4$,
Sara Bladh$^1$
  \\
  $^1$ Dipartimento di Fisica e Astronomia Galileo Galilei,
  Universit\`a di Padova, Vicolo dell'Osservatorio 3, I-35122 Padova, Italy\\
  $^2$ Royal Observatory of Belgium, Ringlaan 3, B-1180 Brussel, Belgium  \\
  $^3$ Osservatorio Astronomico di Padova, Vicolo dell'Osservatorio 5,
  I-35122 Padova, Italy \\
  $^4$ SISSA, via Bonomea 265, I-34136 Trieste, Italy}

\begin{document}

\date{Accepted .  Received ; in original form }

\pagerange{\pageref{firstpage}--\pageref{lastpage}} \pubyear{2016}

\maketitle

\begin{abstract}\label{firstpage}
We present a new approach aimed at constraining the typical size and optical properties of carbon dust grains in Circumstellar envelopes (CSEs) of carbon-rich stars (C-stars) in the Small Magellanic Cloud (SMC).
To achieve this goal, we apply our recent dust growth description, coupled with a radiative transfer code to the CSEs of C-stars evolving along the TP-AGB, for which we compute spectra and colors. Then we compare our modeled colors in the near- and mid-infrared (NIR and MIR) bands with the observed ones, testing different assumptions in our dust scheme and employing several data sets of optical constants for carbon dust available in the literature.  Different assumptions adopted in our dust scheme change the typical size of the carbon grains produced.
We constrain carbon dust properties by selecting the combination of grain size and optical constants which best reproduces several colors in the NIR and MIR at the same time.
The different choices of optical properties and grain size lead to differences in the NIR and MIR colors greater than two magnitudes in some cases.

We conclude that the complete set of observed NIR and MIR colors are best reproduced by small grains, with sizes between $\sim$0.035 and $\sim$0.12~$\mu$m, rather than by large grains between $\sim0.2$ and $0.7$~$\mu$m. The inability of large grains to reproduce NIR and MIR colors seems independent of the adopted optical data set.
We also find a possible trend of the grain size with mass-loss and/or carbon excess in the CSEs of these stars.
\end{abstract}


\section{Introduction}
\label{introduction}
Dust particles in the Circumstellar Envelopes (CSEs) of mass-losing Thermally Pulsing Asymptotic Giant Branch (TP-AGB) stars are able to absorb and scatter the photospheric radiation, redistributing the absorbed energy at wavelengths longer than $\sim1$~$\mu$m.  Since these stars contribute a considerable fraction of the total light emitted by galaxies, the study of dust growth and reprocessing of stellar radiation in the CSEs of TP-AGB stars is essential in order to interpret observations of galaxies in NIR and MIR restframe passbands up to high redshifts.

In this respect, nearby galaxies represent a unique chance for detailed investigations of dusty TP-AGB stars in resolved stellar populations. Particularly important are the Magellanic Clouds (MCs), whose TP-AGB populations, comprising a few tens thousands stars, have been almost completely covered by a series of surveys. Photometric data of the stars in the Small Magellanic Clouds (SMC) are now available in a wide range of wavelengths, including the NIR and MIR provided by Two Micron All Sky Survey \citep[2MASS,][]{skrutskie06} in the J, H, Ks bands and between 3.6-70~$\mu$m by the Spitzer Survey of the Small Magellanic Cloud \citep[S$^{3}$MC,][]{bolatto07}.
The Spitzer Space Telescope Legacy Program entitled ``Surveying the Agents  of Galaxy Evolution in the tidally stripped, low metallicity Small Magellanic Cloud'' \citep[SAGE-SMC,][]{gordon11} is the most complete and spatially uniform set of IR data (3.6-160~$\mu$m) of the evolved stars in the SMC. On the basis of the resulting catalog, \citet{Boyer11} identified and classified about 5800 TP-AGB stars in the SMC.

Studies of resolved stellar populations have been performed in the past by employing stellar isochrones which include an approximated treatment of dusty CSEs \citep{Bressan98, Marigo08}. Dust evolution along the TP-AGB phase has recently been revisited by
\citet{Zhukovska08}, \citet{ventura12, DiCriscienzo13} and by our group
\citep{Nanni13, Nanni14}, following the scheme firstly introduced by \citet{GS99} and developed further by \citet{FG06}.
Our revised dust growth scheme, together with the new updated TP-AGB tracks, has already been shown to successfully reproduce some important observed trends in solar-like environments, such as the expansion velocities as a function of the mass-loss rates, for both oxygen-rich (M-stars) and carbon-rich (C-stars) stars \citep{Nanni13}.

The next step of our work is to compute stellar spectra and colors to be compared with observations of resolved stellar populations.
This kind of comparisons also represents a good way to validate the results of dust growth scheme and of the underlying TP-AGB models.

Recent attempts to interpret the nature of dusty AGB stars in MCs have been performed by \citet{Ventura14, Ventura16, Dellagli15a, Dellagli15b}.
These authors employed the results of their stellar evolutionary models including a self-consistent scheme for dust growth, but they limit their investigations to few color-color and color-magnitude diagrams (CCDs and CMDs), missing a systematic study in which NIR and MIR colors are recovered at the same time.
Moreover, their calculations refer to a particular set of assumed dust optical constants.

In addition to that, in the
studies in which radiative transfer (RT) models are employed to fit the stellar Spectral Energy Distributions (SEDs),
the optical constants and grain size need to be assumed. These assumptions might lead to
considerable differences in the results and large
uncertainties \citep{Groenewegen09, Riebel12,Srinivasan16}. This work provides some constraints.

In this paper we aim at reproducing a large set of NIR and MIR colors simultaneously by employing our latest TP-AGB tracks \citep{marigoetal13} together with the revised version of dust growth scheme \citep{Nanni13, Nanni14}. As we will show, achieving a good agreement between observations and models in several colors at the same time is a complex task, since a specific dusty model can produce a good agreement in a certain CCD but might perform very poorly in another one.

We focus our investigation on C-stars which are particularly relevant for the interpretation of NIR and MIR colors of many galaxies, including the MCs, also as far as the reddest stars are concerned \citep{Woods11, Riebel12}.
We anticipate that this approach sets constraints of the dust optical properties and typical grain size produced in the CSEs of C-stars in the SMC.
In fact, in spite of the importance of C-stars, the nature of carbon dust in their CSEs is extremely uncertain as far as its chemical structure and typical grain size are concerned. A variety of optical data sets for carbon dust, very different one from each other, are available in the literature.
Furthermore, the additional uncertainties in the determination of carbon dust grain sizes, render the modeling of dusty C-stars even more challenging.
A detailed investigation on the dynamical effects produced by different choices of carbon optical data sets has been performed by \citet{Andersen99}.
However, it is still missing a systematic study in which dusty models, consistently coupled to the complete TP-AGB phase, are employed to analyze the properties of carbon dust in evolving CSEs.
This is the main purpose of this work.

This paper is organized as follows.
In Section~\ref{models} we summarize the main characteristics
of the dust condensation and radiative transfer models.
In Section~\ref{results} we discuss how the emerging spectra change by varying different dust parameters. In Section~\ref{effectonTPAGB} we apply our dust formation scheme and radiative transfer to models evolving along the TP-AGB tracks, using different assumptions. In Section~\ref{Cal:section} we use the results of the previous section to select the data sets and grain sizes that best reproduce the observations of C-stars in the SMC. Finally, the results are summarized in Section~\ref{closing}.

\section{Model of dust growth and radiative transfer in C-stars}\label{models}
The dust formation scheme adopted in this work enables us to follow dust production along the TP-AGB phase. Our dust formation description is based on the revised version of the pioneering work of \citet{FG06}, as thoroughly described in \citet{Nanni13, Nanni14}. Here we just recall the basic ingredients and the most useful equations of our formalism.

Our dust formation scheme needs a set of input parameters given by the physical properties of the star, such as the effective temperature ($T_{\rm eff}$), luminosity ($L$), actual stellar mass ($M$), mass-loss rate ($\dot{M}$) and initial elemental abundances in the atmosphere (in particular the C/O ratio).
Such input quantities are provided along the TP-AGB evolution by the stellar tracks computed by the \textsc{parsec} code by \citep{Bressanetal12}, coupled with the \textsc{colibri} code \citep{marigoetal13}.

For each choice of the input quantities, the code integrates a set of differential equations which describes the dust growth coupled with a stationary, spherically symmetric, wind.
The outcome of the calculation characterizes the dust produced in terms of chemistry, dust condensation fractions, grain sizes and condensation temperatures. The most  relevant output concerning the outflow dynamics is the expansion velocity.

In our dust description, dust particles are accreted on pre-existing refractive particles known as ``seed nuclei''.
The number of seed particles is often taken as free parameter \citep{GS99,FG06, ventura12}.
Once the bulk of dust is formed, a dust-driven wind can be accelerated under certain favorable conditions.
The occurrence of the outflow acceleration associated with dust formation, is determined by two competing processes: the radiation pressure of the photons on the dust grains and the gravitational pull of the star. In case the dust species formed in the CSE are abundant and opaque enough, the outflow is accelerated.

It is well known that around 1~$\mu$m, which is roughly the peak of the stellar emission for a TP-AGB star,
the most opaque dust species produced in CSEs of C-stars is amorphous carbon (amC).
The contribution of SiC is negligible in filters which do not include its characteristic feature at 11.3~$\mu m$, since the abundance of SiC is only few percents in mass in the Magellanic Clouds (MCs) \citep{Groenewegen07}. Moreover, the slope of the absorption coefficient of the SiC is similar to the one of amC, thus, a part from the 11.3~$\mu$m feature, it will show a spectral behavior not too different from amC \citep{Groenewegen98}.
Other dust species can be relevant in the MIR bands, as MgS, which produces a feature around 24~$\mu$m. Finally, the presence of iron dust is still a matter of debate, since iron does not produce features in the spectra, but can contribute to the total extinction and emission. However, this dust species is produced in much smaller amount than carbon dust in our C-stars models.
For these reasons, we decide to focus our analysis by only considering amC dust in our models.
Other dust species can be easily added in our dust model, as explained in \citet{Nanni13, Nanni14}.

\subsection{Underlying TP-AGB models}
Stellar evolution is modeled from the pre-main sequence to the development of the first thermal pulse with the \textsc{parsec} code \citep{Bressanetal12}, while the TP-AGB phase is computed by the \textsc{colibri} code \citep{marigoetal13}.
The \textsc{colibri} code includes an accurate on-the-fly computation of the abundances for atomic and $\simeq$500 molecular species and opacities for the atoms and more than 20 molecules performed by the tool \textsc{\ae sopus} \citep{MarigoAringer_09}. This feature allows a consistent coupling between the envelope structure and variations of metal abundances in the stellar atmosphere produced by dredge-up episodes and by Hot Bottom Burning. A key quantity determining the stellar spectral type, opacity, molecular abundances and effective temperature, is the C/O ratio, the variations of which are followed by \textsc{colibri}.
Different dust species are formed according to the C/O ratio.
The treatment for the mass-loss rate has been recently revised by \citet{rosenfield14, Rosenfield16} by using the TP-AGB star counts and luminosity functions in a sample of galaxies from the ACS Nearby Galaxy Survey Treasury.

\subsection{Outflow structure}
The dust-driven wind is described by the spherically symmetric, stationary wind momentum equation.
Under these assumptions, the gas density profile of the outflow is given by the mass conservation equation:
\begin{equation}\label{dens}
\rho(r)=\frac{\dot{M}}{4\pi r^2 v_{\rm exp}},
\end{equation}
where $v_{\rm exp}$ is the expansion velocity and $r$ the distance from the center of the star.
As discussed in \citet{Marigo16}, the profile in Eq.~\ref{dens} is a good approximation for the description of density in the outer regions of dust-driven outflows.

The velocity profile is determined by the differential equation:
\begin{equation}\label{velocity}
v \frac{dv}{dt}=-\frac{G M}{r^2}(1-\Gamma),
\end{equation}
where,
\begin{equation}\label{gamma}
\Gamma=\frac{L}{4\pi c G M} \kappa,
\end{equation}
is the ratio between the radiation pressure and the gravitational pull of the star. The quantity $\kappa$ is the mean opacity of the medium, given by the contribution of gas and amC dust:
\begin{equation}\label{kappa}
\kappa=\kappa_{\rm gas}+ f_{\rm C} \cdot\kappa_{\rm amC},
\end{equation}
as described in \citet{Nanni13}, $\kappa_{\rm amC}$ is the opacity of amC dust computed for the maximum possible condensation of carbon dust. At each distance from the star, we assume dust to be formed by grains of the same size, which varies along the CSEs. We compute $\kappa$ consistently with this hypothesis. The quantity $f_{\rm C}$ is the condensation fraction, defined as the number of atoms of the key-species\footnote{The less abundant between the species available in the gas phase to form that type of dust.} condensed into dust grains over the total initially available. In case of amC in C-stars, the key-element is carbon.
The quantity $f_{\rm C}$ can be expressed as:
\begin{equation}\label{dcond}
 f_{\rm C}=\frac{4 {\rm \pi} (a_{\rm amC}^3-a_0^3)\rho_{\rm d, amC}}{3 m_{\rm amC} \epsilon_{\rm C}} \epsilon_{\rm s,C},
\end{equation}
where $m_{\rm amC}$ is the mass of the dust monomer, $a_{\rm amC}$ the actual grain size at a certain distance from the star, $a_0=10^{-7}$~cm the initial grain size, $\rho_{\rm d, amC}$ is the dust density and
$\epsilon_{\rm C}$, $\epsilon_{\rm s, C}$  are the number densities of the total carbon and initial seed nuclei,
normalized to the number density of hydrogen atoms.

The dust density profile is derived by Eq.~\ref{dens}:
\begin{equation}\label{d_dens}
\bar{\rho}_{\rm amC}(r)=\rho(r)\Psi_{\rm amC},
\end{equation}
where $\Psi_{\rm amC}$ is the dust-to-gas ratio, computed as:
\begin{equation}
 \Psi_{\rm amC}=\frac{X_{\rm C} f_{\rm C} m_{\rm amC}}{m_{\rm C}},
\end{equation}
where $X_{\rm C}$ is the mass fraction of the carbon in the atmosphere.

The gas temperature profile, $T_{\rm gas}$, is
described by a grey and spherically symmetric extended atmosphere \citep{Lucy71, Lucy76}
\begin{equation}\label{T_gas}
 T_{\rm gas}(r)^4=T_{\rm eff}^4 \left[W(r)+\frac{3}{4}\tau_{\rm L}\right],
\end{equation}
where $W(r)$ is given by:
\begin{equation}\label{dilution}
W(r) = \frac{1}{2}\left[1-\sqrt{1-\left(\frac{R_*}{r}\right)^2}\right],
\end{equation}
and the optical depth $\tau_{\rm L}$ is provided by differential equation
\begin{equation}\label{dtaudr}
 \frac{d \tau_L}{d r}=-\rho \kappa \frac{R_{*}^{2}}{r^2},
\end{equation}
with the boundary condition
\begin{equation}\label{taufin}
 \lim_{r\rightarrow\infty}\tau_L=0.
\end{equation}

The dust temperature, $T_{\rm dust}$, is computed from the energy balance between the absorbed and emitted radiation by dust grains, under the optically thin approximation:
\begin{equation}\label{T_dust}
\sigma T_{\rm eff}^4 Q_{\rm abs, P} (a_{\rm amC}, T_{\rm eff}) W(r)=\sigma T_{\rm dust}^4 Q_{\rm abs, P} (a_{\rm amC}, T_{\rm dust}),
\end{equation}
where $W(r)$ is the provided by Eq.~\ref{dilution}.

The quantity $Q_{\rm abs, P}$ is the absorption coefficient as a function of the grain size. In particular, the quantity $Q_{\rm abs, P} (a_{\rm amC}, T_{\rm eff})$ is weighted
with the stellar spectrum at $T_{\rm eff}$.

The spectrum is obtained by interpolating
dust-free spectra of the \textsc{comarcs} grid \citep{Aringer09, Aringer16} in  $T_{\rm eff}$ and carbon excess. The quantity $Q_{\rm abs, P}(a_{\rm amC}, T_{\rm dust})$ is computed through the Planck average performed with a Black Body at the dust temperature, $T_{\rm dust}$.

The quantities $Q_{\rm abs}(a_{\rm amC}, \lambda)$ and $Q_{\rm sca}(a_{\rm amC}, \lambda)$, from which $\kappa_{\rm amC}$ and $Q_{\rm abs, P}$ are calculated, have been pre-computed for a grid of spherical grains of different sizes using the Mie code \textsc{BHMIE} by \citet{Bohren83} starting from the $n, k$ optical constants.

\subsection{Growth of carbon dust}
Amorphous carbon dust is accreted by addition of C$_2$H$_2$ molecules in the gas phase onto the starting seed particles.

The dust growth is described by the balance between the growth and the decomposition rates for amC dust:
\begin{equation}\label{dadt}
\frac{da_{\rm amC}}{dt}=J^{\rm gr}_{\rm amC}-J^{\rm dec}_{\rm amC}.
\end{equation}

The growth rate is computed by taking into account only the efficient collisions of the C$_2$H$_2$ molecules impinging on the grain surface:
\begin{equation}
J^{\rm gr}_{\rm amC, C_2H_2}=2\cdot\alpha n_{\rm C_2H_2} v_{\rm th, C_2H_2},
\end{equation}
where $\alpha$ is the sticking coefficient which equals 1 in this case, $n_{\rm C_2H_2}$ is the number density of C$_2$H$_2$ in the gas phase, and $v_{\rm th, C_2H_2}$ is their thermal velocity. The factor 2 in the equation takes into account that for each molecule of C$_2$H$_2$ sticking on the grain surface, two atoms of carbon dust are formed.

The destruction term $J^{\rm dec}_{\rm amC}$ depends on the temperature and pressure conditions in the CSE,
and it is in general provided by the superimposition of free evaporation of dust grains due to stellar heating and chemisputtering.
This latter term is the destruction rate given by the inverse reaction between H$_2$ molecules in the gas and the grain surface.

In C-stars, the chemisputtering term is assumed to be negligible and the scheme proposed by \citet{Cherchneff92} is usually followed \citep{FG06, ventura12, Nanni13, Nanni14}. In such a framework, carbon dust can accrete only below a certain threshold gas temperature, $T_{\rm gas}=1100$~K, where PAHs can efficiently start to grow, initiating the nucleation and accretion processes \citep{Frenklach89}.

Below this threshold for the gas temperature, sublimation due to dust heating might still be at work. We compute the sublimation rate following the prescription by \citet{Kimura02, Kobayashi09, Kobayashi11}:
\begin{equation}
J^{\rm dec}_{\rm amC}=\alpha v_{\rm th, C}(T_{\rm dust})\frac{p(T_{\rm dust})}{k_B T_{\rm dust}},
\end{equation}
where $p(T_{\rm dust})$ is the saturated vapor pressure at the dust temperature $T_{\rm dust}$ and $v_{\rm th, C}$ is the thermal velocity of the species which evaporates from the dust grain, which is carbon in this case. In the models presented in the following sections, the sublimation process is usually not at work when $T_{\rm gas}$ reaches the threshold temperature of 1100~K.
Before to start growing grains, $T_{\rm dust}$ is computed through Eq.~\ref{T_dust} assuming a grain size of $a_{\rm amC}=a_0=10^{-7}$~cm.

From Eq.~\ref{dadt}, we define the condensation radius, $R_{\rm cond, amC}$, as the distance from the star at which $J^{\rm gr}_{\rm amC}= J^{\rm dec}_{\rm amC}$.

The number of seeds is assumed to be proportional to the carbon excess, $(\epsilon_{\rm C}-\epsilon_{\rm O})$.
By doing so, we assume that nucleation in carbon rich environments is started by C$_2$H$_2$ molecules which form large PAHs \citep{Cherchneff06, Mattsson10}.
We define $\epsilon_{\rm s, C}$ as done in \citet{Nanni13}:
\begin{equation}\label{ns_cex}
\epsilon_{\rm s, C}=\epsilon_{\rm s}\frac{(\epsilon_{\rm C}-\epsilon_{\rm O})}{(\epsilon_{\rm C}-\epsilon_{\rm O})_{\odot}},
\end{equation}
where the quantity $\epsilon_{\rm s}$ is an adjustable model parameter, which determines the final size of the dust grains, as we will discuss in the following.

\subsection{Radiative transfer with MoD}
\label{RT}
The emerging spectra are provided by photospheric spectra reprocessed by dust.

We perform the radiative transfer calculations by means of the code More of \textsc{dusty} \citep[][MoD]{Groenewegen12}, based on \textsc{DUSTY} \citep{Ivezic97}.

The input quantities needed by MoD are: a) the photospheric dust-free spectrum, b) the optical depth at a given wavelength, $\lambda=1$~$\mu$m, that is
situated close to the stellar emission peak ($\tau_1$),  c) dust absorption and scattering coefficients, $Q_{\rm abs}(a_{\rm amC}, \lambda)$ and $Q_{\rm sca}(a_{\rm amC}, \lambda)$, which are a function of the final grain size, d) the dust temperature at the boundary of the dust formation zone, $T_{\rm inn}$ e) the dust density profile.

The code MoD can be either used as an independent routine or be connected with the dust formation model described in the previous section. In the former case the input quantities are specified by the user, while, in the latter, they are provided by the output of our dust models.

In case MoD is connected with our dust scheme, the input quantities are computed as follows.
\begin{enumerate}[label=\alph*)]
\item The photospheric spectra are interpolated in $T_{\rm eff}$ and carbon excess inside the new grid of updated \textsc{comarcs} spectra \citep{Aringer16}.
\item The optical depth at 1~$\mu$m, $\tau_1$, is calculated through the following relation:
\begin{equation}\label{tau}
\tau_1=\int_{r_{\rm c}}^{\infty} \pi a_{\rm amC}^3 \frac{Q_{\rm ext}(a_{\rm amC}, 1\mu m)}{a_{\rm amC}} n_{\rm s, C} dr,
\end{equation}
where $n_{\rm s, C}$ is the number density of the seed nuclei, computed through $\epsilon_{\rm s, C}$ and,
\begin{equation}
Q_{\rm ext}(a_{\rm amC}, 1\mu m)= Q_{\rm abs}(a_{\rm amC}, 1\mu m)+(1-g) Q_{\rm sca}(a_{\rm amC}, 1 \mu m),
\end{equation}
where $g$ is defined as $g =<\cos\theta>$ where $\theta$ is the scattering angle. The quantity $(1-g) Q_{\rm sca}(a_{\rm amC}, \lambda)$ provides the degree of forward scattering.
From Eq.~\ref{tau} it is possible to see that $\tau_1$ is proportional to the total volume of dust formed.
Note that the quantities $Q_{\rm abs}(a_{\rm amC}, 1\mu m)$ and $Q_{\rm sca}(a_{\rm amC}, 1\mu m)$ are self-consistently computed with the final grain size obtained through our dust formation code.
\item  Since MoD only deals with a fixed grain size, we choose to compute the dust temperature for a fully grown grain. Since complete dust growth occurs typically within one stellar radius from $R_{\rm cond}$ in our dust models, we compute $T_{\rm inn}$ through Eq.~\ref{T_dust} at a distance $R_{\rm inn}=R_{\rm cond}+R_{\rm star}$.
However, we check that the calibration presented in this work remains valid if $T_{\rm inn}$ is computed at a distance $R_{\rm cond}$, assuming the grain to be fully grown.
\item  The dust density profile is computed through Eq.~\ref{d_dens}.
\end{enumerate}

The optical depth profile $\tau_\lambda$ is calculated once the quantities $\tau_1$, $Q_{\rm abs}$ and $Q_{\rm sca}$ are provided:
\begin{equation}\label{tau_lambda}
\tau_\lambda=\tau_1\frac{Q_{\rm ext} (a_{\rm amC},\lambda)}{Q_{\rm ext} (a_{\rm amC}, 1\mu m)}.
\end{equation}
Note that a variation in $Q_{\rm ext}$ produces a variation of $\tau_\lambda$, but the value of $\tau_1$ remains fixed in the treatment of MoD.

\section{Preliminary RT calculations: dependence of the emerging spectra on the main dust parameters}
\label{results}
We investigate the dependence of the emerging spectra on the main dust parameters, such as the grain size and the optical data set for amC, as a function of $\tau_1$.
In these computations we do not perform the calculations for the complete dust growth description, but we only employ the RT code described in \ref{RT}, varying the input parameters $\tau_1$, the grain size and optical data set.
Since the typical grain size derived from the best fit of the SEDs in different bands is around $0.1$~$\mu$m for the C-stars in the MCs \citep{Groenewegen07}, we select a grid of grains of $0.05 \le a_{\rm amC}\le 0.4$~$\mu$m, in order to include small to large grains. We assume that dust is formed by grains of the same size, rather than by a distribution of grains.
We keep $T_{\rm inn}=1000$~K fixed and we select an input spectrum with $T_{\rm eff}=3000$~K, C/O~$\sim 1.4$ and an initial set of metallicity typical of the SMC.

We perform two different tests:
\begin{enumerate}
\item We study a set RT models for $10^{-8}\le\tau_1\le 10$ in a grid of grains of $0.05 \le a_{\rm amC}\le 0.4$~$\mu$m. The choice for the range of $\tau_1$ includes stars from almost dust-free ($\tau_1=10^{-8}$) to heavily dust-enshrouded ($\tau_1=10$).
The optical data set assumed for this investigation is taken from \citet{Rouleau91}.

\item We compute a set of models for heavily dust-enshrouded CSEs ($\tau_1=10$), adopting the most commonly used data sets for amC available in the literature, listed in Table~\ref{Table:opacity}. The data sets considered have also been discussed in \citet{Andersen99}. Also for this set of models we perform the computations for a grid of grains of $0.05 \le a_{\rm amC}\le 0.4$~$\mu$m.
\end{enumerate}

\begin{table}
\begin{center}
\caption{Optical data sets for amC available in the literature. The different optical data sets are described in Section~\ref{amCdatasets}.}
\label{opacity_sets}
\label{Table:opacity}
\begin{tabular}{l c c}
\hline
 Designation & $\rho_{\rm d, amC}$ [g/cm$^3$] &  Reference \\
\hline
Jaeger400 & 1.435  &  \cite{Jaeger98} \\
Jaeger600 & 1.670  &  \cite{Jaeger98} \\
Jaeger800 & 1.843  &  \cite{Jaeger98} \\
Jaeger1000 & 1.988  &  \cite{Jaeger98} \\
Zubko1 & 1.87  & \cite{Zubko96}  \\
Zubko2 & 1.87  & \cite{Zubko96}  \\
Zubko3 & 1.87  & \cite{Zubko96}  \\
Rouleau & 1.85  & \cite{Rouleau91}  \\
Hanner & 1.85  & \cite{Hanner88} \\
\hline
\end{tabular}
\end{center}
\end{table}

\subsection{Effect of changing the grain size}
\label{dusty:grain_size}
\begin{figure}
\includegraphics[angle=90, width=0.48\textwidth]{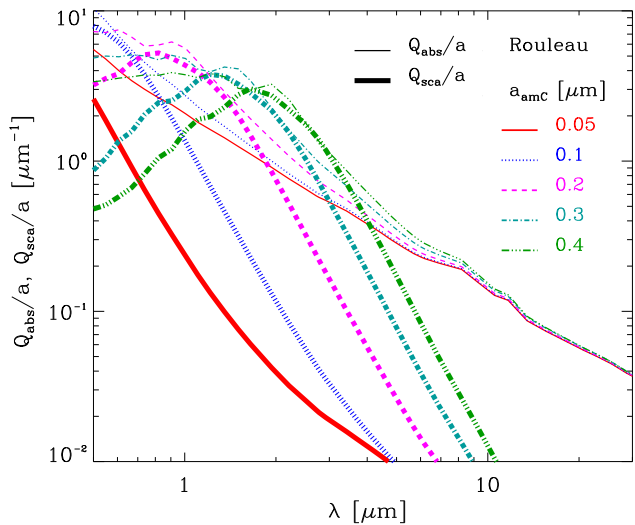}
        \caption{$Q_{\rm abs}(a_{\rm amC}, \lambda)$ (thin lines) and $Q_{\rm sca}(a_{\rm amC}, \lambda)$ (thick lines) as a function of wavelength for different grain sizes, listed in the figure.}
        \label{Qabs_sca}
        \end{figure}

We analyze the effect produced on spectra and colors by only changing the dust grain size.
We start by analyzing the optical properties as a function of the grain size, $Q_{\rm abs}(a_{\rm amC}, \lambda)/a$ and $Q_{\rm sca}(a_{\rm amC}, \lambda)/a$, as plotted in Fig.~\ref{Qabs_sca}.

The optical properties show a strikingly different behavior between small grains, $a_{\rm amC}\leq 0.1$~$\mu$m, and large grains $a_{\rm amC}\geq 0.2$~$\mu$m.
In particular, for the smaller grains considered ($a_{\rm amC}=0.05, 0.1$~$\mu$m) the scattering term is almost negligible for $\lambda\ga0.6$~$\mu$m and the absorption always dominates the total extinction.
For larger grains the scattering term is comparable with the absorption one around $\lambda=1-2$~$\mu$m -- where the exact value depends on the specific grain size considered-- and then it decreases for longer wavelengths.
Around $\lambda\sim2$~$\mu$m, corresponding to the Ks band, $Q_{\rm abs}/a$ and $Q_{\rm sca}/a$ increase with the grain size in the studied range.

\begin{figure}
        \includegraphics[angle=90, width=0.48\textwidth]{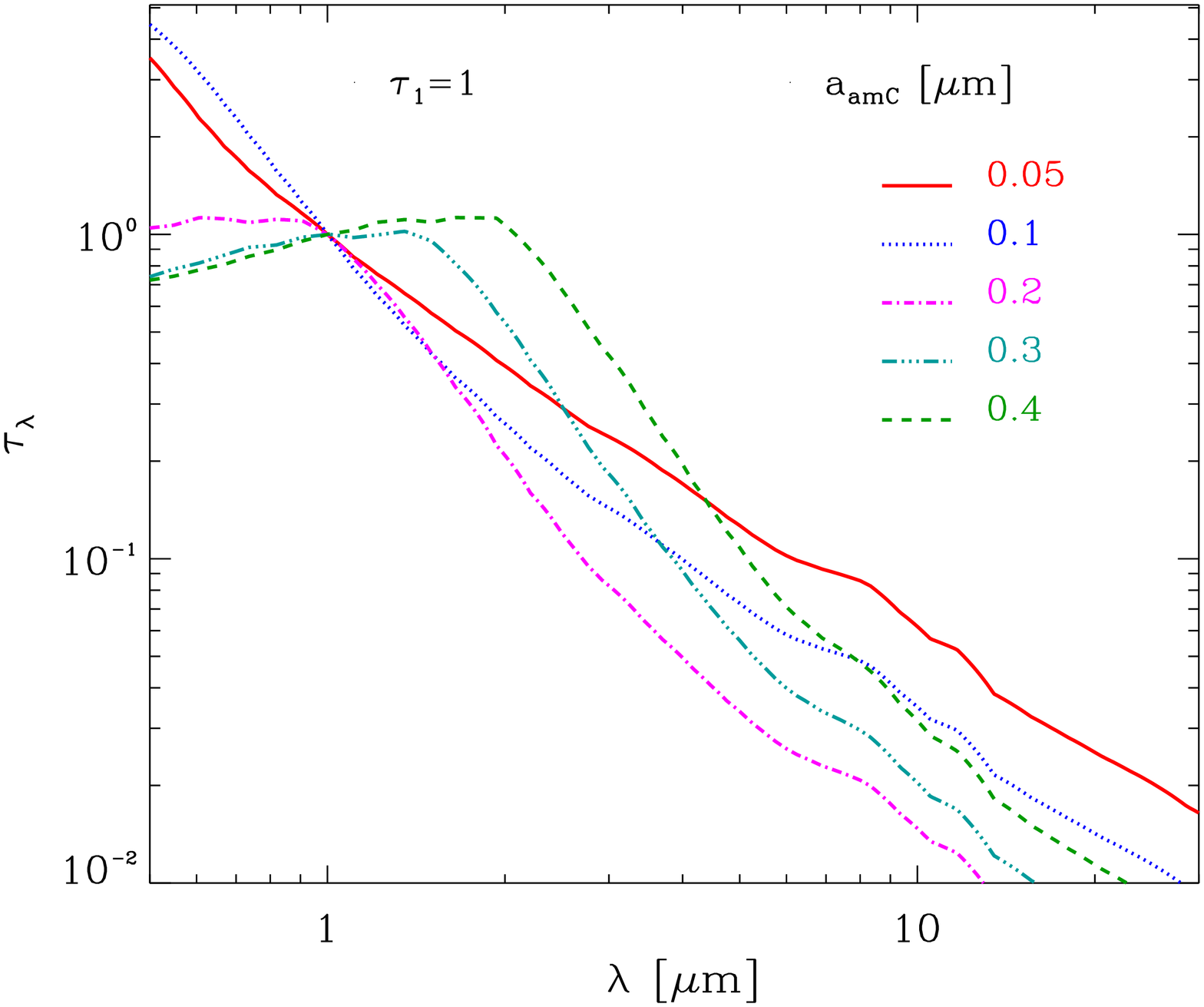}
        \includegraphics[angle=90, width=0.48\textwidth]{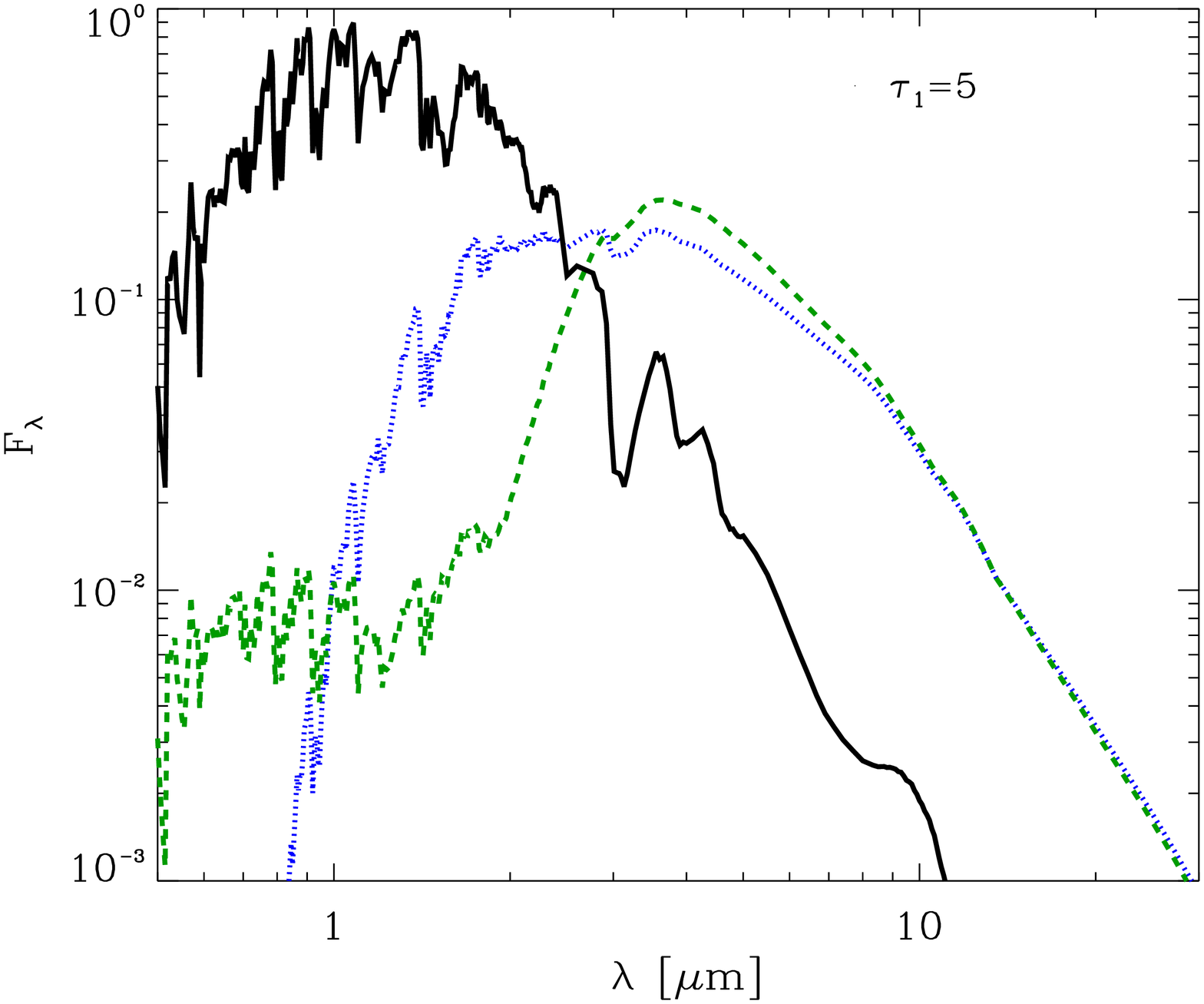}
        \includegraphics[angle=90, width=0.48\textwidth]{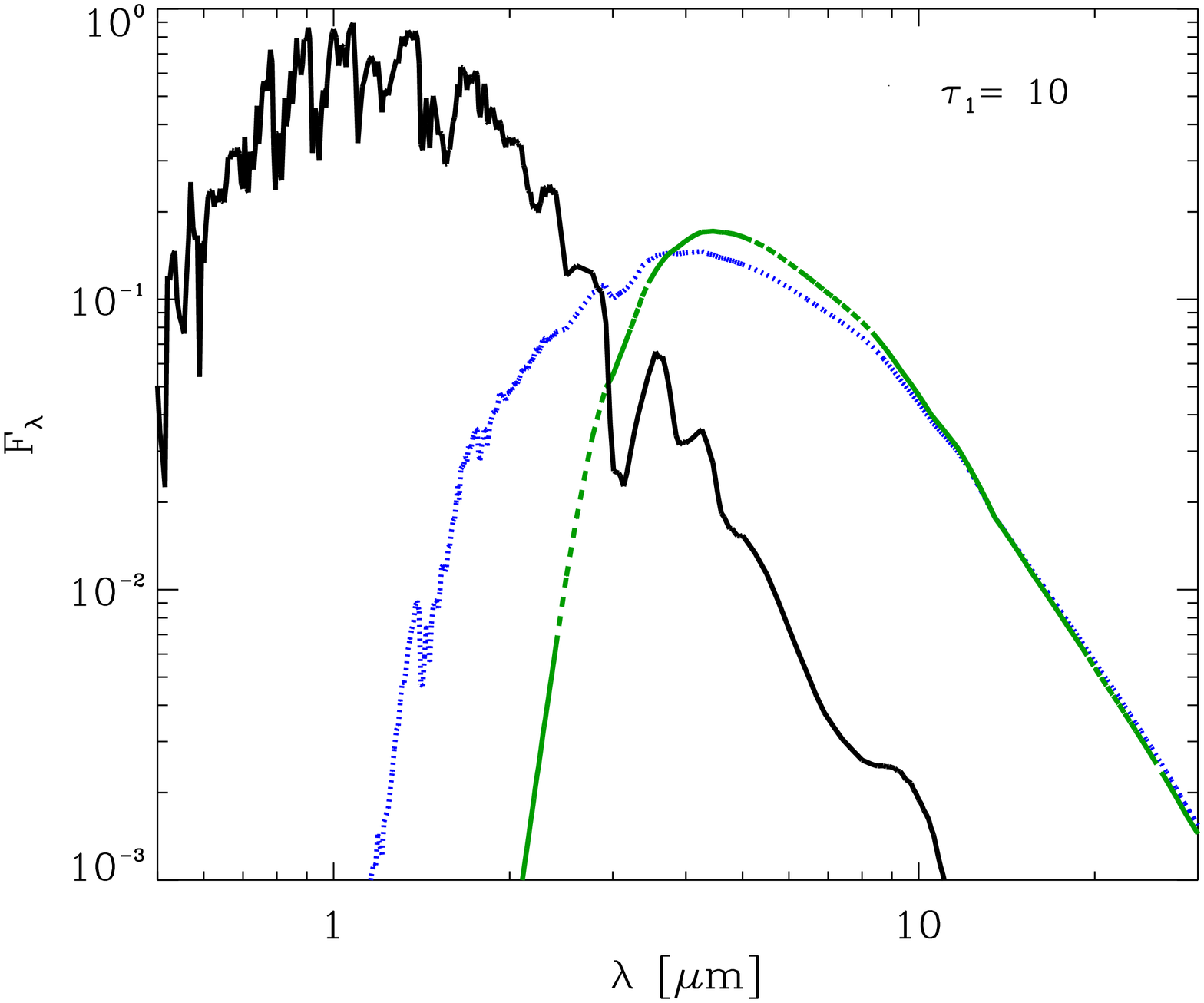}
        \caption{$\tau_\lambda$ profile (Eq.~\ref{tau_lambda}) for $\tau_1=1$ for the grain sizes listed in the top right (top panel). Dust-free spectrum (black line) superimposed with dusty spectra for $\tau_1=5, 10$ (middle and lower panels, respectively) for different choices of the grain size. All the spectra are normalized for the integrated flux.}
        \label{optical_prop}
        \end{figure}

We now study how the shape of the emerging spectra are modified with the grain size, keeping $\tau_1$ constant. We select two values of $\tau_1=5, 10$ corresponding to  dust-enshrouded CSEs.
The choice of keeping $\tau_1$ constant means that the total extinction around the peak of the stellar radiation is the same for all the models considered.
From Eq.~\ref{tau}, we see that changing the optical constants keeping $\tau_1$ constant, also implies that the total volume of dust produced changes.
The results are shown in Fig.~\ref{optical_prop}, where the extinction profiles, $\tau_\lambda$ (top panel) and the spectra for $\tau_1=5, 10$ (middle and bottom panels) are shown for different choices of the grain size.
For sake of clarity we only plot the spectra obtained with $a=0.1, 0.4$~$\mu$m, representative of the behavior of small and large grains, respectively.
For comparison, the dust-enshrouded spectra are plotted together with the corresponding dust-free spectrum.

From the top panel we clearly see that, by modifying the grain size, $\tau_\lambda$ changes according to Eq.~\ref{tau_lambda}, with $\tau_1$ fixed. Here we show the case $\tau_1=1$. The shape of $\tau_\lambda$, however, only depends on the optical properties, as expressed by Eq.~\ref{tau_lambda}. By changing the normalization of $\tau_1$, the curves are all shifted accordingly.
The quantity $\tau_\lambda$ is much steeper for $a_{\rm amC}\leq 0.1$~$\mu$m than for $a_{\rm amC}\geq 0.2$~$\mu$m. In particular, for larger grains, $\tau_\lambda$ shows a much flatter profile than for the smaller grains up to $\lambda\ga1-2$~$\mu$m.

The shape of $\tau_\lambda$ explains why, for $\lambda\la 1$ $\mu$m and $\tau_1=5$, the spectrum is more extincted for $a=0.1$~$\mu$m than for $a_{\rm amC}=0.4$~$\mu$m. Moreover, due to the plateau in $\tau_\lambda$, only present for large grains, the spectrum is more uniformly extincted up to $\lambda\sim2$~$\mu$m for $a_{\rm amC}=0.4$ than for $a_{\rm amC}=0.1$~$\mu$m.
For the same reason, in the most dust-enshrouded model ($\tau_1=10$) the spectrum produced by large grains is heavily extincted up to $\lambda \sim 2$~$\mu$m in contrast to the smaller grain for which extinction is relevant only for $\lambda\la 1$~$\mu$m.

The part of the spectrum affected by dust emission is influenced by the amount of absorbed stellar light and by dust emission properties.
For the choices of $\tau_1$ presented, dust emission is relevant for $\lambda \ga 3$~$\mu$m.
For $\tau_1=5$ dust emission around $3\la\lambda\la5$~$\mu$m is more efficient for $a=0.4$~$\mu$m than for $a_{\rm amC}=0.1$~$\mu$m.
For the case with $\tau_1=10$ and $a_{\rm amC}=0.4$~$\mu$m extinction is relevant in the Ks band, since $\tau_\lambda$ in this band is almost as relevant as in the J band.

\begin{figure}
        \includegraphics[angle=90, width=0.48\textwidth]{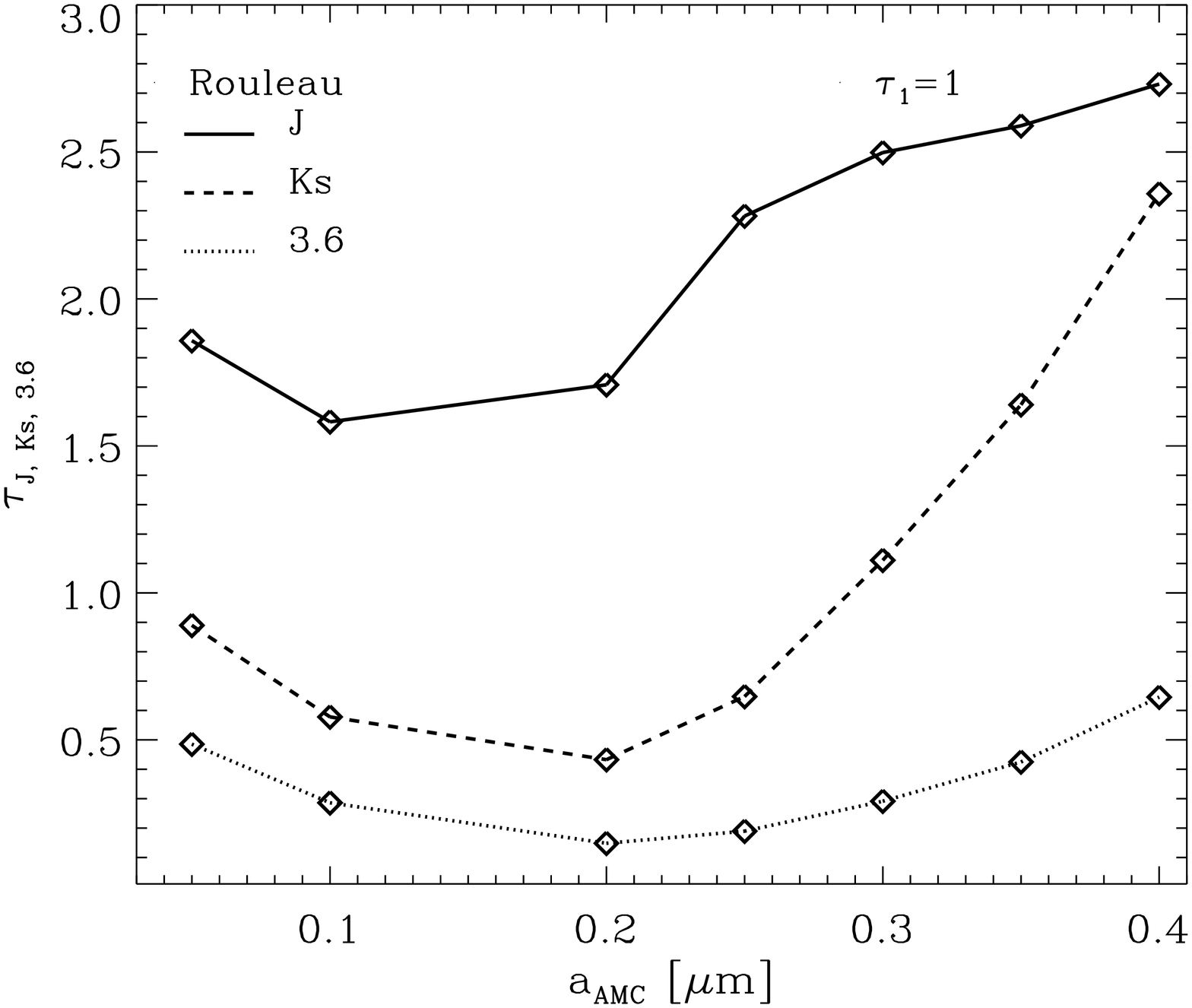}
        \includegraphics[angle=90, width=0.48\textwidth]{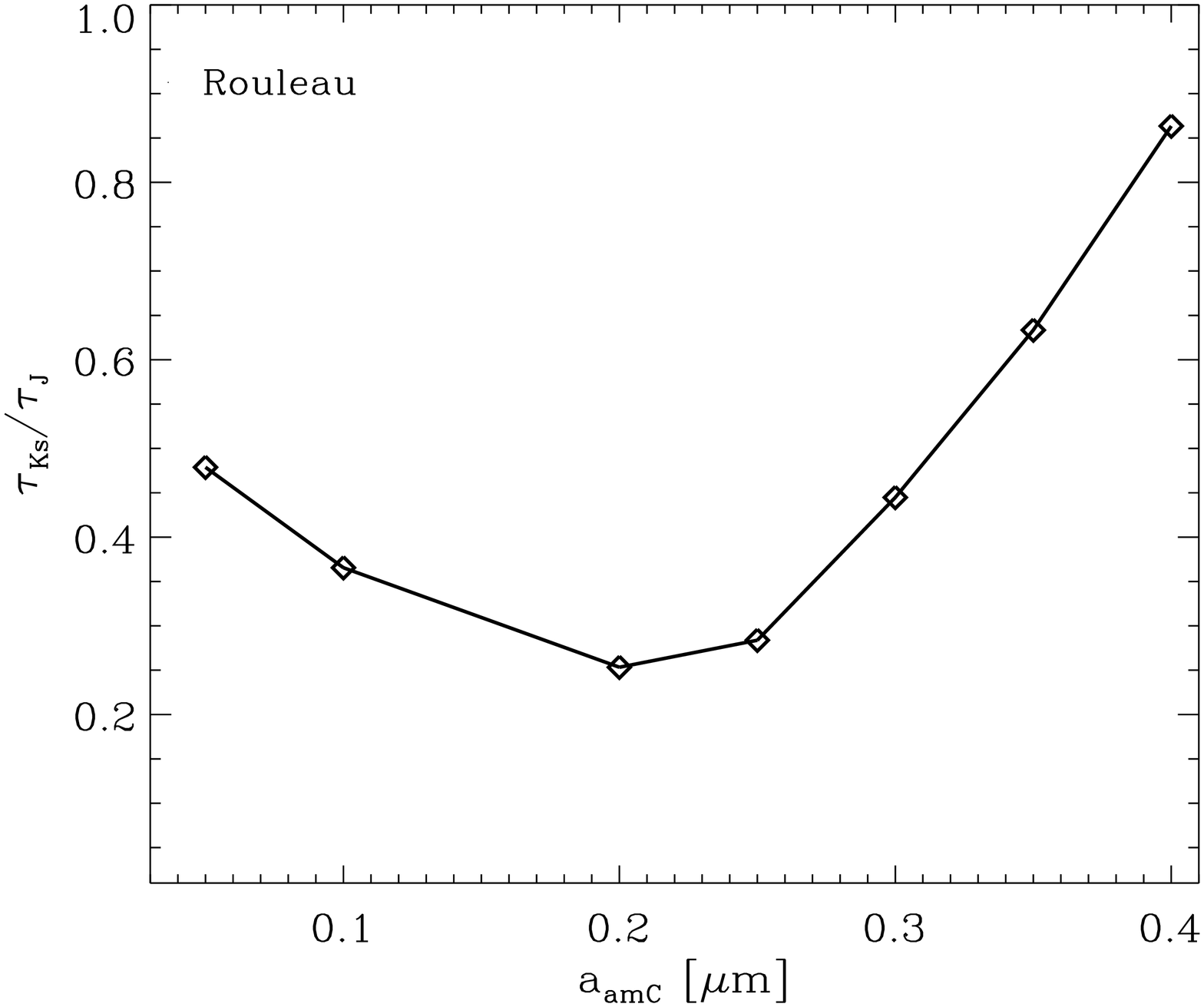}
        \caption{$\tau_{\rm J}$, $\tau_{\rm Ks}$, $\tau_{\rm 3.6}$ (upper panel) and $\tau_{\rm Ks}/\tau_{\rm J}$ (lower panel) as a function of the grain size.}
        \label{tj_tK_rouleau}
        \end{figure}

In the upper panel of Fig.~\ref{tj_tK_rouleau} we show the behavior of $\tau_{\rm J}$, $\tau_{\rm Ks}$ and $\tau_{\rm 3.6}$ as a function of the grain size. We select the case with $\tau_1=1$.
The quantities $\tau_{\rm J}$, $\tau_{\rm Ks}$ and $\tau_{3.6}$ do not behave monotonically with the grain size for all the bands considered. The quantities $\tau_{\rm Ks}$ and $\tau_{3.6}$ show a minimum for $a_{\rm amC}=0.2$~$\mu$m, while for $\tau_{\rm J}$ the minimum is around $a_{\rm amC}=0.1$~$\mu$m.
For highly dust enshrouded CSEs, i. e. $\tau_1=10$, the extinction can be relevant also around $\lambda=3.6$~$\mu$m, since in this case $\tau_{\rm 3.6}\sim 5$ for some combination of grain sizes.

In the lower panel of Fig.~\ref{tj_tK_rouleau}, we show the relative extinction in the Ks and J bands, $\tau_{\rm Ks}/\tau_{\rm J}$. Such a ratio only depends on the optical properties of grains and do not change with $\tau_1$ (Eq.~\ref{tau_lambda}).
This figure shows that $\tau_{\rm Ks}$ becomes comparable to $\tau_{\rm J}$ for large grain sizes.
This behavior can be understood by looking at the change in the optical properties with the grain size shown in Fig.~\ref{Qabs_sca}.

\begin{figure}
        \includegraphics[angle=90, width=0.48\textwidth]{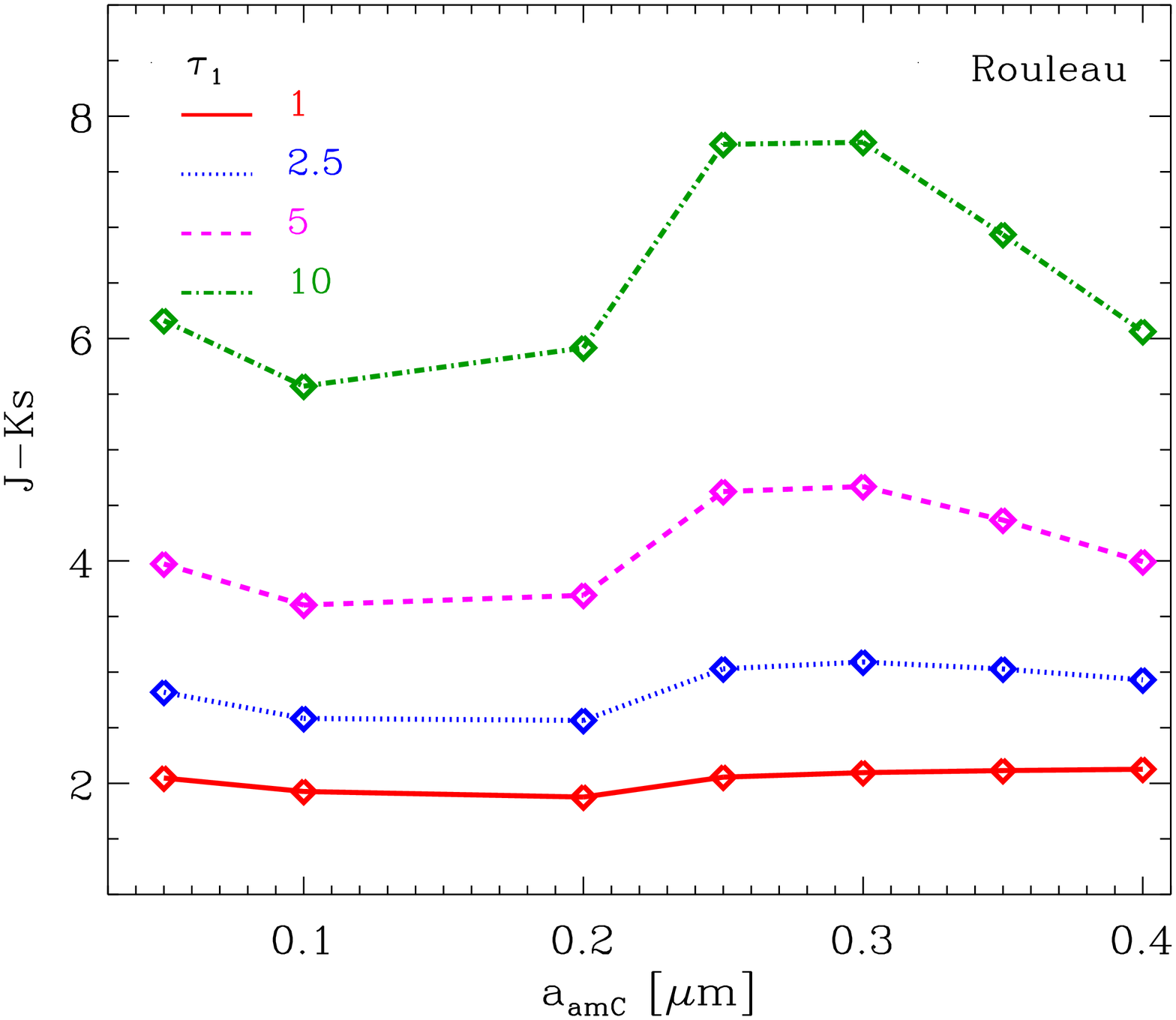}
        \includegraphics[angle=90, width=0.48\textwidth]{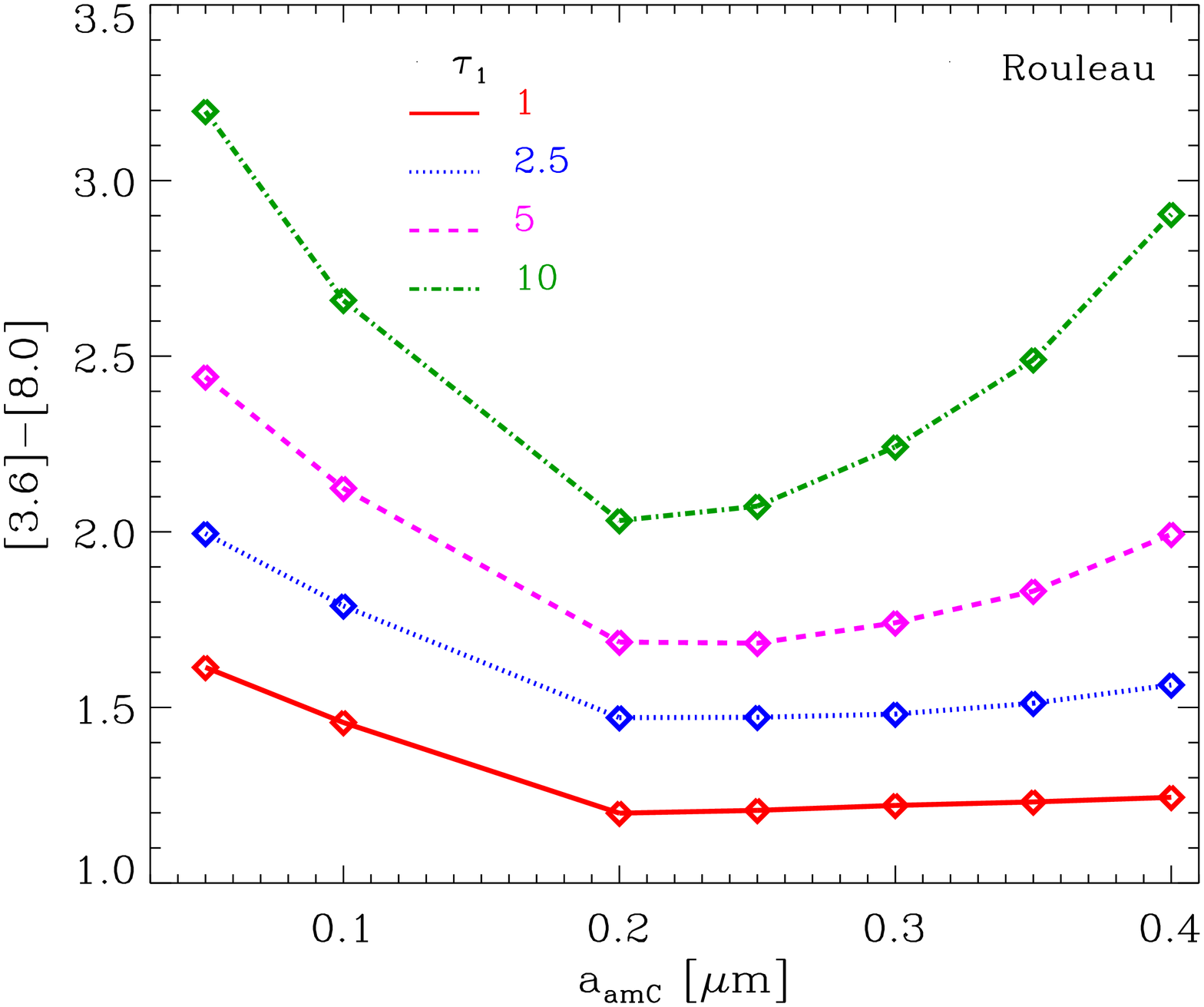}
        \caption{J~$-$~Ks (upper panel) and [3.6]~$-$~[8.0] (lower panel) as a function of the grain size for different values of $\tau_1$, listed in top left. We adopt $T_{\rm inn}=1000$~K, a spectrum for a C-star with $T_{\rm eff}=3000$~K and C/O~$\sim 1.4$, and the optical data set by \citet{Rouleau91}.}
        \label{NIR_rouleau}
        \end{figure}

In Fig.~\ref{NIR_rouleau}, we plot J~$-$~Ks (upper panel) and [3.6]~$-$~[8.0] (lower panel) colors as a function of the grain size, for different choices of $\tau_1$.
The J~$-$~Ks and [3.6]~$-$~[8.0] colors show a weaker trend as a function of the grain size for smaller values of $\tau_1$.
The J~$-$~Ks color is produced by the combination of extinction of the stellar spectrum and emission properties of grains. The results are dependent on the choice of $\tau_1$.
As expected, for a fixed value of the grain size, the J~$-$~Ks increases for increasing values of $\tau_1$. However, for given value of $\tau_1$ this color does not change monotonically as a function of the grain size.
The trend of J~$-$~Ks reflects the dependence of $\tau_{\rm Ks}/\tau_{\rm J}$ on the grain size. In fact, for $0.05 <a_{\rm amC} < 0.25$~$\mu$m the models become redder with increasing grain sizes, because the extinction in the J band gets much larger than the one in the Ks band.
For $a_{\rm amC} \geq 0.25$~$\mu$m, the extinctions in these two bands are comparable, and the models become less red for increasing values of the grain size.
The variations produced only by changing the grain sizes can affect the J~$-$~Ks color considerably, with differences that can be up to two magnitudes in color for the most dust-enshrouded CSEs.

We now analyze the trend of the [3.6]~$-$~[8.0] color plotted in the lower panel of Fig.~\ref{NIR_rouleau}.
As previously mentioned, the most dust-enshrouded models ($\tau_1=5,10$), are dominated in the MIR colors by dust emission. Furthermore, for large values of $\tau_1$, the extinction at 3.6~$\mu$m, described by $\tau_{3.6}$, may become significant and reduce the total flux. As a consequence, the trend of the [3.6]~$-$~[8.0] color reflects the trend of $\tau_{3.6}$. In particular, $\tau_{3.6}$ shows a minimum around $a_{\rm amC}=0.2$~$\mu$m, which corresponds to less red [3.6]~$-$~[8.0] color. On the other hand, for larger values of $\tau_{3.6}$ at the two extremes of grain size range, redder [3.6]~$-$~[8.0] color are produced. For less dust enshrouded models ($\tau_1\leq 2.5$)  $\tau_{3.6}$ is not as large as in the previous cases and the emerging spectra are dependent on dust emission properties.
The variations produced in the [3.6]~$-$~[8.0] color by only changing the grain size can be up to one magnitude in the most extreme case.

We conclude that for the same value of $\tau_1$, variations in the grain size can produce a considerable change in the final colors presented. The differences among models computed with grain sizes are larger for more dust-enshrouded CSEs.

\subsection{Amorphous Carbon optical data sets}
\label{amCdatasets}

\begin{figure}
        \centering
        \includegraphics[angle=90, width=0.48\textwidth]{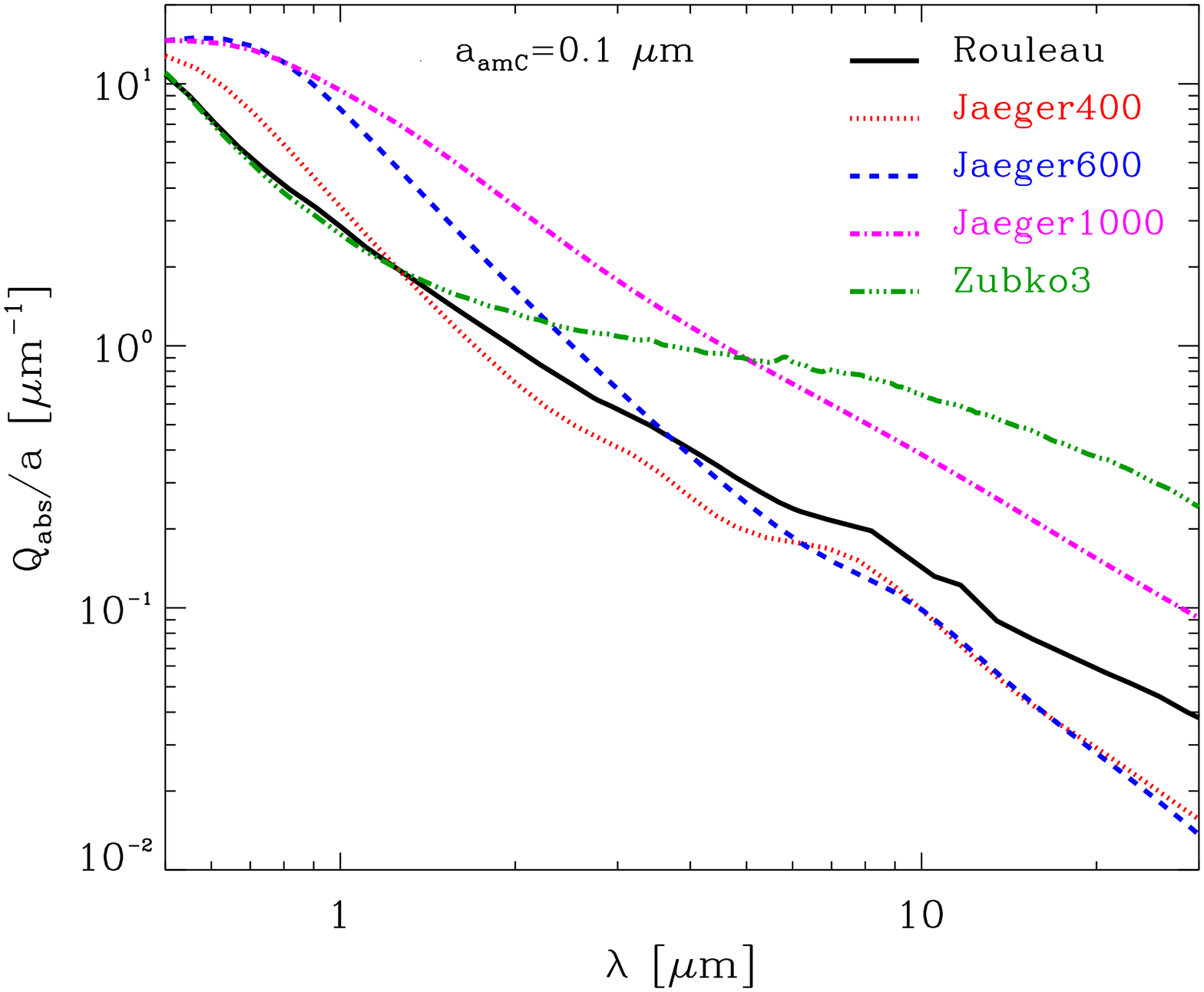}
        \includegraphics[angle=90, width=0.48\textwidth]{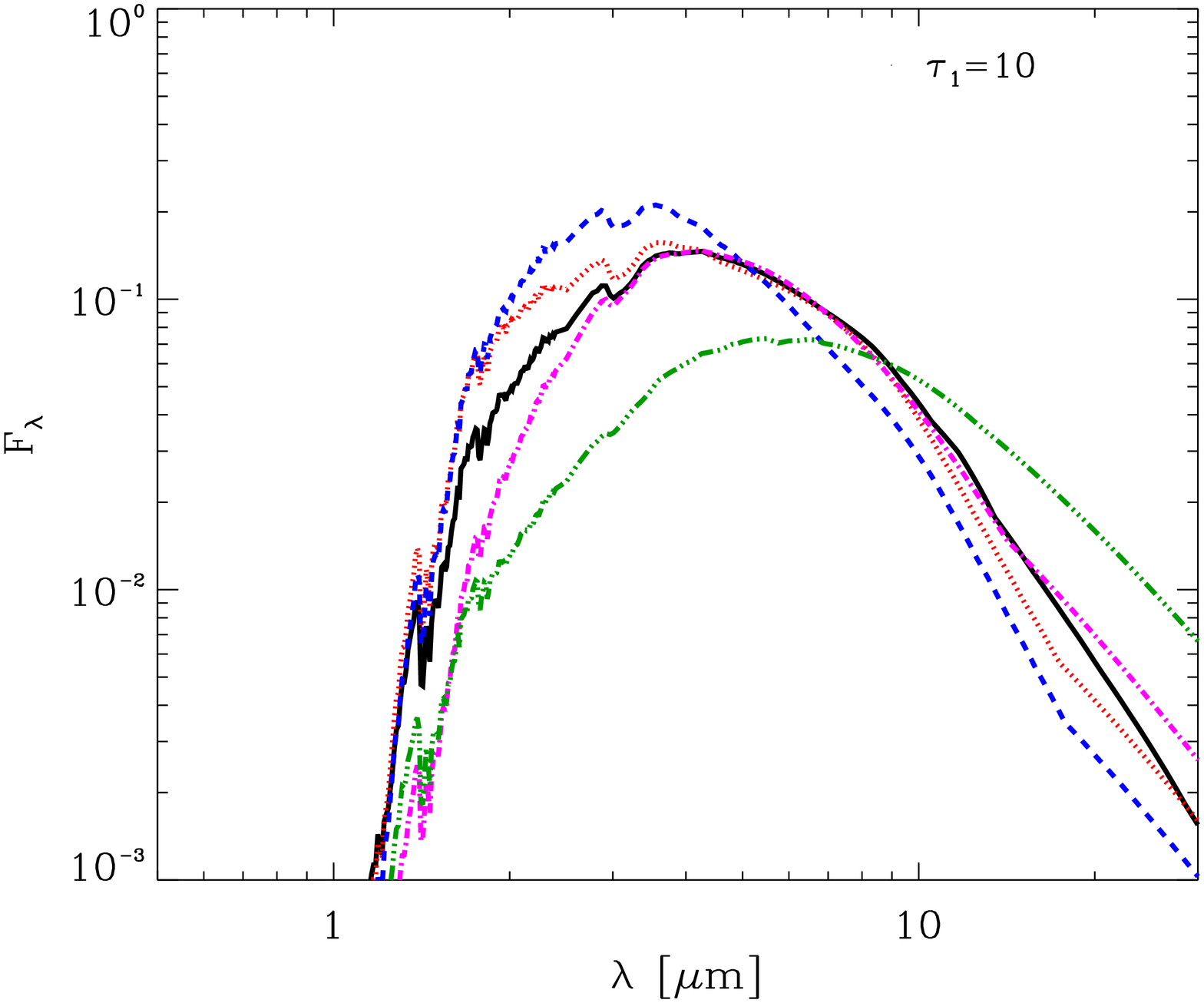}
        \caption{$Q_{\rm abs}/a$ for few selected data sets of amC listed in Table \ref{opacity_sets} for $a_{\rm amC}=0.1$~$\mu$m (upper panel), and the corresponding emerging spectra computed for $\tau_1=10$ (lower panel).}
        \label{opacity_comp}
        \end{figure}

We explore the differences arising in the spectra produced by employing different optical constants for amC dust.

The optical constants have been measured in laboratory by means of different techniques.
Each of the experimental substrate employed to calculate the optical data constants is characterized by a certain density of the material listed in Table~\ref{Table:opacity}.
For an exhaustive analysis of these data sets we refer to \citet{Andersen99}.

Amorphous carbon dust is present in different possible structures ranging from diamond-like (sp$^3$ hybridization) to graphite-like (sp$^1$, sp$^2$ hybridization). Carbon dust closer to a diamond-like structure presents a lower sp$^2$/sp$^3$ ratio, while carbon dust closer to graphite shows an higher sp$^2$/sp$^3$ ratio.
Each of the samples of carbon dust synthesized in laboratory is characterized by a different sp$^2$/sp$^3$ ratio and optical constants.

An example of this is shown in \citet{Jaeger98}. In this investigation amC samples have been obtained by pyrolyzing cellulose material at different temperatures (400, 600, 800, 1000$^\circ$C) and embedding the resulting material in an epoxy resin.
Pyrolization at different temperatures results in different structures of amC dust grains embedded in the substrate.
The samples pyrolized at the highest temperatures ($\geq 800^\circ$C) are characterized by more diamond-like structures than samples pyrolized at lower temperatures ($\leq 600^\circ$C). Therefore the value of sp$^2$/sp$^3$ increases from 1000 to 400$^\circ$C.

Besides pyrolization, \citet{Zubko96} presented three samples of carbon dust, obtained with three different techniques: a) burning benzene in air in normal conditions (BE sample, denoted by Zubko1 in the paper), b) through arc discharge between amC electrodes in a controlled Ar atmosphere (ACAR sample, denoted by Zubko2), c) through the same technique and conditions as a) but in a H$_2$ gas (ACH2 sample, denoted by Zubko3).
The same technique used for producing the BE sample of \citet{Zubko96} was also employed to produce the sample studied by \citet{Rouleau91}.

Other optical constants largely employed in dust modeling along the TP-AGB phase \citep{Nanni13, Nanni14, Ventura14, Dellagli15a, Dellagli15b, Ventura16}, are the ones listed in \citet{Hanner88}. The author of this work adopted the amC optical constants measured in laboratory by \citet{Edoh83} to interpret the IR observations of comets Halley and Wilson.

In the upper panel of Fig.~\ref{opacity_comp} the quantity $Q_{\rm abs}/a$ of the optical data sets listed in Table~\ref{opacity_sets}, is shown for $a_{\rm amC}=0.1$~$\mu$m. The emerging spectra for the different data sets, computed for $\tau_1=10$, are plotted in the lower panel of the same figure.
Since the Jaeger1000, Jaeger800, Zubko1, Zubko2 and Hanner opacity sets exhibit similar absorption coefficients, for the sake of clarity we show only the results for Jaeger1000. Rouleau, Zubko3, Jaeger400 and Jaeger600 show very different absorption coefficients.

\begin{figure}
        \centering
        \includegraphics[angle=90, width=0.48\textwidth]{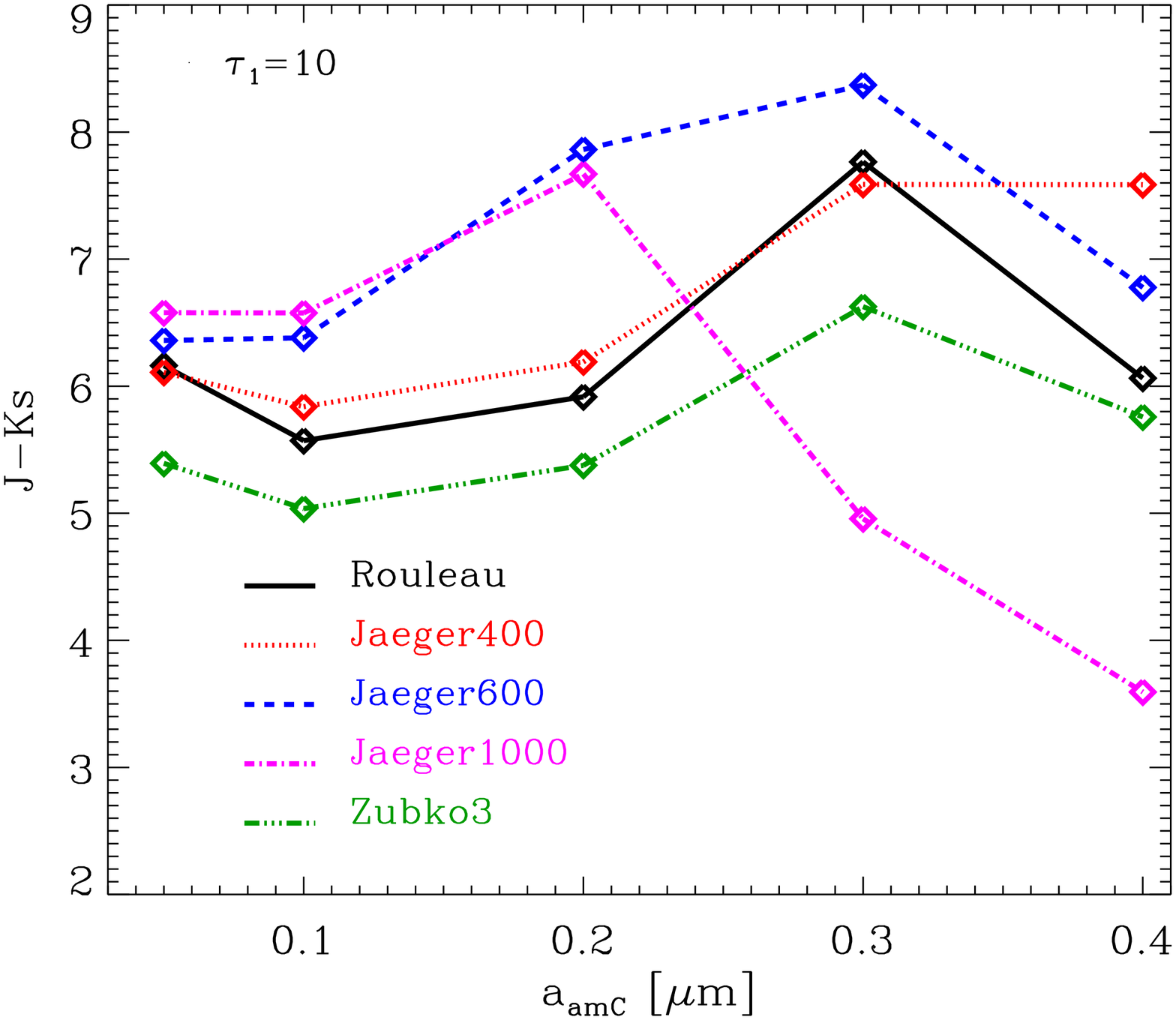}
        \includegraphics[angle=90, width=0.48\textwidth]{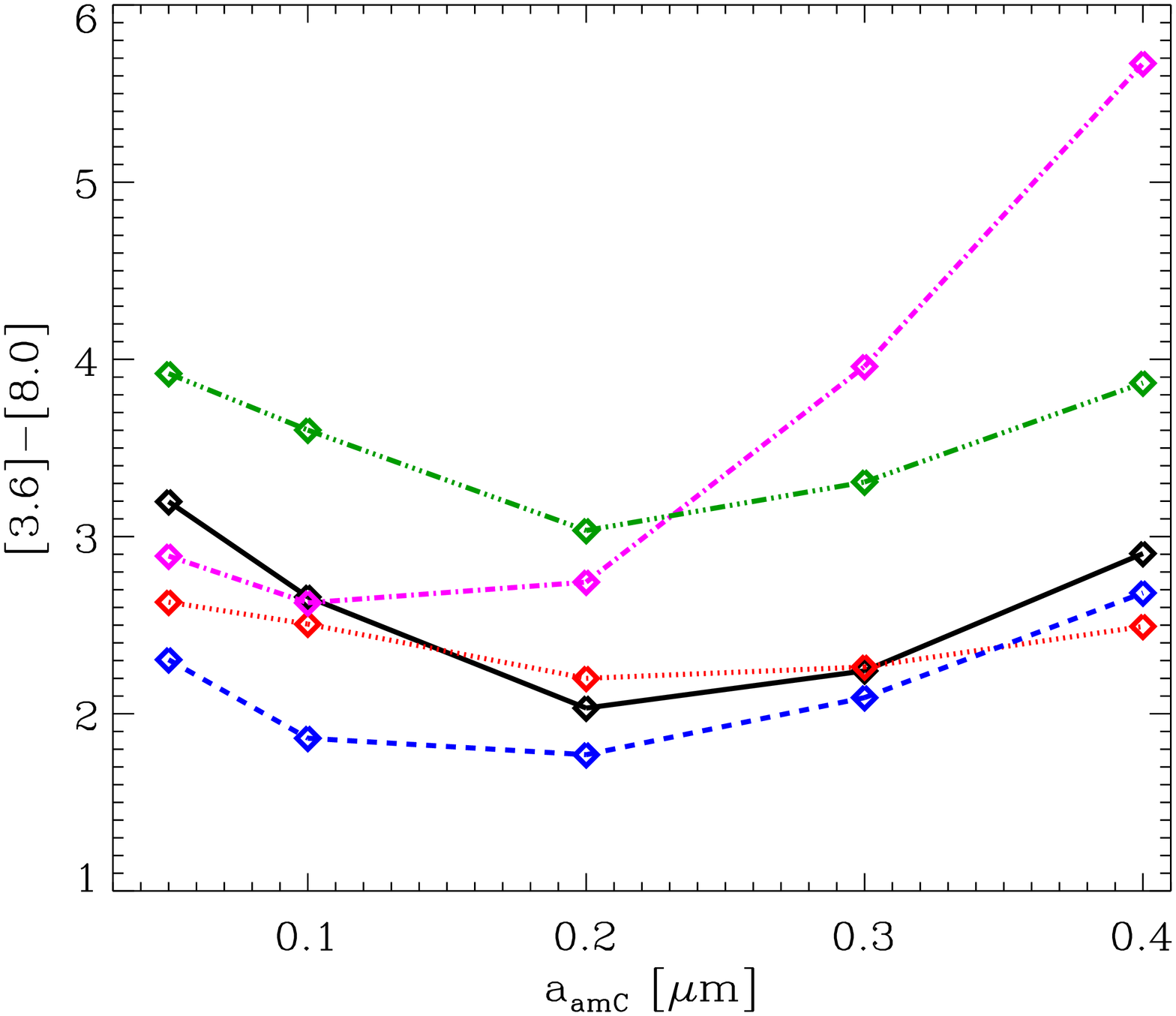}
        \caption{J~$-$~Ks (upper panel) and [3.6]~$-$~[8.0] (lower panel) colors as a function of the grain size for selected data sets of amC listed in Table~\ref{opacity_sets}. The models are computed for $\tau_1=10$.}
        \label{col_allops}
        \end{figure}

In Fig.~\ref{col_allops} J~$-$~Ks (upper panel) and [3.6]~$-$~[8.0] (lower panel) colors are plotted as a function of the grain size for heavily dust-enshrouded CSEs with $\tau_1=10$.
It is impressive how much the colors vary by changing only the optical data set.
In particular, the J~$-$~Ks color, obtained for the largest grain sizes considered, changes by about four magnitudes in color.
Furthermore, for the Jaeger1000 the trend between J~$-$~Ks and grain size shows quite a different behavior with respect to the other data sets.

The variations of the [3.6]~$-$~[8.0] color can be up to four magnitudes for the largest grain size considered. Similarly to the J~$-$~Ks color, Jaeger1000 results in a different trend between [3.6]~$-$~[8.0] and the grain size.

From this analysis we conclude that the variation of colors produced by changing only the optical data sets is remarkable.
Therefore, the choice of the carbon dust optical constants heavily affects the final modeled colors of dust-enshrouded C-stars.

Since there is no way to choose the best set of carbon dust optical constants before comparing the results with observations, we conclude that there is a urgent need for a systematic calibration of the optical properties in order to be able to reproduce all the observed colors in the SMC at the same time. For this calibration we employ some selected stellar tracks and we explore different optical data sets and model assumptions, as explained in the following section.

\section{Dust growth and radiative transfer along the TP-AGB tracks}
\label{effectonTPAGB}
In this Section we investigate the dust growth and RT for selected TP-AGB tracks developing a carbon phase, following the scheme described in Section~\ref{models}.
In a complete simulation of dust growth, the quantities $\tau_1$, $a_{\rm amC}$, $T_{\rm inn}$ and the input spectrum change accordingly to the TP-AGB evolution.
The tracks and the model assumptions are listed in Table~\ref{stars}. The stellar masses at the beginning of the TP-AGB phase and initial metallicity values are the typical ones of carbon stars in the SMC \citep{Marigo07}. Since infrared CCDs are critically shaped by the dust optical constants and grain sizes,
we expect our calibration to be only mildly dependent on the specific choice of the TP-AGB tracks (for reasonable model prescriptions), provided that large enough values of $\tau_1$ are reached in order to reproduce the reddest stars.
For a given optical data set and typical grain size, the fulfillment of this condition depends on the values of the mass-loss rates and carbon excess reached during the TP-AGB evolution. The models along the TP-AGB tracks are sampled in such a way that the variations of the actual stellar mass, mass-loss rate, luminosity, effective temperature and abundances of H $^{12}$C $^{16}$O are all below 0.5 per cent between two adjacent time-steps.

In the following we show the results obtained from our dust growth description applied to the TP-AGB tracks as far as $\tau_1$, $a_{\rm amC}$, and $T_{\rm inn}$ are concerned.

\begin{table}
\caption{TP-AGB tracks and model parameters adopted for the calibration. Models along the TP-AGB tracks are sampled
as explained in Section~\ref{effectonTPAGB}.}
\label{stars}
\centering
\begin{tabular}{l}
\hline
TP-AGB tracks    \\
\hline
Z=0.002; M=1.4, 1.6, 2 [M$_\odot$] \\
Z=0.004; M=2, 2.4 [M$_\odot$] \\
Z=0.006; M=3 [M$_\odot$] \\
\hline
$-15.4<\log(\epsilon_s)<-11$    \\
\hline
Data sets of optical constants in Table~\ref{Table:opacity}\\
\hline
\end{tabular}
\end{table}

\subsection{Typical grain sizes in TP-AGB models}

    \begin{figure}
    \includegraphics[angle=90, width=0.48\textwidth]{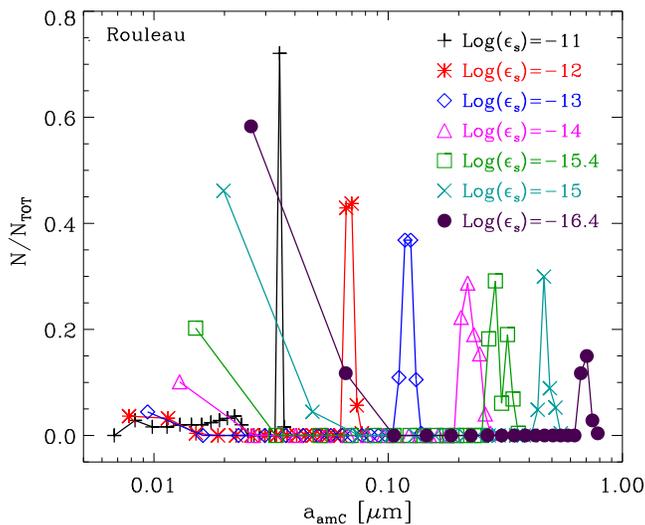}
        \caption{Normalized frequency of the grain size for different choices of the normalized number of seeds, $\epsilon_{\rm s}$, (Eq.~\ref{ns_cex}), listed in the top right. Each symbol along the lines includes the contribution of all the models sampled in the tracks in Table~\ref{stars}.}
        \label{Nrc}
        \end{figure}

            \begin{figure}
    \includegraphics[angle=90, width=0.48\textwidth]{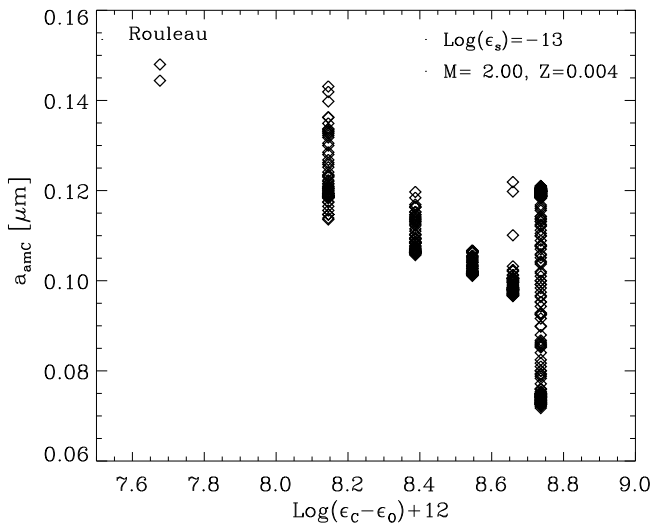}
        \caption{Final grain size as a function of the carbon excess for a selected TP-AGB track with $M=2$~M$_\odot$ and Z=0.004 computed with Rouleau data set and normalized number of seeds $\log(\epsilon_s)=-13$.}
        \label{avsco}
        \end{figure}

    \begin{figure}
    \includegraphics[angle=90, width=0.48\textwidth]{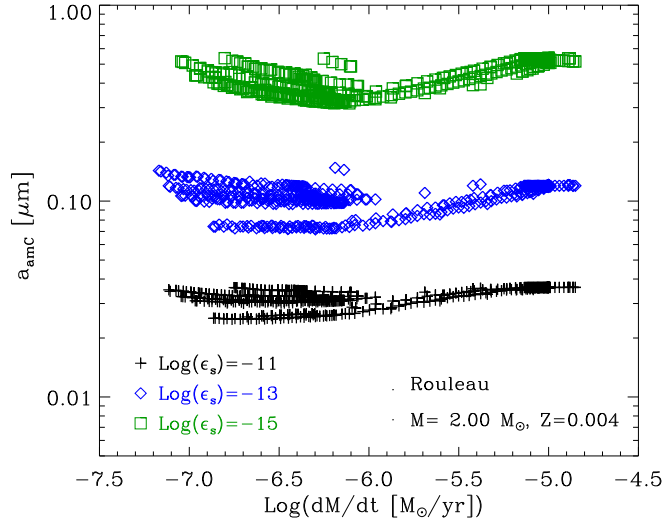}
        \caption{Final grain size as a function of the mass-loss rate for a selected TP-AGB track with $M=2$~M$_\odot$ and Z=0.004 computed with Rouleau data set and for different choices of the normalized number of seeds, $\epsilon_{\rm s}$, listed in the figure.}
        \label{avsmloss}
        \end{figure}

In Fig.~\ref{Nrc} we plot the normalized frequency of grain sizes obtained by applying our dust formation description to the TP-AGB tracks listed in Table~\ref{stars} for different choices of the normalized number of available seeds, $\epsilon_{\rm s}$, in Eq.~\ref{ns_cex}. The optical data set adopted is the one by \citet{Rouleau91}, but we check the results to be independent of the optical data set adopted.
All our simulations show a tail of small grains formed in passively expanding envelopes which fail to produce a dust-driven wind.
In these models dust is formed in CSEs expanding with constant velocity $v_{\rm exp}=4$~km~s$^{-1}$. Such models are nearly dust-free envelopes.

In dust-enshrouded models, where the wind is accelerated, the typical grain size increases for decreasing value of $\epsilon_{\rm s}$, from very small grains $a_{\rm amC}\sim 0.035$~$\mu$m for $\log(\epsilon_{\rm s})=-11$ to very large, $a_{\rm amC} \sim 0.7$~$\mu$m, for $\log(\epsilon_{\rm s})=-15.4$.
The distribution of grains shown in Fig.~\ref{Nrc} is determined by our choice of $\epsilon_{\rm s}$ and by assuming the number of seeds to be proportional to the carbon excess.

The trend found -- smaller grains for increasing values of $\epsilon_{\rm s}$-- can be understood
from Eqs.~\ref{dcond} and \ref{ns_cex} from which the relation between the grain size and the number of seeds is:
\begin{equation}\label{avseps}
a_{\rm amC}\propto\left[\frac{f_{\rm C}}{\epsilon_{\rm s} (\epsilon_{\rm C}-\epsilon_{\rm O})}\right]^{1/3},
\end{equation}

The maximum change of the final condensed fraction $f_{\rm C}$ obtained by changing the value of $\epsilon_{\rm s}$ is a factor four for models producing a dust-driven wind, whereas we vary $\epsilon_{\rm s}$ by order of magnitudes in the tests we performed.
As a consequence, for a given model with a certain carbon excess, the grain size is dominated by the choice of $\epsilon_{\rm s}$ rather than by the variation of the condensed fraction. The smaller $\epsilon_{\rm s}$ is, the larger we expect the grain size to be. In fact, by decreasing the number of the starting seeds, the molecules available in the gas phase accrete on a smaller number of particles, producing larger grains.
For example, by decreasing $\epsilon_{\rm s}$ in Eq.~\ref{avseps}, from $\log\epsilon_{\rm s}=-13$ to $\log\epsilon_{\rm s}=-15$, we expect the typical grain size to increase by a factor $100^{1/3}\sim 4.6$. According to Fig.~\ref{Nrc}, the trend recovered from our complete simulation and the simple estimate of Eq.~\ref{avseps} is in good agreement. Even if there are differences between the results obtained from the simple formula and the full calculation, the general trend is similar.

Since Eq.~\ref{dcond} shows that the fraction of carbon condensed into dust is proportional to $\epsilon_{\rm s}$, lower condensation fractions are obtained at the beginning of the condensation process, when the grains are still small, for lower values $\epsilon_{\rm s}$. As a consequence, from Eqs.~\ref{velocity}, \ref{gamma} and \ref{kappa} it follows that the models which fail to produce a dust-driven wind in C-stars are more numerous for low values of $\epsilon_{\rm s}$. Indeed, these models show a larger frequency of small grains in Fig.~\ref{Nrc}.

We now study the dependence of the final grain size on stellar parameters, especially on the carbon excess and mass-loss rate.
As far as the dependence of the grain size on the carbon excess is concerned, two scenarios are possible, depending on the underlying assumptions of the dust growth scheme.
If $\epsilon_{\rm s, C}$ is not proportional to the carbon excess ($\epsilon_{\rm s, C}=\epsilon_{\rm s}=\rm const$), the typical size of carbon grains tends to be larger for larger values of the carbon excess, since more C$_2$H$_2$ molecules are available to form dust \citep{Ventura14, Dellagli15a, Dellagli15b, Ventura16}.
On the other hand, if $\epsilon_{\rm s, C}$ is proportional to the carbon excess as assumed in our description (Eq.~\ref{ns_cex}), the grain growth process for large carbon excess is counterbalanced by a larger number of seed particles.
In Fig.~\ref{avsco} the final grain sizes as a function of the carbon excess taken from our dust formation models are shown for a selected TP-AGB track of $M=2$~M$_\odot$, Z=0.004, with $\log\epsilon_{\rm s}=-13$ computed for the optical data set by \citet{Rouleau91}.
As can be seen, there is some correlation between the grain size and carbon excess but the variation of the grain size is smaller than the one produced by changing the parameter $\epsilon_{\rm s}$.

In Fig.~\ref{avsmloss} the grain size as a function of the mass-loss rate is shown for different choices of $\epsilon_{\rm s}$.  For a given choice of $\epsilon_{\rm s}$, there is not a clear trend between the grain size and the mass-loss rate.
The lack of a clear trend is in good agreement with the findings by complete hydrodynamical calculations for models computed with different choices of the carbon excess by \citet{Mattsson10}. Such computations are based on \citet{Hofner03} and include nucleation theory in its classical formulation.
The small sensitivity of the final grain size to the mass-loss rate and carbon excess, for a given $\epsilon_{\rm s}$, is not found if a constant number of seeds is assumed \citep{Ventura14, Dellagli15a, Dellagli15b, Ventura16}.

Therefore, in the framework of our dust formation scheme, the typical grain size obtained is essentially determined by the choice of $\epsilon_{\rm s}$ in Eq.~\ref{ns_cex}, with a milder dependence on the variation of stellar parameters (see Figs.~\ref{Nrc}-\ref{avsmloss}).

Since the typical grain size is physically more meaningful than the number of seeds, we will, from now on, refer to the typical carbon grain size obtained in our models by employing the different values of $\epsilon_{\rm s}$, as listed in Table~\ref{size}.

\begin{table}
\caption{Typical grain size obtained for different choices of $\epsilon_{\rm s}$ in our dust models (Eq.~\ref{ns_cex}) for the TP-AGB tracks listed in Table~\ref{stars}.}
\label{size}
\begin{tabular}{l r r r r r r r r}
\hline
$\log\epsilon_{\rm s}$ & $-11$  & $-12$  & $-13$ & $-14$ &  $-14.5$ & $-15$ & $-15.4$\\
$a_{\rm amC}$ [$\mu m$] &   0.035 &  0.07    &    0.12  &  0.2 & 0.3 & 0.5 & 0.7\\
\hline
\end{tabular}
\end{table}

\subsection{Dust temperature at the inner boundary of the dusty zone along the TP-AGB phase}
\begin{figure}
  \includegraphics[angle=90, width=0.48\textwidth]{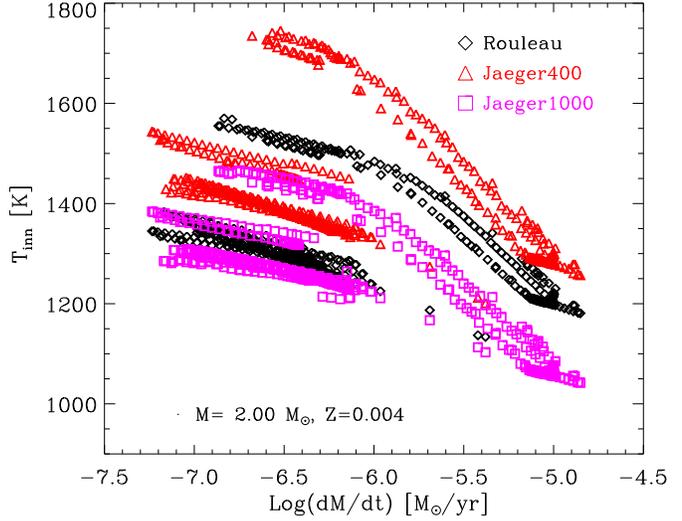}
        \caption{$T_{\rm inn}$ as a function of mass-loss rate for different optical data sets, listed in the figure. We show a TP-AGB track with $M=2$~M$_\odot$ and Z=0.004.}
        \label{Tdust_dmdt}
        \end{figure}

In Fig.~\ref{Tdust_dmdt} we show the evolution of $T_{\rm inn}$ with the mass-loss rate for a track with $M=2$~M$_\odot$ and Z=0.004.
For all the optical data sets considered, the quantity $T_{\rm inn}$ clearly shows the same qualitative, decreasing trend with larger mass-loss rate.
This means that more dust-enshrouded stars will typically be characterized by lower $T_{\rm inn}$ with respect to stars with a smaller amount of dust in their CSEs.
This trend is qualitatively in agreement with the one found by \citet{Groenewegen09}, who obtained a good SED fitting by employing lower $T_{\rm inn}$ for stars with larger mass-loss rates.

The differences in temperature between models with the same input parameters, except for different optical data sets, can be up to $\sim 300$~K.
Furthermore, different ranges in temperature are covered by models with different optical data sets. The Rouleau data set ranges between $\sim 1200$ and $\sim 1600$~K, Jaeger1000 between $\sim 1000$ and $\sim 1500$~K and Jaeger400 covers the range between $\sim 1200$ and $\sim 1800$~K.
For different choices of the optical data set, the inner radius of dust zone and the final grain size remain approximately the same for a given choice of the stellar input parameters.
Therefore, we can conclude that the differences in $T_{\rm inn}$ computed for the same TP-AGB models are essentially determined by different choices of the optical data set.

\subsection{Optical depth at 1~$\mu$m}
We explore how $\tau_1$ changes with the optical data set and typical final grain size.

\begin{figure}
   \includegraphics[angle=90, width=0.48\textwidth]{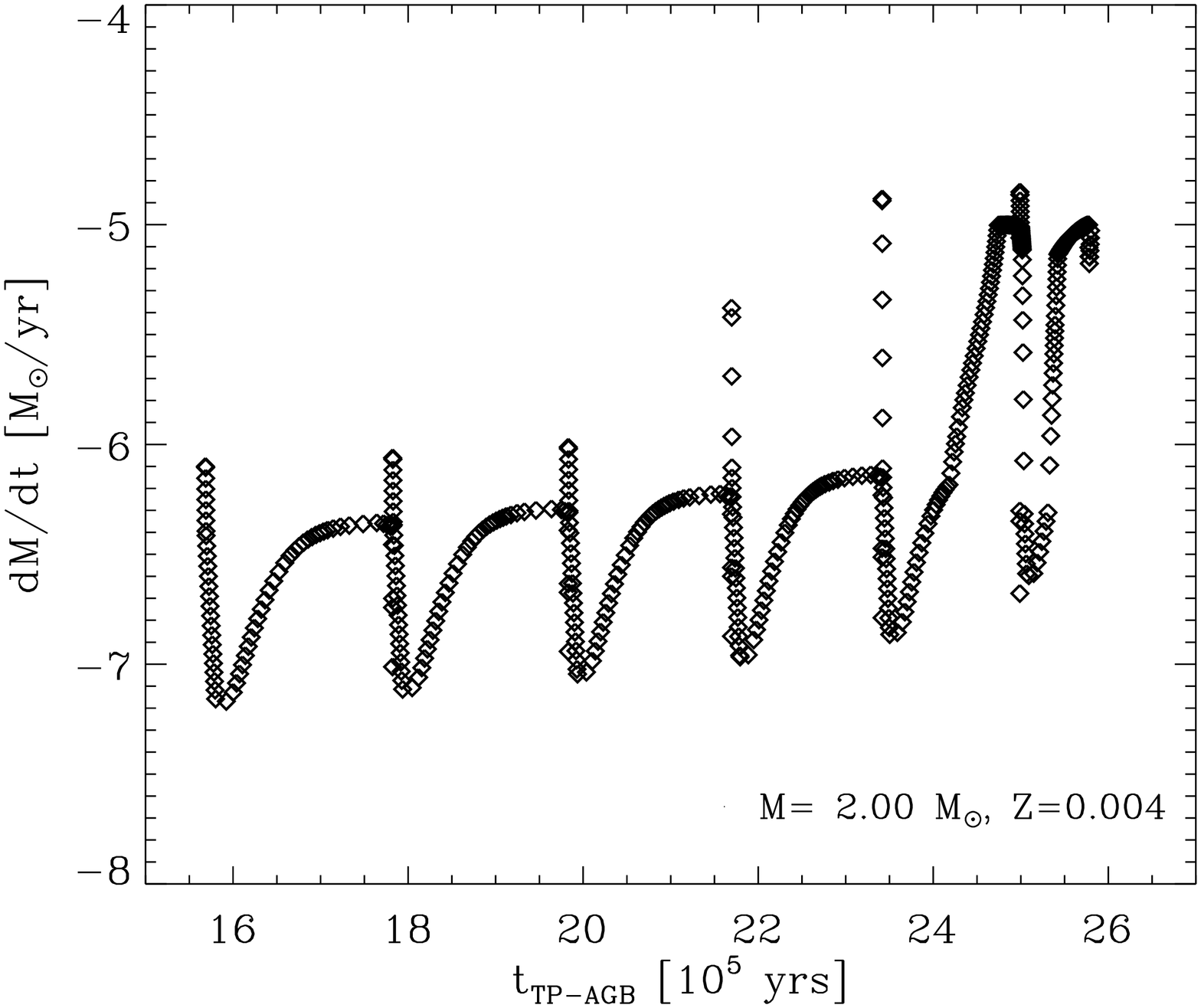}
   \includegraphics[angle=90, width=0.48\textwidth]{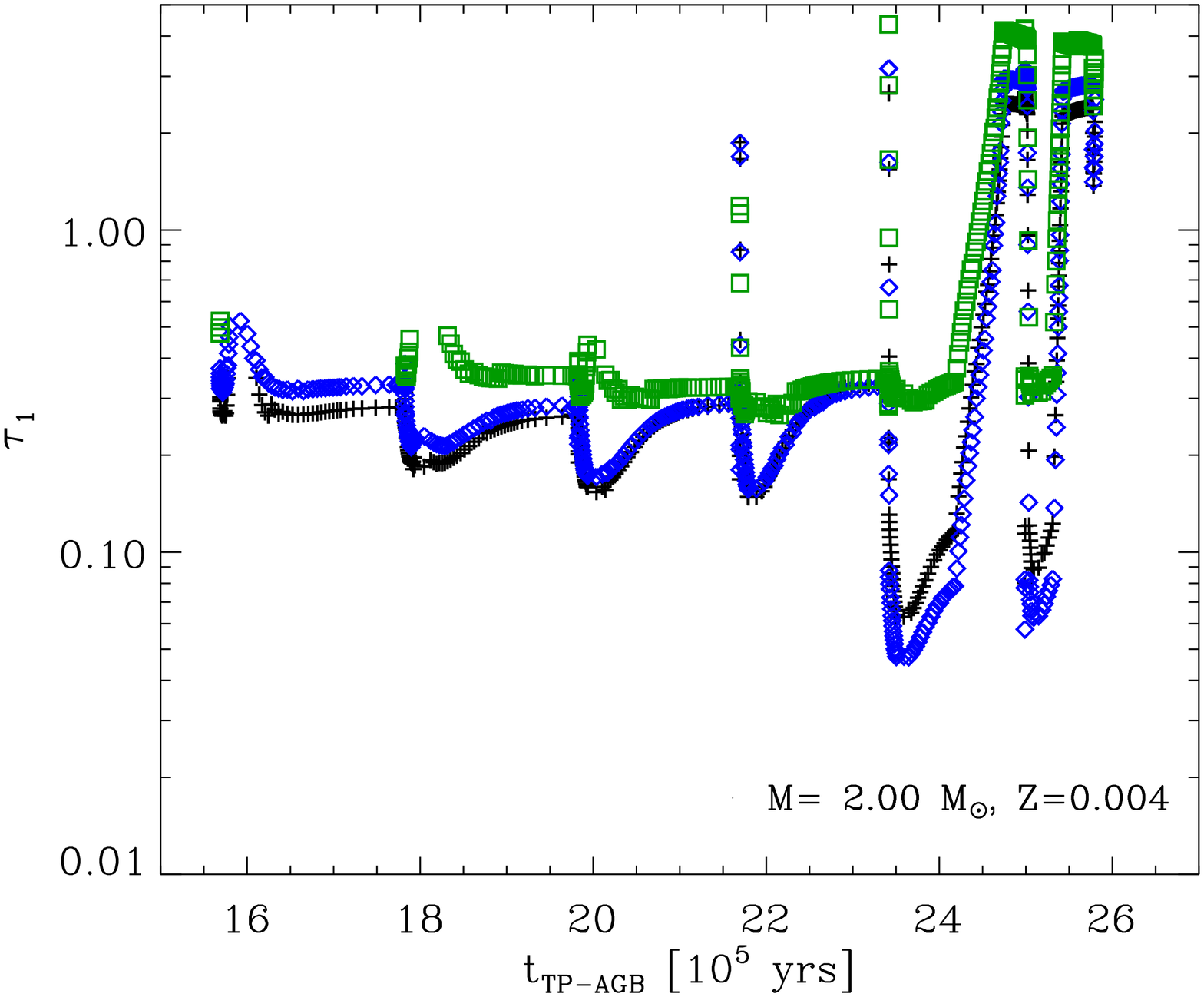}
   \includegraphics[angle=90, width=0.48\textwidth]{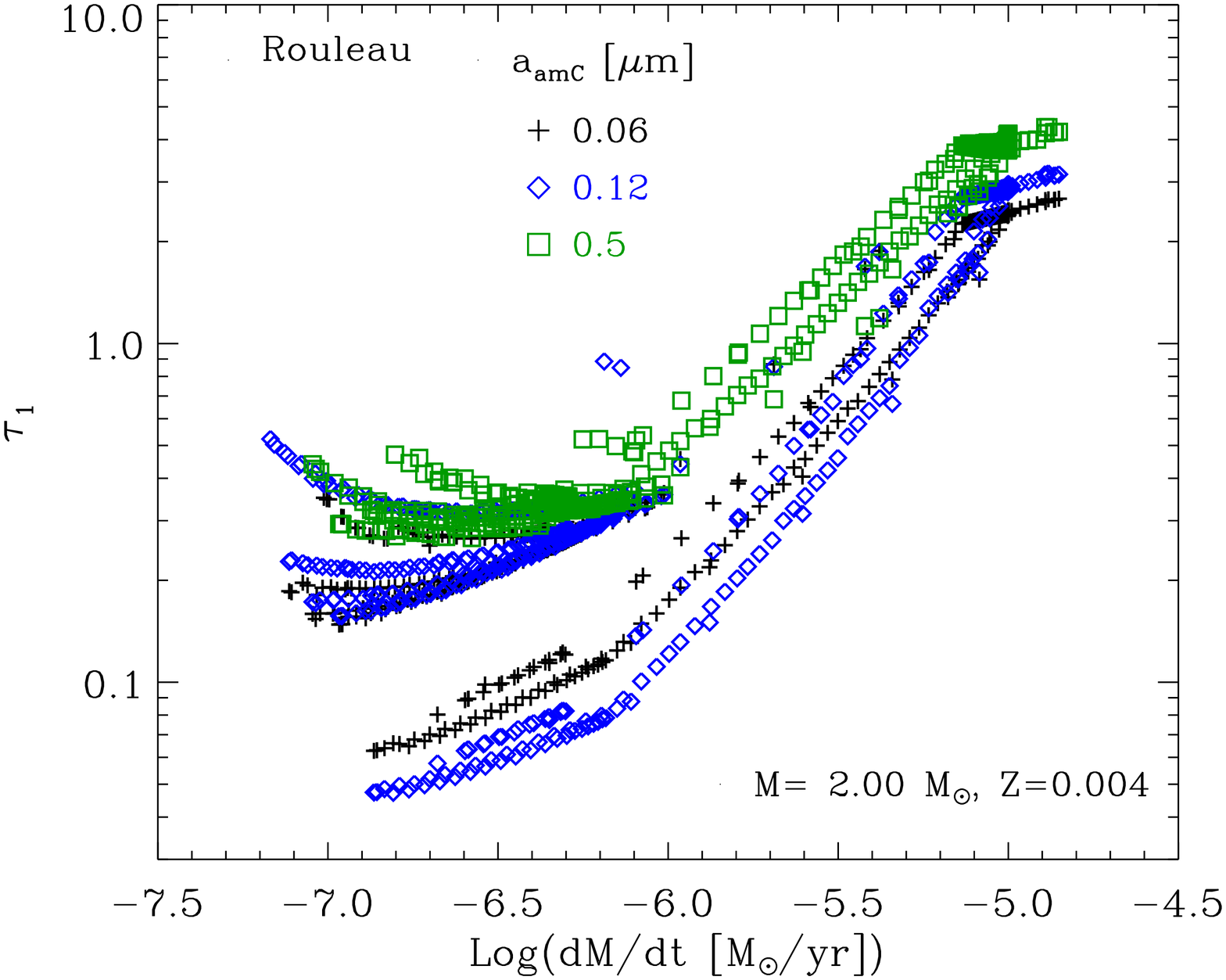}
        \caption{Mass-loss rate (top panel) and $\tau_1$ (middle panel) as a function of the time elapsed since the beginning of the TP-AGB phase. The quantity $\tau_1$ is computed for different grain sizes with the optical data set by Rouleau. In the lower panel, $\tau_1$ vs the mass-loss rate is shown. The models are plotted for one selected TP-AGB track in the carbon phase with $M=2$~M$_\odot$, Z=0.004.}
        \label{tau1_size}
        \end{figure}

In the upper panel of Fig.~\ref{tau1_size} we show the evolution of the mass-loss rate over the last six thermal pulse cycles experienced by the $M=2$~M$_\odot$, Z=0.004 TP-AGB model, in the carbon rich phase.
We see that typical super-wind mass-loss rate, $\log(\dot{M}[M_{\odot yr^{-1}}])\ga-6$, are reached only during the last two thermal pulses, while earlier stages are characterized by much lower values. The corresponding temporal evolution of $\tau_1$ is displayed in the middle panel for three choices of the carbon dust grain size for
the optical data set by Rouleau.
The models shown are only the ones able to accelerate the wind.
In all the cases the trend of $\tau_1$ follows the one of the mass-loss but attaining different values for different $a_{\rm amC}$ and other stellar parameters.
This can be better appreciated by looking at the lower panel of Fig.~\ref{tau1_size}.

At a fixed grain size, $\tau_1$ increases with the mass-loss until it reaches a sort of plateau for $\log(\dot{M}[M_\odot yr^{-1}])\ga -5$.
At a given value of the mass-loss rate, a significant dispersion of $\tau_1$ is predicted for different values of the grain size depending on the evolutionary stage of the pulse cycle. Such a dispersion tends to decrease for larger values of the mass-loss rate.

\begin{figure}
   \includegraphics[angle=90, width=0.48\textwidth]{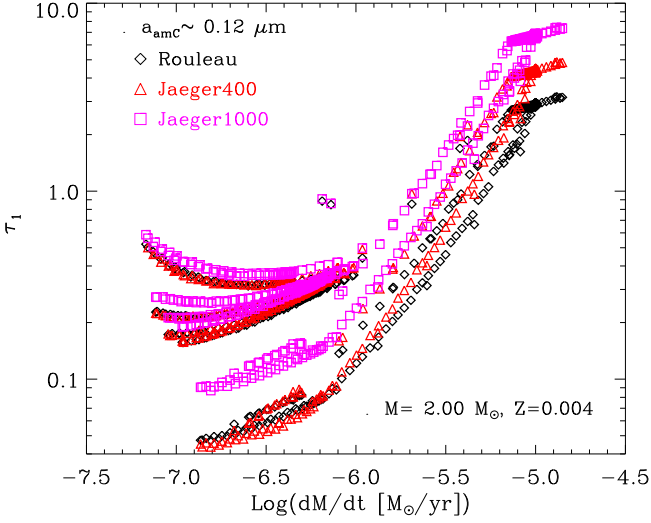}
        \caption{The same as the lower panel of Fig.~\ref{tau1_size} but for different optical data sets and $a_{\rm amC}\sim 0.1$~$\mu$m.}
        \label{tau1_op}
        \end{figure}

In Fig.~\ref{tau1_op} $\tau_1$ is plotted as a function of the mass-loss rate for different choices of the optical data set, but the same typical grain size $a_{\rm amC} \sim 0.12$~$\mu$m. For different choices of the optical data set, larger values of $\tau_1$ are produced in models with $-7<\log(\dot{M}[M_{\odot}yr^{-1}])<-6$ for Jaeger1000, while models obtained with Rouleau and Jaeger400 data sets produce approximately the same values of $\tau_1$, except for the largest mass-loss rate for which $\tau_1\sim 3.2$ for Rouleau, $\tau_1\sim 5$ for Jaeger400, and $\tau_1\sim 7.9$ for Jaeger1000.
These differences may impact on the evolution of dust-enshrouded stars in the CCDs (see Section~\ref{results}).

\section{Calibration of carbon dust optical data set and typical grain size}
\label{Cal:section}
\subsection{Method}
\label{Cal:method}
We employ our dust growth description and RT computed along the TP-AGB phase in order to constrain the combination of the most suitable opacity set and final typical grain size for carbon dust which best reproduces the observed CCDs. We also quantify the deviation between our models and the observations.
We perform this study employing several colors since they are independent of the intrinsic stellar luminosity and distance and they sample different parts of the SED of C-stars.

The observed sample of TP-AGB stars selected for the comparison with our models is taken from the catalog of cool evolved stars in the SMC presented by \citet{Boyer11}. The photometric data used by \citet{Boyer11} is a co-addition of two epochs of SAGE-SMC data, together with a third epoch from the \textit{Spitzer} Survey of the SMC (S$^{3}$MC) where the coverage overlaps. The SAGE-SMC survey has uniformly imaged the SMC bar, wing and tail regions and the resulting catalog includes optical to far-IR photometry: UBVI photometry from the Magellanic Cloud Photometric Survey \citep[MCPS,][]{Zaritsky02}, JHKs photometry from 2MASS and the InfraRed Survey Facility \citep[IRSF,][]{Kato07}, MIR photometry from Spitzer's IRAC \citep{Fazio04}, and far-IR photometry from MIPS \citep{Rieke04}.

According to a set of color-magnitude cuts in NIR and MIR CMDs, \citet{Boyer11} classified the sample of TP-AGB stars in C-stars, M-stars, anomalous oxygen-rich and extreme (x-) stars. The majority of x-stars is probably carbon-rich \citep{vanLoon97, vanLoon06, vanLoon08, Matsuura09}.
C-stars and M-stars are selected using color-magnitude cuts in the J~$-$~Ks vs Ks CMD, following the same approach of \citet{cioni06a, cioni06b}.
Due to the dust extinction, the population of x-stars is identified through MIR colors. In particular, the sources classified as x-stars are brighter than the 3.6~$\mu$m tip of the Red Giant Branch and redder than J~$-$~[3.6]~=~3.1 mag. The most dust-enshrouded sources with no NIR detection are included in the sample of x-stars if they are brighter than the 3.6~$\mu$m tip of the Red Giant Branch and their [3.6]~$-$~[8] color is redder than 0.8~mag. We refer to Section~3 in \citet{Boyer11} for a detailed description of the classification scheme and the criteria adopted to minimize the contamination from Young Stellar Objects and unresolved background galaxies.
For the present study we select C- and x-stars.

The selected colors for the calibration are J~$-$~Ks, [3.6]~$-$~[8.0], J~$-$~[3.6], J~$-$~[8.0], Ks~$-$~[3.6] and Ks~$-$~[8.0] in the entire J~$-$~Ks range, plus [5.8]~$-$~[8.0] and [3.6]~$-$~[4.5] for stars with J~$-$~Ks$\ga 2.5$, dominated by dust emission. We select the observed stars for which the photometry is available in all these bands.
We include the [5.8]~$-$~[8.0] and [3.6]~$-$~[4.5] colors only in the dust-dominated cases because the spectra of less dust-rich stars are more affected by C$_3$ absorption features between 4 and 6~$\mu$m \citep{Sloan15}. These features are not reproduced by the available opacity data \citep[see Fig.~10 of][]{Jorgensen00}. A thorough study of C-stars spectra in the wavelength range between 4 and 6~$\mu$m will be the subject of an upcoming paper (Aringer et al.~in preparation).
We exclude colors the V band from the study, since this band experiences large variability during the pulsation cycle of the star and depends on the epoch of observation \citep{Nowotny11}.
We also exclude from the present investigation colors including the 24~$\mu$m flux since our calibration is only focused on carbon dust and does not take into account MgS dust which might be relevant at this wavelength \citep{Hony02, Lombaert12}.
The emission properties and colors at 24~$\mu$m will be analyzed separately in a forthcoming paper (Nanni et al.~in preparation).
The NIR and MIR colors of our models are computed for the 2MASS and IRAC filters.

In order to compare the results of the simulations with the observations, we divide the observed sample of C- and x-stars into five bins according to their J~$-$~Ks color, for which the average values are (J~$-$~Ks)$_{\rm av}\sim 1.5, 2.2, 3.0, 3.7, 4.5$. The number of observed stars in the five bins are $N_{\rm obs, stars}=$1630, 212, 117, 43, 10. Each observed star in a certain J~$-$~Ks color bin, occupies a given position in the space of parameters defined by the colors considered. Therefore, for each bin in J~$-$~Ks we compute the average values of the other colors for the observed stars.
In each J~$-$~Ks bin and for each color, we therefore compute the deviation of the TP-AGB models from the average observed value, normalized by the dispersion of the observed data, $\sigma_{\rm c, obs}$ :
\begin{equation}\label{sigma_star}
\sigma_{\rm c}=\sqrt{\frac{\sum_{\rm model} \frac{\big(x_{\rm model}-x_{\rm av}\big)^2}{\sigma^2_{\rm c, obs}}}{N_{\rm model}}},
\end{equation}

where the sampled models along the TP-AGB tracks are equally weighed in the calculation.
The total deviation of the simulated points for all the colors will be:
\begin{equation}\label{tot_sigma}
\sigma=\frac{\sum_c\sigma_{\rm c}}{N_{\rm c}},
\end{equation}

where $N_{\rm c}$ is the number of colors considered. This value will be shown in the Figs.~\ref{sigmavsa} and \ref{sigma}.
Clearly, the best performing models are the ones with low values of the normalized $\sigma$ and in any case they should not be far from $\sigma=1$.

\subsection{Results}
 \begin{figure}
    \includegraphics[angle=90, width=0.48\textwidth]{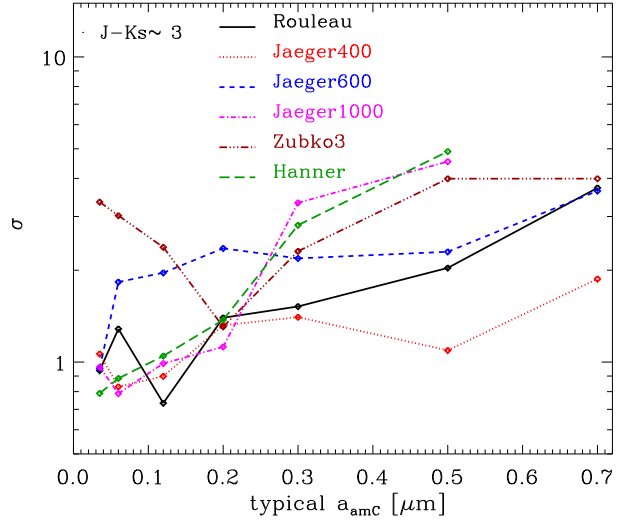}
        \caption{Deviation between observed data and models computed through Eqs.~\ref{sigma_star} and \ref{tot_sigma} as a function of the typical grain size obtained for the TP-AGB models listed in Table~\ref{stars} in the J~$-$~Ks$\sim 3$ bin. The results are not plotted for Jaeger1000 and Hanner with $a_{\rm amC}\sim 0.7$ $\mu$m, because, for these combinations of the parameters, none of the TP-AGB models fall in the J~$-$~Ks$\sim 3$ bin.}
        \label{sigmavsa}
        \end{figure}

 \begin{figure}
    \includegraphics[angle=90, width=0.48\textwidth]{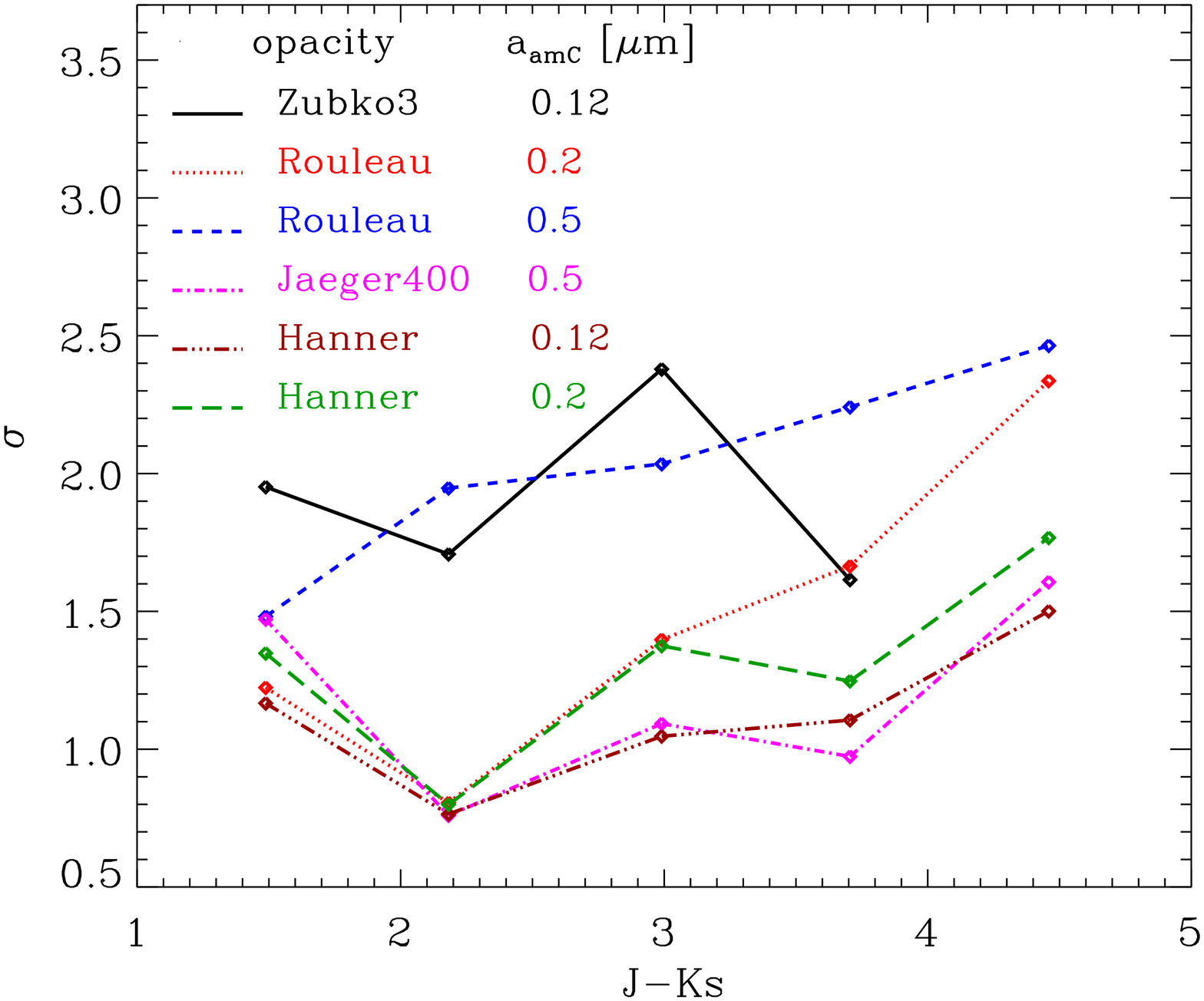}
    \includegraphics[angle=90, width=0.48\textwidth]{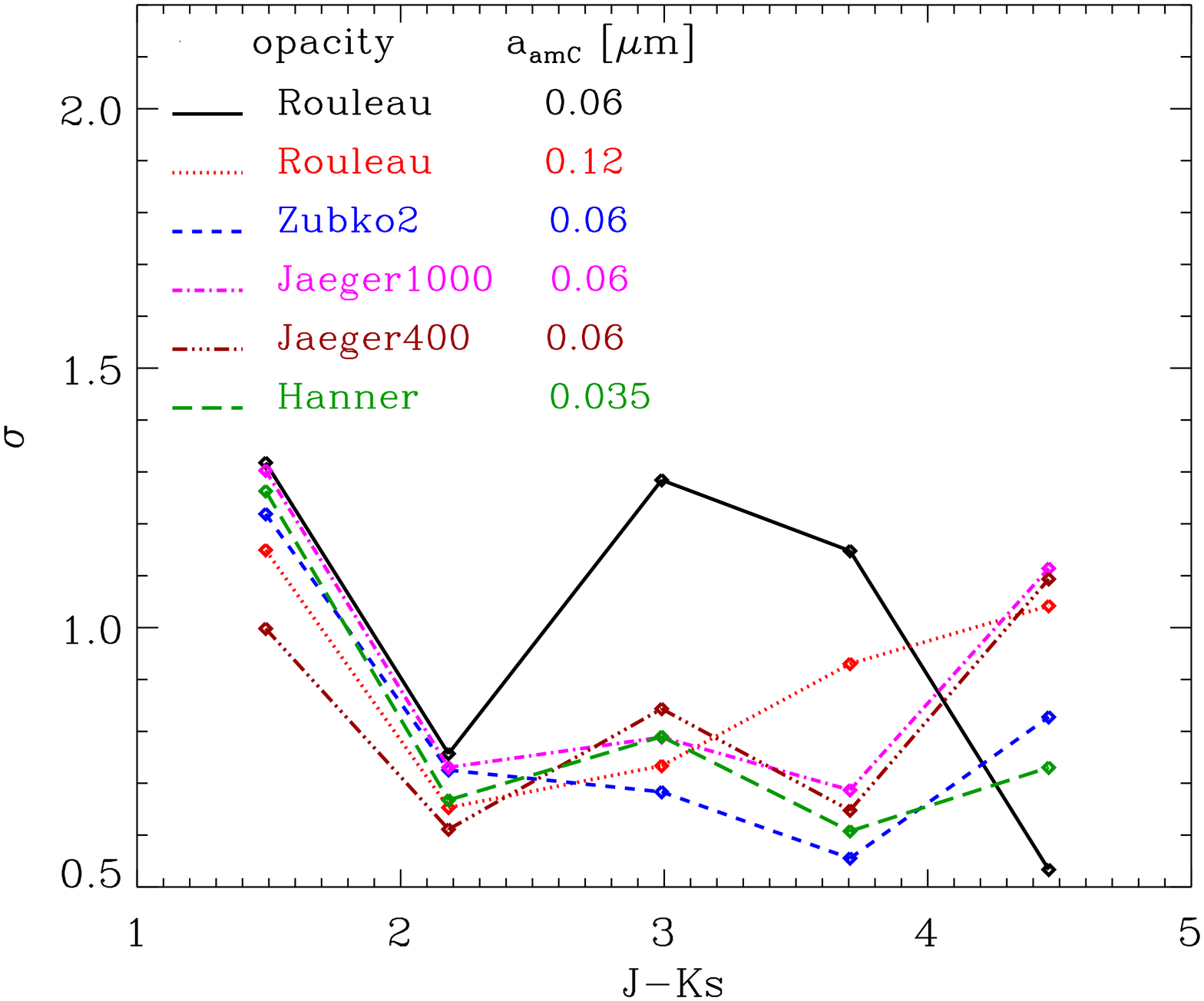}
        \caption{Deviations between observed data and models computed through Eqs.~\ref{sigma_star} and \ref{tot_sigma} as a function of J~$-$~Ks. The value of $\sigma$ is computed for TP-AGB models listed in Table~\ref{stars}, selected along the TP-AGB tracks as described in Section~\ref{effectonTPAGB}. In the upper panel some combinations of optical data sets and grain sizes which poorly reproduce the observations are shown, while in the lower panel some of the well performing combinations are plotted.}
        \label{sigma}
        \end{figure}

    \begin{figure}
    \includegraphics[angle=90, width=0.48\textwidth]{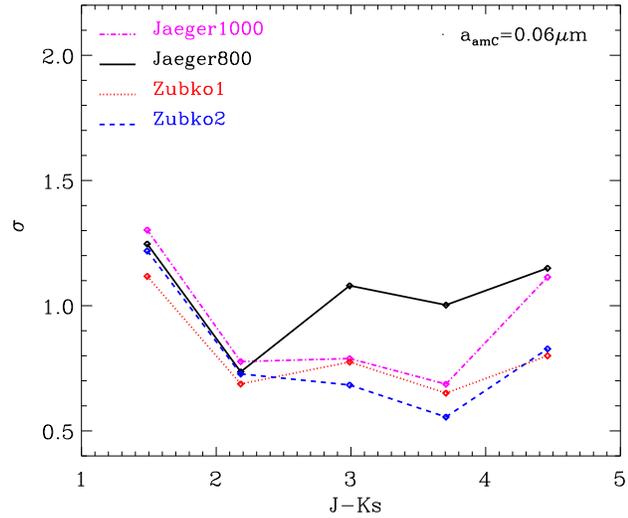}
        \caption{The same as in the lower panel of Fig.~\ref{sigma}, but for the optical data sets similar to Jaeger1000 and $a_{\rm amC}\sim 0.06$~$\mu$m.}
        \label{sigma_jaeger1000}
        \end{figure}

        \begin{figure}
    \includegraphics[angle=90, width=0.48\textwidth]{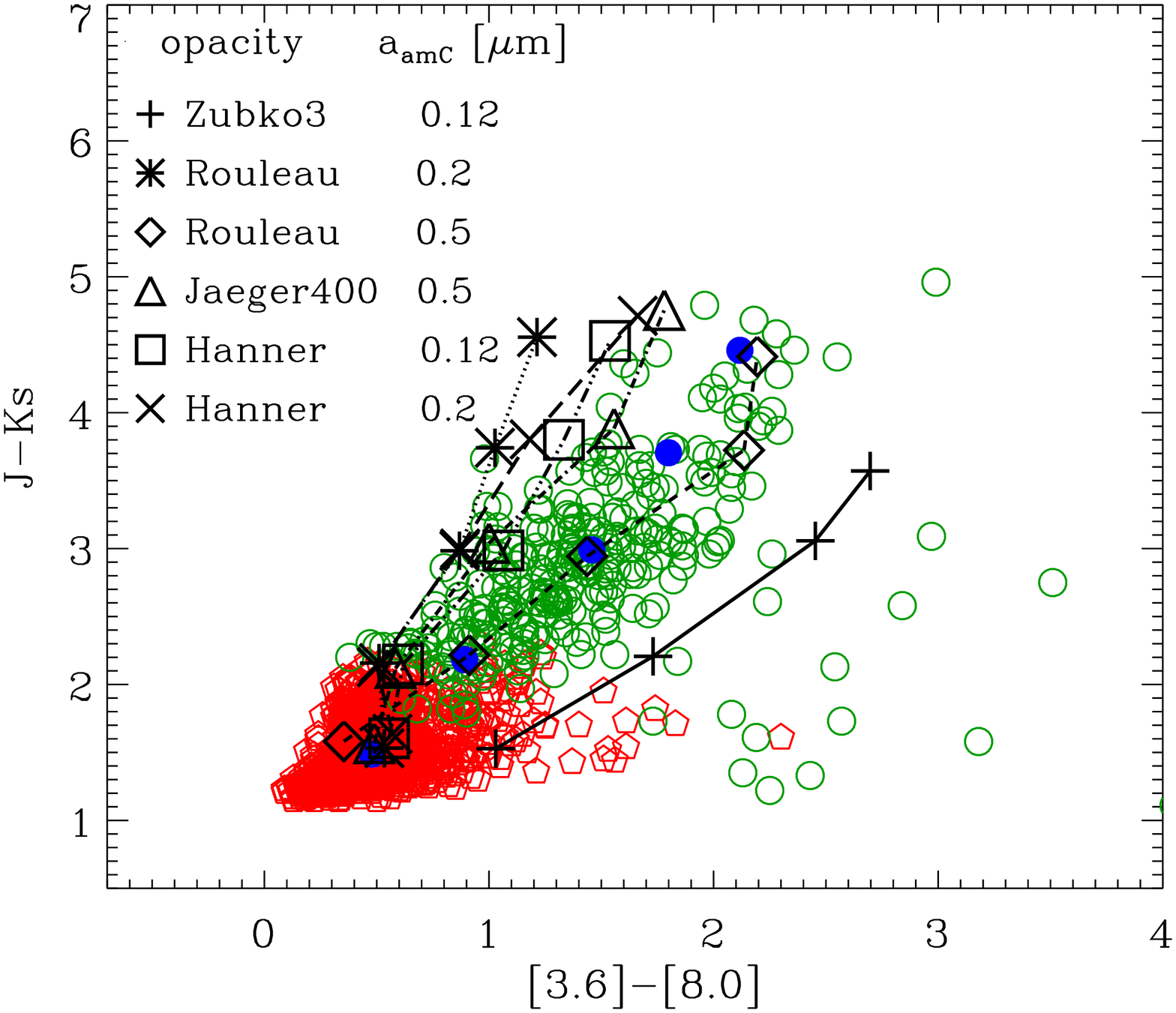}
    \includegraphics[angle=90, width=0.48\textwidth]{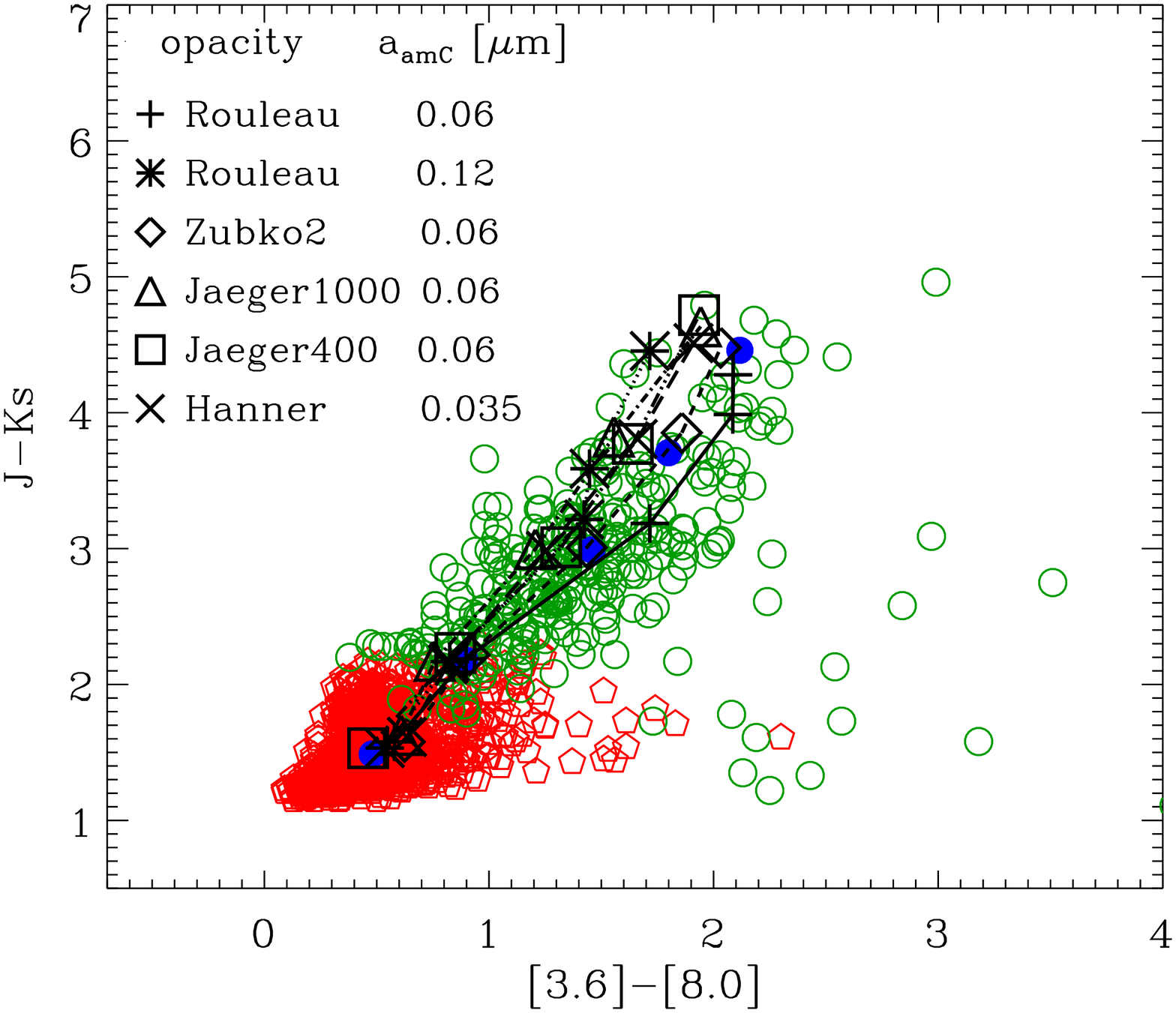}
        \caption{J~$-$~Ks vs [3.6]~$-$~[8.0] CCDs for the observed sample of C- (red pentagons) and x-stars (open green circles) superimposed with the average value for models obtained with different optical data sets and grain size, listed in the top left. The computations have been performed by employing the TP-AGB tracks listed in Table~\ref{stars}. Full blue circles represent the average values of the observed stars. The combinations of optical data sets and grain sizes are the same of Fig.~\ref{sigma}.}
        \label{CCDs_368}
        \end{figure}

 \begin{figure}
  \includegraphics[angle=90, width=0.48\textwidth]{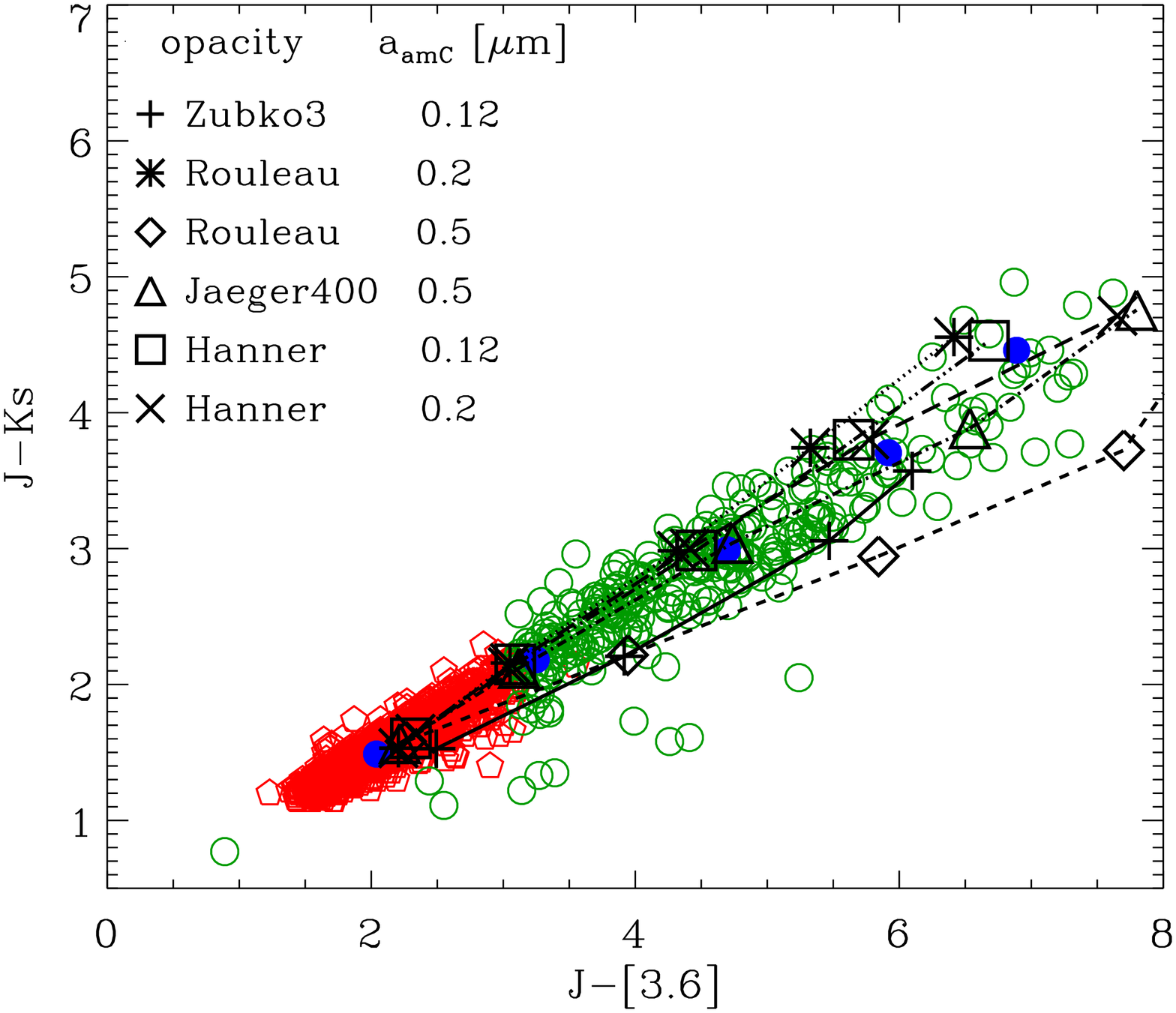}
   \includegraphics[angle=90, width=0.48\textwidth]{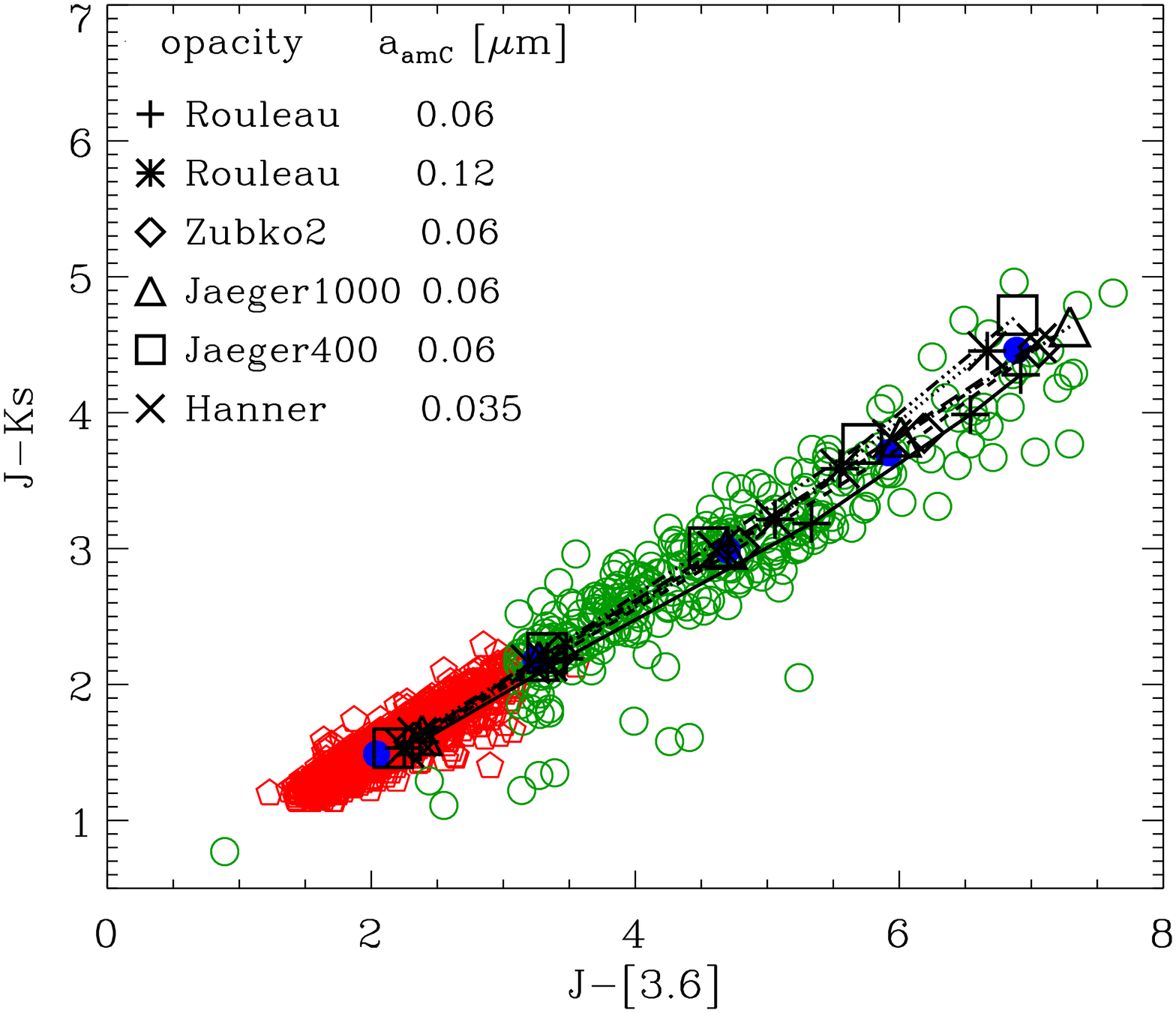}
        \caption{The same as in Fig.~\ref{CCDs_368} but for J~$-$~Ks vs J~$-$~[3.6].}
        \label{CCDs_j36}
        \end{figure}

 \begin{figure}
  \includegraphics[angle=90, width=0.48\textwidth]{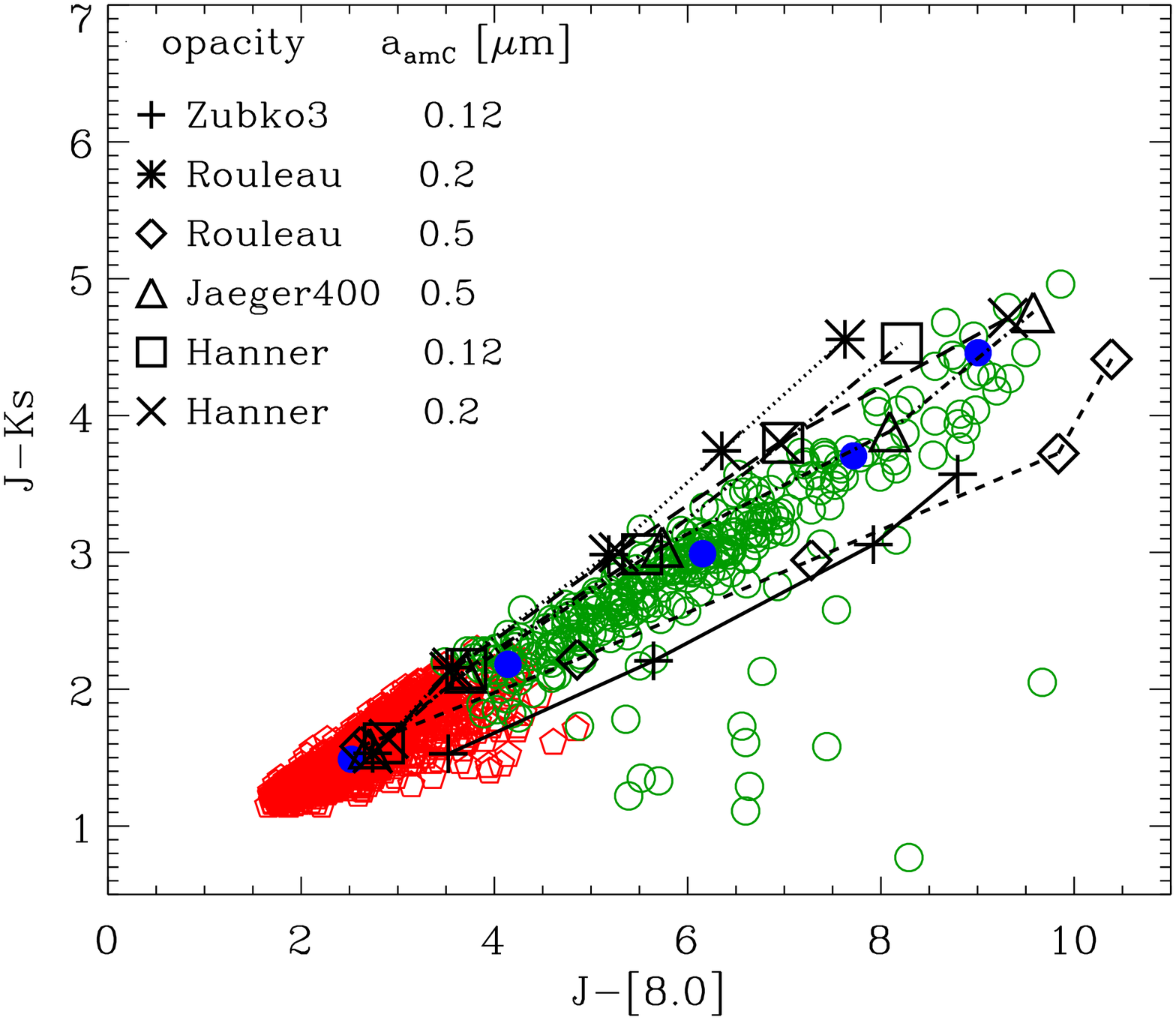}
   \includegraphics[angle=90, width=0.48\textwidth]{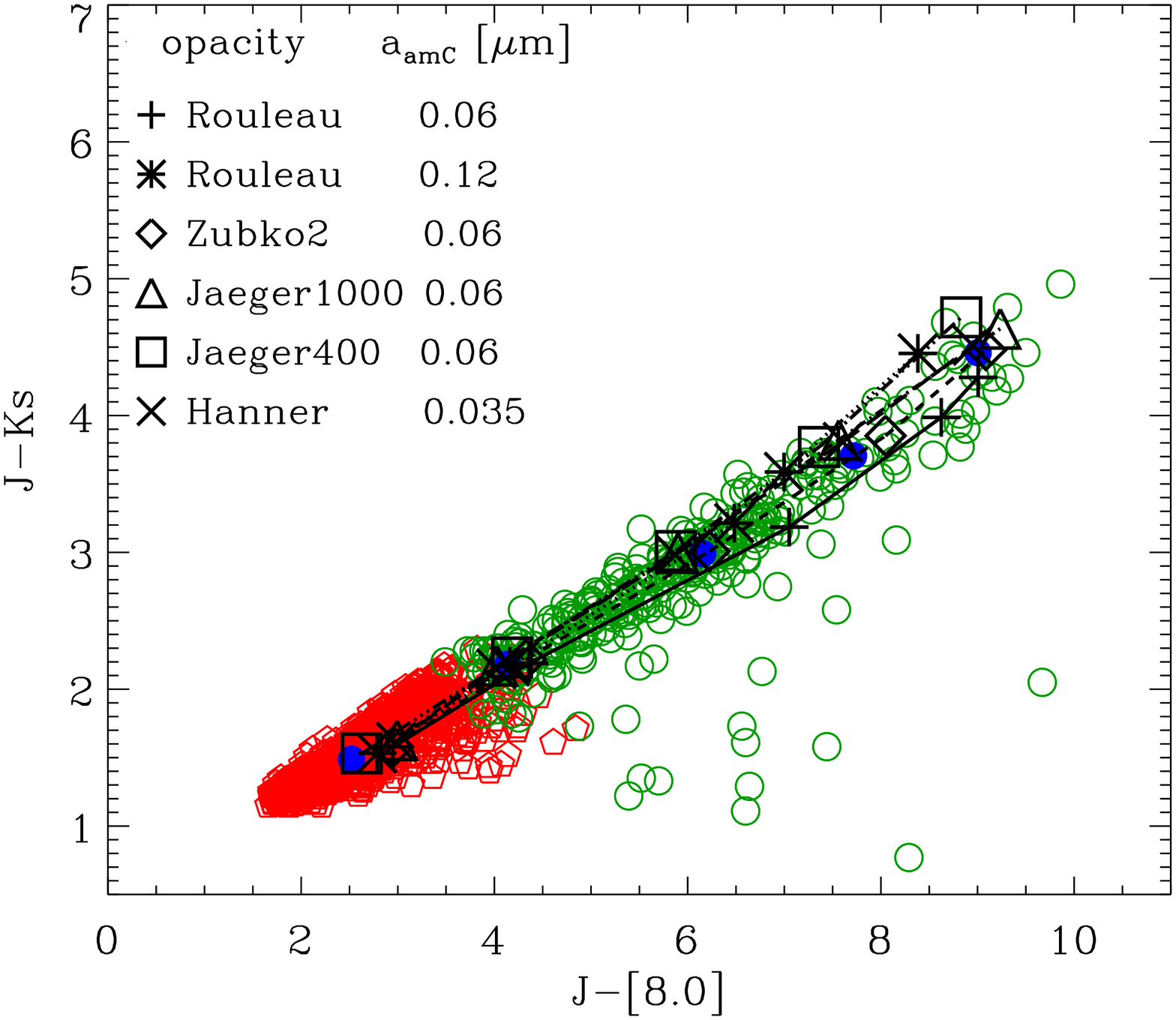}
        \caption{The same as in Fig.~\ref{CCDs_368} but for J~$-$~Ks vs J~$-$~[8.0].}
        \label{CCDs_j8}
        \end{figure}

 \begin{figure}
  \includegraphics[angle=90, width=0.48\textwidth]{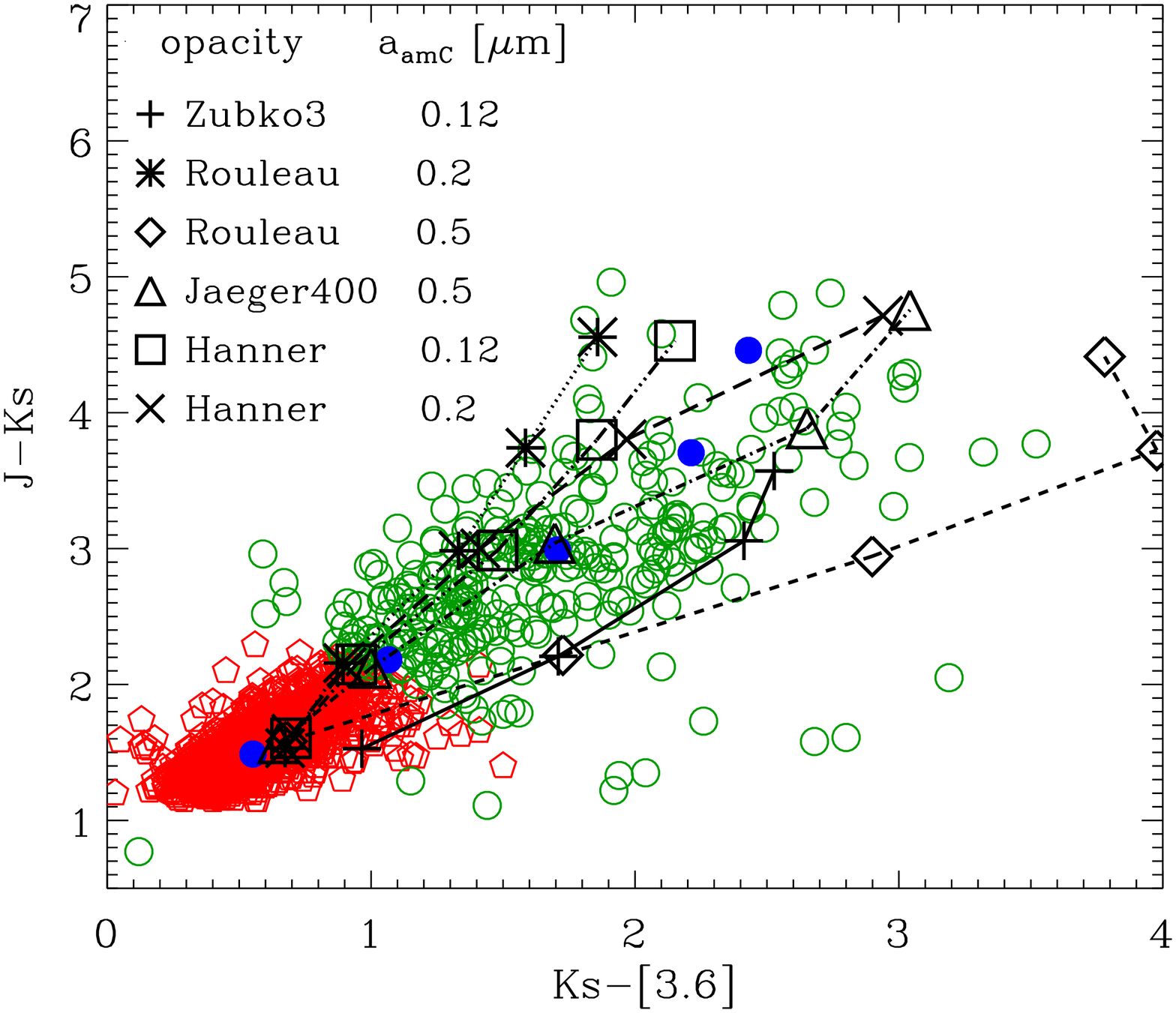}
   \includegraphics[angle=90, width=0.48\textwidth]{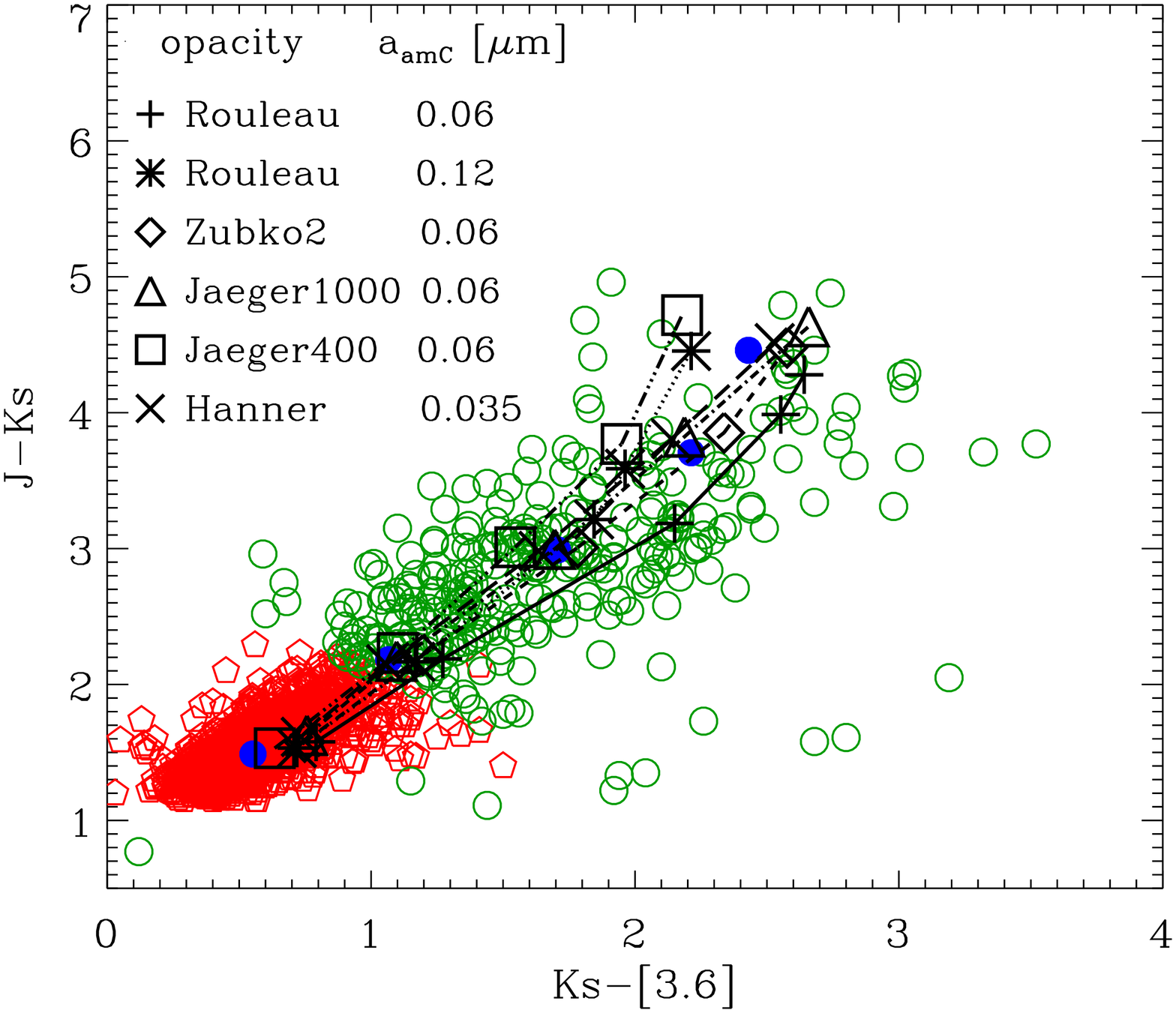}
        \caption{The same as in Fig.~\ref{CCDs_368} but for J~$-$~Ks vs Ks~$-$~[3.6].}
        \label{CCDs_k36}
        \end{figure}

 \begin{figure}
  \includegraphics[angle=90, width=0.48\textwidth]{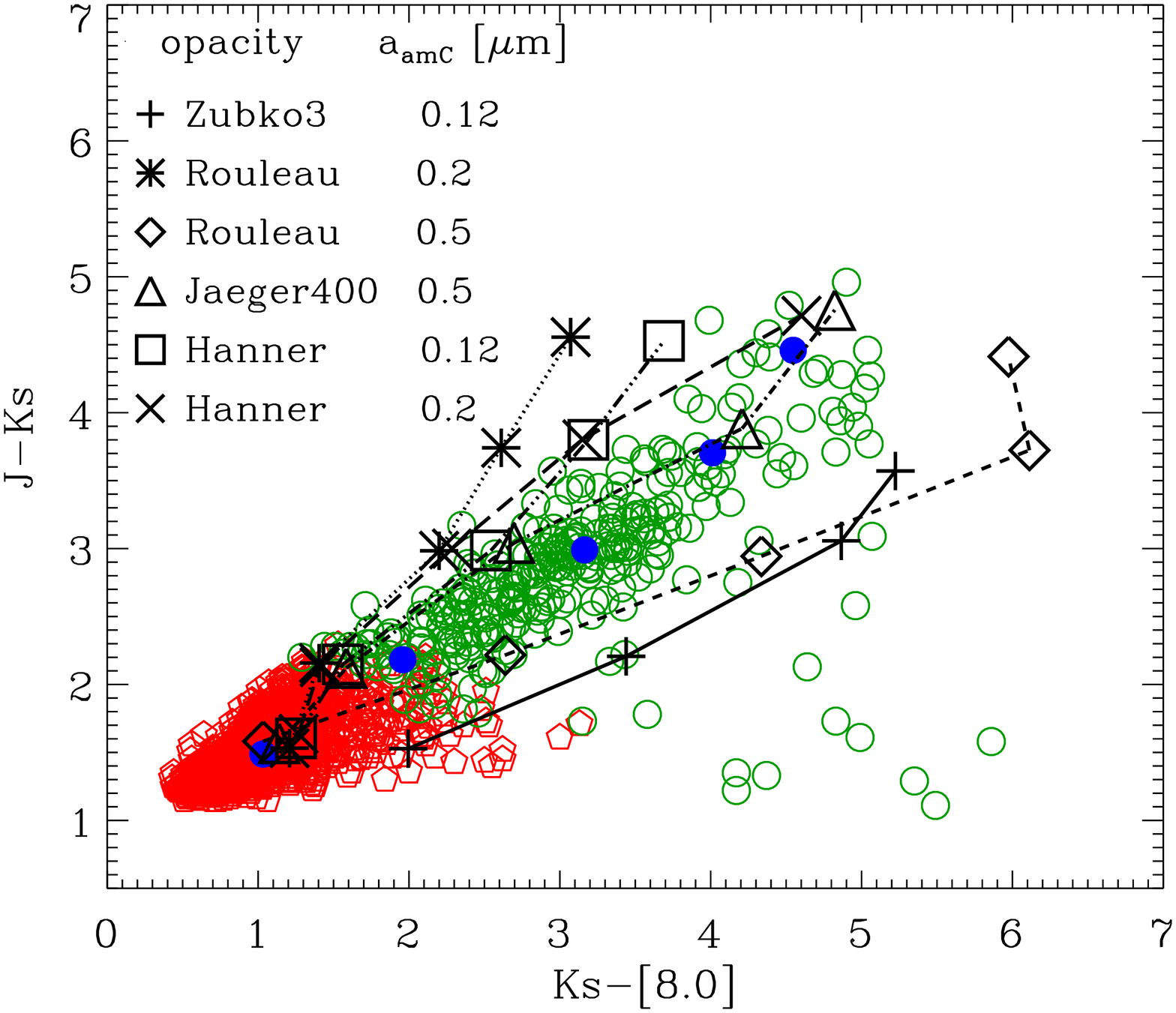}
   \includegraphics[angle=90, width=0.48\textwidth]{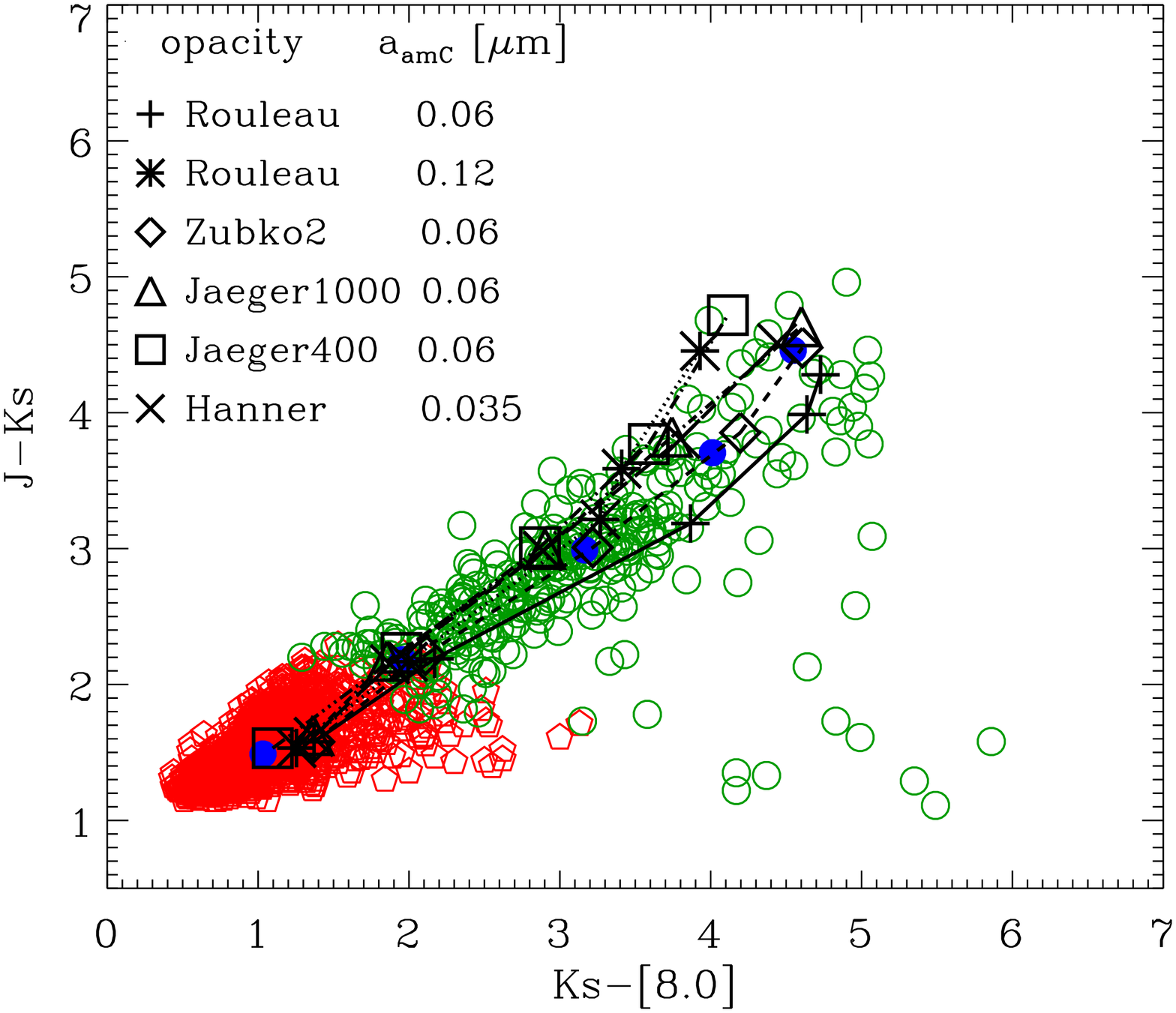}
        \caption{The same as in Fig.~\ref{CCDs_368} but for J~$-$~Ks vs Ks~$-$~[8.0].}
        \label{CCDs_k8}
        \end{figure}

         \begin{figure}
 \includegraphics[angle=90, width=0.48\textwidth]{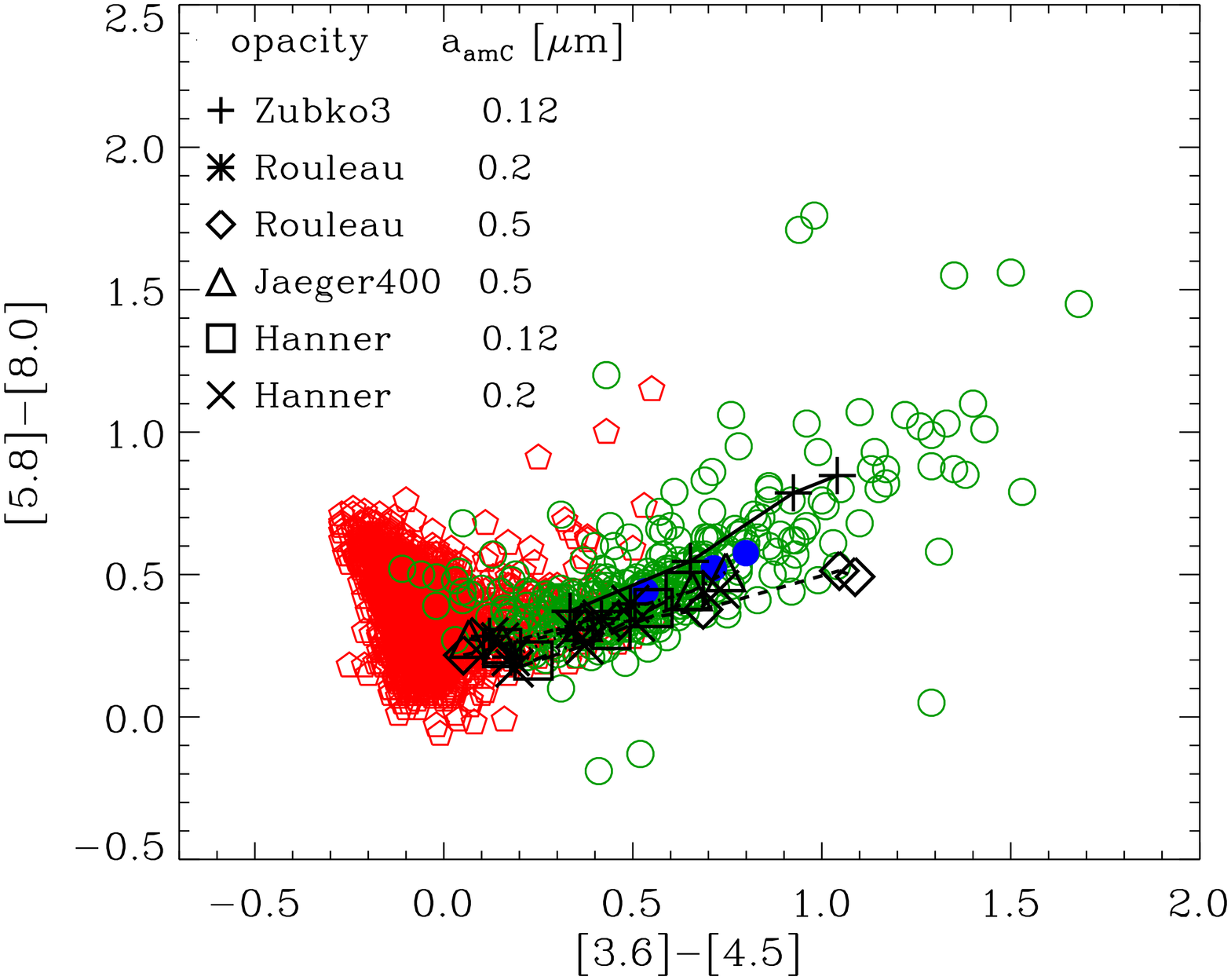}
 \includegraphics[angle=90, width=0.48\textwidth]{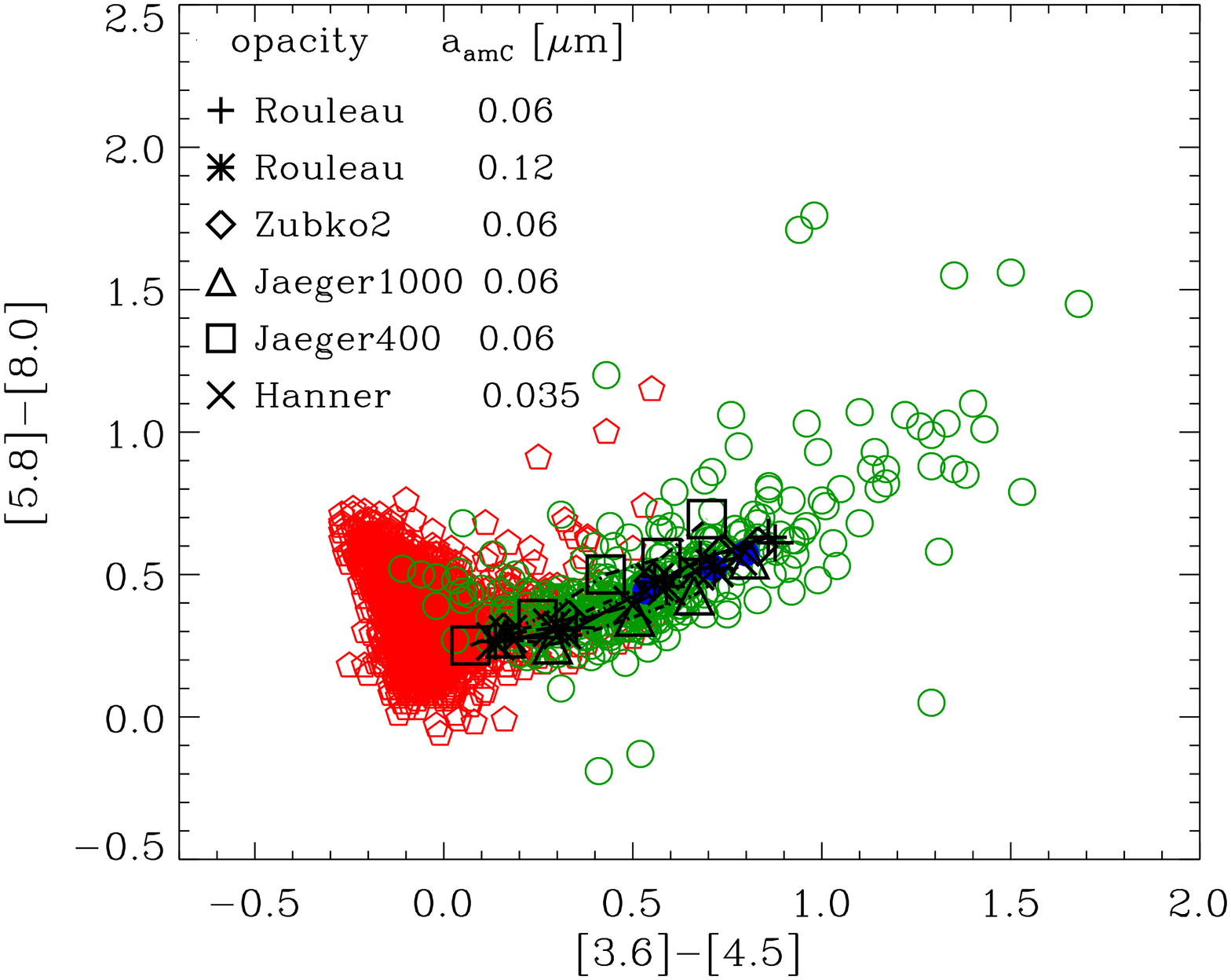}
        \caption{The same as in Fig.~\ref{CCDs_368} but for [5.8]~$-$~[8.0] vs [3.6]~$-$~[4.5].}
        \label{CCDs_5883645}
        \end{figure}

The deviations between observed data and models for some selected data sets are summarized in the Tables of the Appendix. The data sets listed are Rouleau, Jaeger400, Jaeger600, Jaeger1000 and Hanner with $0.035\la a_{\rm amC}\la0.7$~$\mu$m.

Larger values of $\sigma$ are usually obtained for $a_{\rm amC}\ga 0.2$~$\mu$m for all the values of J~$-$~Ks, especially for J~$-$~Ks$\ga 3$.
In Fig.~\ref{sigmavsa} the trend of $\sigma$ as a function of the grain size obtained for the TP-AGB tracks listed in Table~\ref{stars} is shown for different optical data sets. As an example, the results are plotted for heavily dust-enshrouded CSEs (J~$-$Ks~$\sim 3$).
The deviations between observed data and models are usually large for Zubko3 optical data set, independently of the final grain size.

For all the data sets shown, the values of $\sigma$ are larger for $a_{\rm amC}\ga 0.2$~$\mu$m than for smaller grains.
In general, the deviation between observations and models tends to be larger for increasing values of the grain size for any choice of the optical data set, except for Jaeger400.
However, Jaeger400 with $a_{\rm amC}\sim 0.5, 0.7$~$\mu$m does not well reproduce the observations for stars with J~$-$Ks~$\ga 4$ (see also Figs.~\ref{CCDs_368}~-~\ref{CCDs_5883645} and Table~\ref{Table:results2} in the Appendix). Moreover, it performs worse than grains of $a_{\rm amC}\sim 0.06$~$\mu$m.

Now we briefly discuss the performance of the individual optical data sets, referring to the tables in the Appendix.

For the Rouleau data set, the best agreement between observations and models is obtained for $a_{\rm amC}\sim 0.12$~$\mu$m for all the values of J~$-$~Ks, except for J~$-$Ks~$\ga 4$. For this large value of the J~$-$Ks, the best agreement between observations and models is obtained with grains of $a_{\rm amC}\sim 0.06$~$\mu$m ($\sigma\sim 0.5$). The typical value of the grain size reproducing the colors in almost the entire J~$-$~Ks range, $a_{\rm amC}\sim 0.12$~$\mu$m, is in good agreement with the one obtained in complete hydrodynamical simulations by \citet{Mattsson10}, in which the same optical data has been used.

For Jaeger400 the best agreement is usually achieved for grains of $a_{\rm amC}\sim 0.06$~$\mu$m for all the values of J~$-$~Ks.
For larger grain sizes the results are worse in the reddest bin, for which the best agreement is obtained with $a_{\rm amC}\sim 0.035$~$\mu$m, with $\sigma\sim 1$.

For Jaeger600, the best performance is usually obtained for grains of $a_{\rm amC}\sim 0.035$~$\mu$m, for all the value of J~$-$~Ks, except for J~$-$~Ks~$\sim1.5$.
However, for the reddest bin in J~$-$~Ks the Jaeger600 data set shows $\sigma\ga 1.8$ for all the grain sizes.

For Jaeger1000 the observations are well reproduced by  $a_{\rm amC}\sim 0.06$~$\mu$m for all the values of  J~$-$~Ks.
However, for J~$-$~Ks$\ga 4$ there are not large differences in the value of $\sigma$ computed with $0.035 \la a_{\rm amC}\la 0.12$~$\mu$m, but we obtain always that $\sigma>1$.

For the Hanner data set, a good agreement is obtained for grains of $a_{\rm amC}\sim 0.035$~$\mu$m for all the values of J~$-$~Ks.

From the above description we can derive a few relevant trends.
We confirm, as also anticipated in Section~\ref{results}, that differences in the final grain size mostly affect the colors corresponding to the largest  J~$-$~Ks.
The stars in the reddest bin (J~$-$~Ks$\ga 4$) are reproduced well only by some combinations of optical data sets and grain sizes. Namely, Rouleau with $a_{\rm amC}\sim 0.06$~$\mu$m ($\sigma\sim 0.5$) and Hanner with $a_{\rm amC}\sim 0.035$~$\mu$m ($\sigma\sim 0.7$). We also note that the Rouleau data set with $a_{\rm amC}\sim 0.12$~$\mu$m performs well for all the other values of J~$-$~Ks.

In summary, we may draw the conclusion that the majority of carbon grains is likely to have a grains of $0.035 \la a_{\rm amC}\la 0.12$~$\mu$m, except for the reddest stars, for which the grains tend to be smaller ($0.035 \la a_{\rm amC}\la 0.06$~$\mu$m).

The colors of stars with J~$-$~Ks$\la 1.5$ are less sensitive to the size of the dust grains.

The results obtained with the Rouleau data set, for which the reddest stars form smaller grains, suggest the possible existence of a trend between the mass-loss rate and/or the carbon excess with the grain size.
The dependence of the number of seeds on the mass-loss rate and/or the carbon excess might reflect the competition between nucleation and accretion on dust grains, for increasingly dense CSEs in which smaller and more numerous dust grains might be produced.
We suggest that our adopted linear relation between the number of seeds and carbon excess, expressed by Eq.~\ref{ns_cex}, might have to be revised introducing  a power law of the carbon excess and/or including a dependence on the mass-loss rate.

The values of $\sigma$ as a function of J~$-$~Ks for some selected combinations of optical data sets and grain sizes are compared in the two panels of Fig.~\ref{sigma}. The bad performing combinations are plotted in the upper panel, while the ones in good agreement with the observations are shown in the lower panel.
For J~$-$~Ks$\la 2.2$, the performances of the data sets in the two panels are roughly comparable, except for Zubko3 with $a_{\rm amC}\sim 0.12$~$\mu$m and Rouleau with $a_{\rm amC}\sim 0.5$~$\mu$m, for which $\sigma>1.5$ for all the values of J~$-$~Ks.
For the reddest bins (J~$-$~Ks$\ga 3$) the values of $\sigma$ for the combinations in the upper panel are always larger than the ones obtained for the models in the lower panel for which $\sigma<1.5$ for all the J~$-$~Ks.

In the lower panel of Fig.~\ref{sigma}, Jaeger400 data set, characterized by high sp$^2$/sp$^3$ ratio, always shows a better agreement with the observations than Jaeger1000 for J~$-$~Ks$\la 2$. This trend suggests that carbon grains with more graphite-like structure might be preferred to diamond-like ones, for low values of mass-loss rate.
In Fig.~\ref{sigma_jaeger1000} we compare the optical data sets similar to Jaeger1000. Among these data sets, a good agreement between observations and models is obtained for Zubko1 and Zubko2 with $a_{\rm amC}\sim 0.06$~$\mu$m for all the values of J~$-$~Ks. However, also for these two data sets the reddest stars show larger values of $\sigma$ ($\sim 0.8$) than for the Rouleau data set with $a_{\rm amC}\sim 0.06$~$\mu$m.

The indication we derive in this investigation -- smaller grains in more evolved and redder stars--  is not in line with the results obtained in other works, in which the optical constants by \citet{Hanner88} are adopted a priori, without a calibration of the most suitable data set and grain size \citep[][and references therein]{Ventura16}.
In particular, their typical size of carbon grains is larger for more evolved, redder stars (up to 0.2~$\mu$m or 0.3~$\mu$m, for heavily dust-enshrouded C-stars in the SMC and LMC, respectively).
These differences are likely due to the different assumptions regarding the starting number of seeds,
which does not scale with the carbon excess in Ventura's works, and to different prescriptions of the TP-AGB phase.

In Figs. from \ref{CCDs_368} to \ref{CCDs_5883645} we show some representative observed CCDs superimposed with the average values (arithmetic means) obtained for our sampled TP-AGB models in each bin in J~$-$~Ks. Average values of the observed data in the different J~$-$~Ks bins are also shown (filled blue circles). We employ the same combinations of optical data sets and typical grain sizes as in Fig.~\ref{sigma}, producing poor (upper panel) and good (lower panel) agreement with the observations.

Even if only some selected examples of CCDs are shown, the results obtained from the analysis of the $\sigma$ values are confirmed by all the CCDs included in our study. In fact, the methodology we described in the previous section reproduces the selected colors simultaneously.

Some combinations of the optical data set and grain sizes show a good agreement with the observed data in some CCDs, but fail to reproduce the observations in others. For example, models computed with Hanner data set and with $a_{\rm amC}\sim 0.12, 0.2$~$\mu$m show a reasonably fair agreement with the observations in the J~$-$~Ks vs J~$-$~[3.6] CCD (Fig.~\ref{CCDs_j36}), but show a very poor agreement in the J~$-$~Ks vs [3.6]~$-$~[8.0] and J~$-$~Ks vs J~$-$~[8.0] CCDs (Figs.~\ref{CCDs_368} and \ref{CCDs_j8}).
Another example are the models computed with Zubko3 and $a_{\rm}\sim 0.12$~$\mu$m, which seem to very well reproduce the observed data in the [5.8]$-$[8.0] vs [3.6]~$-$~[4.5] CCD (Fig.~\ref{CCDs_5883645}), but perform very poorly in several other CCDs (J~$-$~Ks vs [3.6]~$-$~[8.0], J~$-$~Ks vs J~$-$~[8.0], J~$-$~Ks vs Ks~$-$~[8.0], in Figs.~\ref{CCDs_368}, \ref{CCDs_j8}, \ref{CCDs_k8}).

The examples reported, highlight again the importance of considering many colors to be reproduced simultaneously, without restricting the analysis to individual CCDs.

All the combinations with $a_{\rm amC} \ga 0.2$~$\mu$m (upper panels of the figures) show  significant differences between synthetic and observed colors in at least one of the CCDs.
Models computed for the Hanner and Rouleau data sets with $a_{\rm amC}\sim 0.2$~$\mu$m, show too low values of the [3.6]~$-$~[8.0], J~$-$~[3.6], J~$-$~[8.0], Ks~$-$~[3.6], Ks~$-$~[8.0] colors for all the values of J~$-$~Ks (Figs.~\ref{CCDs_368}-\ref{CCDs_k8}). On the other hand, the results for Zubko3 with $a_{\rm amC}\sim 0.5$~$\mu$m show too red colors in these bands for a given value of J~$-$~Ks. Jaeger400 with $a_{\rm amC}\sim 0.5$~$\mu$m performs reasonably well in some of the CCDs but it is not reproducing the bulk of data in the J~$-$~Ks vs Ks~$-$~[8.0] and J~$-$~Ks vs [3.6]~$-$~[8.0], for which it shows too low values of the [3.6]~$-$~[8.0] and Ks~$-$~[8.0] colors for a given J~$-$~Ks (Figs.~\ref{CCDs_368} and \ref{CCDs_k8}).

Conversely, in the lower panels of the figures, all the CCDs show a good agreement between the selected models and the observations. The J~$-$~Ks vs J~$-$~[3.6] and J~$-$~[8.0] CCDs in Figs.~\ref{CCDs_j36} and \ref{CCDs_j8} are characterized by a particularly tight correlation which is very nicely reproduced by the models shown.

For the models obtained by employing the Rouleau data set with $a_{\rm amC}\sim 0.12$~$\mu$m, J~$-$~Ks vs [3.6]~$-$~[8.0] in Fig.~\ref{CCDs_368} shows too small values of [3.6]~$-$~[8.0] color for J~$-$~Ks$\sim 4.5$, which is instead well reproduced by models with smaller grains and the same optical data set.
The shift to larger values of the [3.6]~$-$~[8.0] color obtained by decreasing the grain size for the Rouleau data set, can be qualitatively explained by referring to our initial investigation shown in Fig.~\ref{NIR_rouleau}.
Considering a fixed value of $\tau_1\sim 5$ with J~$-$~Ks$\sim 4$, a variation in the grain size from $\sim 0.1$ to $0.05$~$\mu$m produces a shift in the [3.6]~$-$~[8.0] color of about 0.4 magnitudes, which is required to better reproduce the observed trend in the J~$-$~Ks vs [3.6]~$-$~[8.0] CCD.

All the trends found in our calculations can be qualitatively understood through the analysis presented in Section~\ref{dusty:grain_size}. As an example, we focus on the J~$-$~Ks vs [3.6]~$-$~[8.0] CCDs in Fig.~\ref{CCDs_368} for the reddest models, and we compare the results for two different data sets (Rouleau and Zubko3) with the same grain size $a_{\rm amC}\sim 0.12$~$\mu$m. In Fig.~\ref{col_allops}, for $a_{\rm amC}=0.1$~$\mu$m and $\tau_1=10$, the J~$-$~Ks color is redder for Rouleau than for Zubko3 data set, but the [3.6]~$-$~[8.0] colors are similar. As a consequence, for this grain size, we expect redder colors in the J~$-$~Ks for Rouleau than for Zubko3 for the same [3.6]~$-$~[8.0].

For Jeager400 the [3.6]~$-$[8.0] color remains about the same for all the grain sizes shown in Fig.~\ref{col_allops}, while the J~$-$~Ks color is larger for larger grains. This results in a corresponding shift of the models from $a_{\rm amC}\sim 0.5$ to $a_{\rm amC}\sim 0.06$~$\mu$m to lower values of J~$-$~Ks for a given [3.6]~$-$~[8.0].

From Fig.~\ref{NIR_rouleau} it is possible to understand the trend with the grain size for the Rouleau data set. For the reddest models ($\tau_1=10$), from $a_{\rm amC}=0.05$ to 0.2~$\mu$m the J~$-$~Ks colors remain approximately the same, while the models become less red in the [3.6]~$-$~[8.0] color. As a consequence in Fig.~\ref{CCDs_368}, for a given value of the J~$-$~Ks color, we find that the [3.6]~$-$~[8.0] color becomes less red from $a_{\rm amC}\sim 0.06$ to 0.2~$\mu$m. On the other hand, for $a_{\rm amC}=0.4$~$\mu$m, J~$-$~Ks has about the same value as for $a_{\rm amC}=0.2$~$\mu$m, but the [3.6]~$-$~[8.0] color is almost one magnitude larger for $a_{\rm amc}=0.4$~$\mu$m. Therefore, in the upper panel of Fig.~\ref{CCDs_368}, for the same value of J~$-$~Ks, the colors are shifted to redder [3.6]~$-$~[8.0] from $a_{\rm amC}\sim 0.2$~$\mu$m to 0.5~$\mu$m for the Rouleau data set.

\begin{figure}
\includegraphics[angle=90, width=0.48\textwidth]{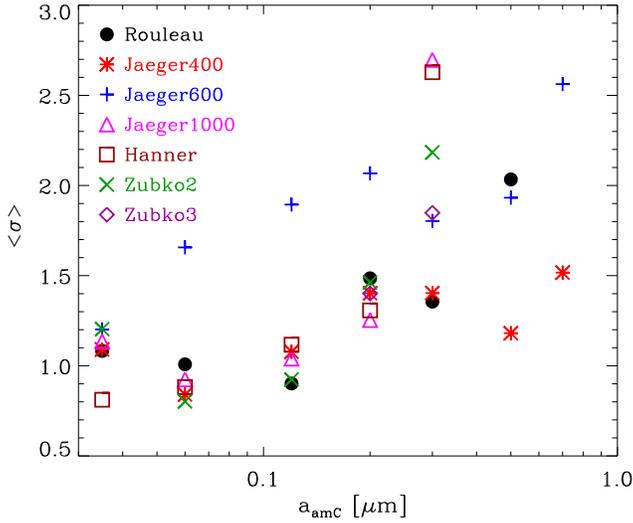}
        \caption{Deviations between models and observations for the different combinations of grain sizes and optical data sets. The value of $<\sigma>$ is the average of $\sigma$ in all the bins.}
        \label{Summary_fig}
        \end{figure}

In Fig.~\ref{Summary_fig} the deviations between models and observations averaged in all the bins, $<\sigma>$, are shown for the different combinations of grain sizes and optical data sets. This figure summarizes which combinations best reproduce the selected colors simultaneously for all the values of J~$-$~Ks. The combinations of optical data sets and grain size which do not produce models at least in one of the bins have not been plotted (see also the tables in the Appendix). The results plotted in Fig.~\ref{Summary_fig} are in agreement with the conclusions we draw in the previous discussion. In particular, the overall agreement between observations and models for all the values of J~$-$~Ks is reached for some optical data sets with $a_{\rm amC}\la 0.1$~$\mu$m, while the discrepancy between observations and models tend to increase with the grain size for a given optical data set. Some of the optical data sets perform comparably well, i.e.
Zubko2, Hanner, Jaeger400, Jaeger1000 and Rouleau data sets with $a_{\rm amC}\la 0.06$~$\mu$m. On the other hand, some of the data sets are never in good agreement with the observations (see for example Jaeger600).

To test the sensitivity of our results to the adopted bin sampling we perform the same analysis dividing the J~$-$~Ks range in five logarithmic bins.  The average values of J~$-$~Ks are (J~$-$~Ks)$_{\rm av}\sim 1.3, 1.7, 2.4, 3.1, 4.1$ with $N_{\rm obs, stars}= 864, 849, 138, 130, 31$, respectively. Doing that, a larger number of observed stars populates now the reddest bins (J~$-$~Ks$\ga 3$). Nevertheless, the previously described trends are recovered and no significant difference arises.

Our best performing combinations of optical constants and grain sizes will be tested in future works by employing complete simulations of stellar populations rather than models selected along the TP-AGB tracks. By employing complete stellar populations syntheses, the quantity $\sigma$ will be correctly weighted for the number of TP-AGB stars in the different bins.

\section{Summary and conclusions}
\label{closing}
This paper is aimed at putting further astrophysical constraints on the sets of carbonaceous dust grain optical properties
found in the literature.
For this purpose we couple our recent TP-AGB models \citep{Bressanetal12, marigoetal13, rosenfield14} with dust
formation \citep{Nanni13, Nanni14} and a RT code \citep{Groenewegen12}. For the first time, we carry out
a systematic analysis of the performance of different optical data sets and grain sizes
compared with observations of C- and x-stars in the SMC \citep{Boyer11}.

The main conclusions of our analysis of dusty CSEs are summarized as follows:
\begin{itemize}
\item For a fixed value of $\tau_1$, different
grain sizes produce a considerable change in the final colors (up to two magnitudes), which are larger for more dust-enshrouded CSEs.
\item For a fixed value of $\tau_1$, the differences in the final colors obtained by selecting different
optical data sets is remarkable (up to four magnitudes in the most extreme case analyzed), especially for the reddest stars.
\item The assumption underlying our dust formation scheme yields a
typical grain size which is mostly determined by the choice
of the initial number of available seeds, with a modest dependence on the variation of stellar
parameters. This result is in agreement with hydrodynamical simulations by \citet{Mattsson10}.
\item The dust temperature at the boundary of the dust formation zone is strongly affected by the choice of the optical data set.
\item The final value of $\tau_1$ is also affected by the final grain size and optical data set.
\end{itemize} 

Using a least squares minimization method we derived the combinations of optical data sets and final grain sizes that best reproduce simultaneously the observed NIR and MIR colors of the models along the TP-AGB tracks.
As expected, the performance of different optical grain properties is tightly coupled
with their size. In particular, we conclude what follows.
\begin{itemize}
\item Some of the combinations
of data sets and grain sizes performing well in a specific CCD can
yield poor results in other diagrams.
\item The colors considered are best reproduced by small grains of $0.035 \la a_{\rm amC}\la 0.12$~$\mu$m rather than by large ones $a_{\rm amC}\ga 0.2$~$\mu$m.
Independently of the optical data set adopted, larger grains are never able to reproduce all the observed colors for all ranges considered, which makes our results robust.
\item Models computed by employing the optical data set by \citet{Rouleau91} and with grain size $a_{\rm amC}\sim 0.12$~$\mu$m reproduce the observed colors for almost all the ranges well, except for the reddest bin (with J~$-$Ks~$\ga$4). For these heavily dust-enshrouded objects, the best agreement with data is achieved for a smaller grain size of $a_{\rm amC}\sim 0.06$~$\mu$m.

Also the other data sets tend to show a better agreement with the observed stars in the reddest bin if the grain size becomes smaller.
This finding suggests a possible inverse trend between the carbon grain size and the mass-loss rate and/or the carbon excess.
\item Models computed with the optical data sets Jaeger400, Jaeger1000, Zubko1 and Zubko2 with $a_{\rm amC}\sim 0.06$~$\mu$m are also in good agreement with observations in the entire color ranges. However, for the reddest stars, they perform worse than models computed with Rouleau data set and with $a_{\rm amC}\sim 0.06$~$\mu$m and Hanner with $a_{\rm amC}\sim 0.035$~$\mu$m.

\item The better agreement of the models computed with Jaeger400 rather than Jaeger1000 for J~$-$~Ks$\la 3$ suggests that carbon grains might be characterized by the presence of more graphite-like bounds in their structure in stars with low mass-loss.
\end{itemize}

In the future investigations we will extend our analysis to other samples of stars in galaxies with different metallicity and to different spectral types, as M-stars (Nanni et al.~in preparation).

Further comparisons between modeled and observed expansion velocities of TP-AGB stars will help in constraining the most suitable carbon data set and final grain size of amC produced in C-stars.

The revised dust formation model here presented will be an essential ingredient in complete stellar population simulations performed by the TRILEGAL code, and will soon be included in our theoretical isochrones (Marigo et al.~submitted).

\subsection*{Acknowledgements}
This work was supported by the ERC Consolidator Grant funding scheme
({\em project STARKEY}, G.A. n.~615604).

\bibliographystyle{mn2e/mn2e_new}
\bibliography{nanni}

\begin{thebibliography}{66}
\expandafter\ifx\csname natexlab\endcsname\relax\def\natexlab#1{#1}\fi

\bibitem[{{Andersen}, {Loidl} \& {H{\"o}fner}(1999){Andersen}, {Loidl}, \&
  {H{\"o}fner}}]{Andersen99}
{Andersen} A.~C., {Loidl} R., {H{\"o}fner} S., 1999, \aap, 349, 243

\bibitem[{{Aringer} {et~al}\mbox{.}(2016){Aringer}, {Girardi}, {Nowotny},
  {Marigo}, \& {Bressan}}]{Aringer16}
{Aringer} B., {Girardi} L., {Nowotny} W., {Marigo} P., {Bressan} A., 2016,
  \mnras

\bibitem[{{Aringer} {et~al}\mbox{.}(2009){Aringer}, {Girardi}, {Nowotny},
  {Marigo}, \& {Lederer}}]{Aringer09}
{Aringer} B., {Girardi} L., {Nowotny} W., {Marigo} P., {Lederer} M.~T., 2009,
  \aap, 503, 913

\bibitem[{{Bohren}, {Huffman} \& {Kam}(1983){Bohren}, {Huffman}, \&
  {Kam}}]{Bohren83}
{Bohren} C.~F., {Huffman} D.~R., {Kam} Z., 1983, \nat, 306, 625

\bibitem[{{Bolatto} {et~al}\mbox{.}(2007){Bolatto}, {Simon},
  {Stanimirovi{\'c}}, {van Loon}, {Shah}, {Venn}, {Leroy}, {Sandstrom},
  {Jackson}, {Israel}, {Li}, {Staveley-Smith}, {Bot}, {Boulanger}, \&
  {Rubio}}]{bolatto07}
{Bolatto} A.~D. {et~al.}, 2007, \apj, 655, 212

\bibitem[{{Boyer} {et~al}\mbox{.}(2011){Boyer}, {Srinivasan}, {van Loon},
  {McDonald}, {Meixner}, {Zaritsky}, {Gordon}, {Kemper}, {Babler}, {Block},
  {Bracker}, {Engelbracht}, {Hora}, {Indebetouw}, {Meade}, {Misselt},
  {Robitaille}, {Sewi{\l}o}, {Shiao}, \& {Whitney}}]{Boyer11}
{Boyer} M.~L. {et~al.}, 2011, \aj, 142, 103

\bibitem[{{Bressan}, {Granato} \& {Silva}(1998){Bressan}, {Granato}, \&
  {Silva}}]{Bressan98}
{Bressan} A., {Granato} G.~L., {Silva} L., 1998, \aap, 332, 135

\bibitem[{{Bressan} {et~al}\mbox{.}(2012){Bressan}, {Marigo}, {Girardi},
  {Salasnich}, {Dal Cero}, {Rubele}, \& {Nanni}}]{Bressanetal12}
{Bressan} A., {Marigo} P., {Girardi} L., {Salasnich} B., {Dal Cero} C.,
  {Rubele} S., {Nanni} A., 2012, \mnras, 427, 127

\bibitem[{{Cherchneff}(2006)}]{Cherchneff06}
{Cherchneff} I., 2006, \aap, 456, 1001

\bibitem[{{Cherchneff}, {Barker} \& {Tielens}(1992){Cherchneff}, {Barker}, \&
  {Tielens}}]{Cherchneff92}
{Cherchneff} I., {Barker} J.~R., {Tielens} A.~G.~G.~M., 1992, \apj, 401, 269

\bibitem[{{Cioni} {et~al}\mbox{.}(2006{\natexlab{a}}){Cioni}, {Girardi},
  {Marigo}, \& {Habing}}]{cioni06a}
{Cioni} M.-R.~L., {Girardi} L., {Marigo} P., {Habing} H.~J.,
  2006{\natexlab{a}}, \aap, 448, 77

\bibitem[{{Cioni} {et~al}\mbox{.}(2006{\natexlab{b}}){Cioni}, {Girardi},
  {Marigo}, \& {Habing}}]{cioni06b}
---, 2006{\natexlab{b}}, \aap, 452, 195

\bibitem[{{Dell'Agli} {et~al}\mbox{.}(2015{\natexlab{a}}){Dell'Agli},
  {Garc{\'{\i}}a-Hern{\'a}ndez}, {Ventura}, {Schneider}, {Di Criscienzo}, \&
  {Rossi}}]{Dellagli15b}
{Dell'Agli} F., {Garc{\'{\i}}a-Hern{\'a}ndez} D.~A., {Ventura} P., {Schneider}
  R., {Di Criscienzo} M., {Rossi} C., 2015{\natexlab{a}}, \mnras, 454, 4235

\bibitem[{{Dell'Agli} {et~al}\mbox{.}(2015{\natexlab{b}}){Dell'Agli},
  {Ventura}, {Schneider}, {Di Criscienzo}, {Garc{\'{\i}}a-Hern{\'a}ndez},
  {Rossi}, \& {Brocato}}]{Dellagli15a}
{Dell'Agli} F., {Ventura} P., {Schneider} R., {Di Criscienzo} M.,
  {Garc{\'{\i}}a-Hern{\'a}ndez} D.~A., {Rossi} C., {Brocato} E.,
  2015{\natexlab{b}}, \mnras, 447, 2992

\bibitem[{{Di Criscienzo} {et~al}\mbox{.}(2013){Di Criscienzo}, {Dell'Agli},
  {Ventura}, {Schneider}, {Valiante}, {La Franca}, {Rossi}, {Gallerani}, \&
  {Maiolino}}]{DiCriscienzo13}
{Di Criscienzo} M. {et~al.}, 2013, \mnras, 433, 313

\bibitem[{{Edoh}(1983)}]{Edoh83}
{Edoh} O., 1983, PhD thesis, THE UNIVERSITY OF ARIZONA.

\bibitem[{{Fazio} {et~al}\mbox{.}(2004){Fazio}, {Hora}, {Allen}, {Ashby},
  {Barmby}, {Deutsch}, {Huang}, {Kleiner}, {Marengo}, {Megeath}, {Melnick},
  {Pahre}, {Patten}, {Polizotti}, {Smith}, {Taylor}, {Wang}, {Willner},
  {Hoffmann}, {Pipher}, {Forrest}, {McMurty}, {McCreight}, {McKelvey},
  {McMurray}, {Koch}, {Moseley}, {Arendt}, {Mentzell}, {Marx}, {Losch},
  {Mayman}, {Eichhorn}, {Krebs}, {Jhabvala}, {Gezari}, {Fixsen}, {Flores},
  {Shakoorzadeh}, {Jungo}, {Hakun}, {Workman}, {Karpati}, {Kichak}, {Whitley},
  {Mann}, {Tollestrup}, {Eisenhardt}, {Stern}, {Gorjian}, {Bhattacharya},
  {Carey}, {Nelson}, {Glaccum}, {Lacy}, {Lowrance}, {Laine}, {Reach},
  {Stauffer}, {Surace}, {Wilson}, {Wright}, {Hoffman}, {Domingo}, \&
  {Cohen}}]{Fazio04}
{Fazio} G.~G. {et~al.}, 2004, \apjs, 154, 10

\bibitem[{{Ferrarotti} \& {Gail}(2006)}]{FG06}
{Ferrarotti} A.~S., {Gail} H.-P., 2006, \aap, 447, 553

\bibitem[{{Frenklach} \& {Feigelson}(1989)}]{Frenklach89}
{Frenklach} M., {Feigelson} E.~D., 1989, \apj, 341, 372

\bibitem[{{Gail} \& {Sedlmayr}(1999)}]{GS99}
{Gail} H.-P., {Sedlmayr} E., 1999, \aap, 347, 594

\bibitem[{{Gordon} {et~al}\mbox{.}(2011){Gordon}, {Meixner}, {Meade},
  {Whitney}, {Engelbracht}, {Bot}, {Boyer}, {Lawton}, {Sewi{\l}o}, {Babler},
  {Bernard}, {Bracker}, {Block}, {Blum}, {Bolatto}, {Bonanos}, {Harris},
  {Hora}, {Indebetouw}, {Misselt}, {Reach}, {Shiao}, {Tielens}, {Carlson},
  {Churchwell}, {Clayton}, {Chen}, {Cohen}, {Fukui}, {Gorjian}, {Hony},
  {Israel}, {Kawamura}, {Kemper}, {Leroy}, {Li}, {Madden}, {Marble},
  {McDonald}, {Mizuno}, {Mizuno}, {Muller}, {Oliveira}, {Olsen}, {Onishi},
  {Paladini}, {Paradis}, {Points}, {Robitaille}, {Rubin}, {Sandstrom}, {Sato},
  {Shibai}, {Simon}, {Smith}, {Srinivasan}, {Vijh}, {Van Dyk}, {van Loon}, \&
  {Zaritsky}}]{gordon11}
{Gordon} K.~D. {et~al.}, 2011, \aj, 142, 102

\bibitem[{{Groenewegen}(2012)}]{Groenewegen12}
{Groenewegen} M.~A.~T., 2012, \aap, 543, A36

\bibitem[{{Groenewegen} {et~al}\mbox{.}(2009){Groenewegen}, {Sloan},
  {Soszy{\'n}ski}, \& {Petersen}}]{Groenewegen09}
{Groenewegen} M.~A.~T., {Sloan} G.~C., {Soszy{\'n}ski} I., {Petersen} E.~A.,
  2009, \aap, 506, 1277

\bibitem[{{Groenewegen} {et~al}\mbox{.}(1998){Groenewegen}, {Whitelock},
  {Smith}, \& {Kerschbaum}}]{Groenewegen98}
{Groenewegen} M.~A.~T., {Whitelock} P.~A., {Smith} C.~H., {Kerschbaum} F.,
  1998, \mnras, 293, 18

\bibitem[{{Groenewegen} {et~al}\mbox{.}(2007){Groenewegen}, {Wood}, {Sloan},
  {Blommaert}, {Cioni}, {Feast}, {Hony}, {Matsuura}, {Menzies}, {Olivier},
  {Vanhollebeke}, {van Loon}, {Whitelock}, {Zijlstra}, {Habing}, \&
  {Lagadec}}]{Groenewegen07}
{Groenewegen} M.~A.~T. {et~al.}, 2007, \mnras, 376, 313

\bibitem[{{Hanner}(1988)}]{Hanner88}
{Hanner} M., 1988, {Grain optical properties}. Tech. rep.

\bibitem[{{H{\"o}fner} {et~al}\mbox{.}(2003){H{\"o}fner}, {Gautschy-Loidl},
  {Aringer}, \& {J{\o}rgensen}}]{Hofner03}
{H{\"o}fner} S., {Gautschy-Loidl} R., {Aringer} B., {J{\o}rgensen} U.~G., 2003,
  \aap, 399, 589

\bibitem[{{Hony}, {Waters} \& {Tielens}(2002){Hony}, {Waters}, \&
  {Tielens}}]{Hony02}
{Hony} S., {Waters} L.~B.~F.~M., {Tielens} A.~G.~G.~M., 2002, \aap, 390, 533

\bibitem[{{Ivezic} \& {Elitzur}(1997)}]{Ivezic97}
{Ivezic} Z., {Elitzur} M., 1997, \mnras, 287, 799

\bibitem[{{Jager}, {Mutschke} \& {Henning}(1998){Jager}, {Mutschke}, \&
  {Henning}}]{Jaeger98}
{Jager} C., {Mutschke} H., {Henning} T., 1998, \aap, 332, 291

\bibitem[{{J{\o}rgensen}, {Hron} \& {Loidl}(2000){J{\o}rgensen}, {Hron}, \&
  {Loidl}}]{Jorgensen00}
{J{\o}rgensen} U.~G., {Hron} J., {Loidl} R., 2000, \aap, 356, 253

\bibitem[{{Kato} {et~al}\mbox{.}(2007){Kato}, {Nagashima}, {Nagayama},
  {Kurita}, {Koerwer}, {Kawai}, {Yamamuro}, {Zenno}, {Nishiyama}, {Baba},
  {Kadowaki}, {Haba}, {Hatano}, {Shimizu}, {Nishimura}, {Nagata}, {Sato},
  {Murai}, {Kawazu}, {Nakajima}, {Nakaya}, {Kandori}, {Kusakabe}, {Ishihara},
  {Kaneyasu}, {Hashimoto}, {Tamura}, {Tanab{\'e}}, {Ita}, {Matsunaga},
  {Nakada}, {Sugitani}, {Wakamatsu}, {Glass}, {Feast}, {Menzies}, {Whitelock},
  {Fourie}, {Stoffels}, {Evans}, \& {Hasegawa}}]{Kato07}
{Kato} D. {et~al.}, 2007, \pasj, 59, 615

\bibitem[{{Kimura} {et~al}\mbox{.}(2002){Kimura}, {Mann}, {Biesecker}, \&
  {Jessberger}}]{Kimura02}
{Kimura} H., {Mann} I., {Biesecker} D.~A., {Jessberger} E.~K., 2002, \icarus,
  159, 529

\bibitem[{{Kobayashi} {et~al}\mbox{.}(2011){Kobayashi}, {Kimura}, {Watanabe},
  {Yamamoto}, \& {M{\"u}ller}}]{Kobayashi11}
{Kobayashi} H., {Kimura} H., {Watanabe} S.-i., {Yamamoto} T., {M{\"u}ller} S.,
  2011, Earth, Planets, and Space, 63, 1067

\bibitem[{{Kobayashi} {et~al}\mbox{.}(2009){Kobayashi}, {Watanabe}, {Kimura},
  \& {Yamamoto}}]{Kobayashi09}
{Kobayashi} H., {Watanabe} S.-i., {Kimura} H., {Yamamoto} T., 2009, \icarus,
  201, 395

\bibitem[{{Lombaert} {et~al}\mbox{.}(2012){Lombaert}, {de Vries}, {de Koter},
  {Decin}, {Min}, {Smolders}, {Mutschke}, \& {Waters}}]{Lombaert12}
{Lombaert} R., {de Vries} B.~L., {de Koter} A., {Decin} L., {Min} M.,
  {Smolders} K., {Mutschke} H., {Waters} L.~B.~F.~M., 2012, \aap, 544, L18

\bibitem[{{Lucy}(1971)}]{Lucy71}
{Lucy} L.~B., 1971, \apj, 163, 95

\bibitem[{{Lucy}(1976)}]{Lucy76}
---, 1976, \apj, 205, 482

\bibitem[{{Marigo} \& {Aringer}(2009)}]{MarigoAringer_09}
{Marigo} P., {Aringer} B., 2009, A\&A, 508, 1539

\bibitem[{{Marigo} {et~al}\mbox{.}(2013){Marigo}, {Bressan}, {Nanni},
  {Girardi}, \& {Pumo}}]{marigoetal13}
{Marigo} P., {Bressan} A., {Nanni} A., {Girardi} L., {Pumo} M.~L., 2013,
  \mnras, 434, 488

\bibitem[{{Marigo} \& {Girardi}(2007)}]{Marigo07}
{Marigo} P., {Girardi} L., 2007, \aap, 469, 239

\bibitem[{{Marigo} {et~al}\mbox{.}(2008){Marigo}, {Girardi}, {Bressan},
  {Groenewegen}, {Silva}, \& {Granato}}]{Marigo08}
{Marigo} P., {Girardi} L., {Bressan} A., {Groenewegen} M.~A.~T., {Silva} L.,
  {Granato} G.~L., 2008, \aap, 482, 883

\bibitem[{{Marigo} {et~al}\mbox{.}(2016){Marigo}, {Ripamonti}, {Nanni},
  {Bressan}, \& {Girardi}}]{Marigo16}
{Marigo} P., {Ripamonti} E., {Nanni} A., {Bressan} A., {Girardi} L., 2016,
  \mnras, 456, 23

\bibitem[{{Matsuura} {et~al}\mbox{.}(2009){Matsuura}, {Barlow}, {Zijlstra},
  {Whitelock}, {Cioni}, {Groenewegen}, {Volk}, {Kemper}, {Kodama}, {Lagadec},
  {Meixner}, {Sloan}, \& {Srinivasan}}]{Matsuura09}
{Matsuura} M. {et~al.}, 2009, \mnras, 396, 918

\bibitem[{{Mattsson}, {Wahlin} \& {H{\"o}fner}(2010){Mattsson}, {Wahlin}, \&
  {H{\"o}fner}}]{Mattsson10}
{Mattsson} L., {Wahlin} R., {H{\"o}fner} S., 2010, \aap, 509, A14

\bibitem[{{Nanni} {et~al}\mbox{.}(2013){Nanni}, {Bressan}, {Marigo}, \&
  {Girardi}}]{Nanni13}
{Nanni} A., {Bressan} A., {Marigo} P., {Girardi} L., 2013, \mnras, 434, 2390

\bibitem[{{Nanni} {et~al}\mbox{.}(2014){Nanni}, {Bressan}, {Marigo}, \&
  {Girardi}}]{Nanni14}
---, 2014, \mnras

\bibitem[{{Nowotny} {et~al}\mbox{.}(2011){Nowotny}, {Aringer}, {H{\"o}fner}, \&
  {Lederer}}]{Nowotny11}
{Nowotny} W., {Aringer} B., {H{\"o}fner} S., {Lederer} M.~T., 2011, \aap, 529,
  A129

\bibitem[{{Riebel} {et~al}\mbox{.}(2012){Riebel}, {Srinivasan}, {Sargent}, \&
  {Meixner}}]{Riebel12}
{Riebel} D., {Srinivasan} S., {Sargent} B., {Meixner} M., 2012, \apj, 753, 71

\bibitem[{{Rieke} {et~al}\mbox{.}(2004){Rieke}, {Young}, {Engelbracht},
  {Kelly}, {Low}, {Haller}, {Beeman}, {Gordon}, {Stansberry}, {Misselt},
  {Cadien}, {Morrison}, {Rivlis}, {Latter}, {Noriega-Crespo}, {Padgett},
  {Stapelfeldt}, {Hines}, {Egami}, {Muzerolle}, {Alonso-Herrero}, {Blaylock},
  {Dole}, {Hinz}, {Le Floc'h}, {Papovich}, {P{\'e}rez-Gonz{\'a}lez}, {Smith},
  {Su}, {Bennett}, {Frayer}, {Henderson}, {Lu}, {Masci}, {Pesenson}, {Rebull},
  {Rho}, {Keene}, {Stolovy}, {Wachter}, {Wheaton}, {Werner}, \&
  {Richards}}]{Rieke04}
{Rieke} G.~H. {et~al.}, 2004, \apjs, 154, 25

\bibitem[{{Rosenfield} {et~al}\mbox{.}(2014){Rosenfield}, {Marigo}, {Girardi},
  {Dalcanton}, {Bressan}, {Gullieuszik}, {Weisz}, {Williams}, {Dolphin}, \&
  {Aringer}}]{rosenfield14}
{Rosenfield} P. {et~al.}, 2014, \apj, 790, 22

\bibitem[{{Rosenfield} {et~al}\mbox{.}(2016){Rosenfield}, {Marigo}, {Girardi},
  {Dalcanton}, {Bressan}, {Williams}, \& {Dolphin}}]{Rosenfield16}
{Rosenfield} P., {Marigo} P., {Girardi} L., {Dalcanton} J.~J., {Bressan} A.,
  {Williams} B.~F., {Dolphin} A., 2016, \apj, 822, 73

\bibitem[{{Rouleau} \& {Martin}(1991)}]{Rouleau91}
{Rouleau} F., {Martin} P.~G., 1991, \apj, 377, 526

\bibitem[{{Skrutskie} {et~al}\mbox{.}(2006){Skrutskie}, {Cutri}, {Stiening},
  {Weinberg}, {Schneider}, {Carpenter}, {Beichman}, {Capps}, {Chester},
  {Elias}, {Huchra}, {Liebert}, {Lonsdale}, {Monet}, {Price}, {Seitzer},
  {Jarrett}, {Kirkpatrick}, {Gizis}, {Howard}, {Evans}, {Fowler}, {Fullmer},
  {Hurt}, {Light}, {Kopan}, {Marsh}, {McCallon}, {Tam}, {Van Dyk}, \&
  {Wheelock}}]{skrutskie06}
{Skrutskie} M.~F. {et~al.}, 2006, \aj, 131, 1163

\bibitem[{{Sloan} {et~al}\mbox{.}(2015){Sloan}, {Lagadec}, {Kraemer}, {Boyer},
  {Srinivasan}, {McDonald}, \& {Zijlstra}}]{Sloan15}
{Sloan} G.~C., {Lagadec} E., {Kraemer} K.~E., {Boyer} M.~L., {Srinivasan} S.,
  {McDonald} I., {Zijlstra} A.~A., 2015, in Astronomical Society of the Pacific
  Conference Series, Vol. 497, Why Galaxies Care about AGB Stars III: A Closer
  Look in Space and Time, {Kerschbaum} F., {Wing} R.~F., {Hron} J., eds., p.
  429

\bibitem[{{Srinivasan} {et~al}\mbox{.}(2016){Srinivasan}, {Boyer}, {Kemper},
  {Meixner}, {Sargent}, \& {Riebel}}]{Srinivasan16}
{Srinivasan} S., {Boyer} M.~L., {Kemper} F., {Meixner} M., {Sargent} B.~A.,
  {Riebel} D., 2016, \mnras, 457, 2814

\bibitem[{{van Loon} {et~al}\mbox{.}(2008){van Loon}, {Cohen}, {Oliveira},
  {Matsuura}, {McDonald}, {Sloan}, {Wood}, \& {Zijlstra}}]{vanLoon08}
{van Loon} J.~T., {Cohen} M., {Oliveira} J.~M., {Matsuura} M., {McDonald} I.,
  {Sloan} G.~C., {Wood} P.~R., {Zijlstra} A.~A., 2008, \aap, 487, 1055

\bibitem[{{van Loon} {et~al}\mbox{.}(2006){van Loon}, {McDonald}, {Oliveira},
  {Evans}, {Boyer}, {Gehrz}, {Polomski}, \& {Woodward}}]{vanLoon06}
{van Loon} J.~T., {McDonald} I., {Oliveira} J.~M., {Evans} A., {Boyer} M.~L.,
  {Gehrz} R.~D., {Polomski} E., {Woodward} C.~E., 2006, \aap, 450, 339

\bibitem[{{van Loon} {et~al}\mbox{.}(1997){van Loon}, {Zijlstra}, {Whitelock},
  {Waters}, {Loup}, \& {Trams}}]{vanLoon97}
{van Loon} J.~T., {Zijlstra} A.~A., {Whitelock} P.~A., {Waters} L.~B.~F.~M.,
  {Loup} C., {Trams} N.~R., 1997, \aap, 325, 585

\bibitem[{{Ventura} {et~al}\mbox{.}(2012){Ventura}, {Criscienzo}, {Schneider},
  {Carini}, {Valiante}, {D'Antona}, {Gallerani}, {Maiolino}, \&
  {Tornamb{\'e}}}]{ventura12}
{Ventura} P. {et~al.}, 2012, \mnras, 424, 2345

\bibitem[{{Ventura} {et~al}\mbox{.}(2014){Ventura}, {Dell'Agli}, {Schneider},
  {Di Criscienzo}, {Rossi}, {La Franca}, {Gallerani}, \&
  {Valiante}}]{Ventura14}
{Ventura} P., {Dell'Agli} F., {Schneider} R., {Di Criscienzo} M., {Rossi} C.,
  {La Franca} F., {Gallerani} S., {Valiante} R., 2014, \mnras, 439, 977

\bibitem[{{Ventura} {et~al}\mbox{.}(2016){Ventura}, {Karakas}, {Dell'Agli},
  {Garc{\'{\i}}a-Hern{\'a}ndez}, {Boyer}, \& {Di Criscienzo}}]{Ventura16}
{Ventura} P., {Karakas} A.~I., {Dell'Agli} F., {Garc{\'{\i}}a-Hern{\'a}ndez}
  D.~A., {Boyer} M.~L., {Di Criscienzo} M., 2016, \mnras, 457, 1456

\bibitem[{{Woods} {et~al}\mbox{.}(2011){Woods}, {Oliveira}, {Kemper}, {van
  Loon}, {Sargent}, {Matsuura}, {Szczerba}, {Volk}, {Zijlstra}, {Sloan},
  {Lagadec}, {McDonald}, {Jones}, {Gorjian}, {Kraemer}, {Gielen}, {Meixner},
  {Blum}, {Sewi{\l}o}, {Riebel}, {Shiao}, {Chen}, {Boyer}, {Indebetouw},
  {Antoniou}, {Bernard}, {Cohen}, {Dijkstra}, {Galametz}, {Galliano}, {Gordon},
  {Harris}, {Hony}, {Hora}, {Kawamura}, {Lawton}, {Leisenring}, {Madden},
  {Marengo}, {McGuire}, {Mulia}, {O'Halloran}, {Olsen}, {Paladini}, {Paradis},
  {Reach}, {Rubin}, {Sandstrom}, {Soszy{\'n}ski}, {Speck}, {Srinivasan},
  {Tielens}, {van Aarle}, {van Dyk}, {van Winckel}, {Vijh}, {Whitney}, \&
  {Wilkins}}]{Woods11}
{Woods} P.~M. {et~al.}, 2011, \mnras, 411, 1597

\bibitem[{{Zaritsky} {et~al}\mbox{.}(2002){Zaritsky}, {Harris}, {Thompson},
  {Grebel}, \& {Massey}}]{Zaritsky02}
{Zaritsky} D., {Harris} J., {Thompson} I.~B., {Grebel} E.~K., {Massey} P.,
  2002, \aj, 123, 855

\bibitem[{{Zhukovska}, {Gail} \& {Trieloff}(2008){Zhukovska}, {Gail}, \&
  {Trieloff}}]{Zhukovska08}
{Zhukovska} S., {Gail} H.-P., {Trieloff} M., 2008, \aap, 479, 453

\bibitem[{{Zubko} {et~al}\mbox{.}(1996){Zubko}, {Mennella}, {Colangeli}, \&
  {Bussoletti}}]{Zubko96}
{Zubko} V.~G., {Mennella} V., {Colangeli} L., {Bussoletti} E., 1996, \mnras,
  282, 1321

\end{thebibliography}

\begin{appendix}
\begin{table*}
\begin{center}
\caption{Deviations between observations and models for the Rouleau data set and different grain sizes. The deviations are listed for the individual colors (Eq.~\ref{sigma_star}) and for the all the colors ($\sigma$ in Eq.~\ref{tot_sigma}). Hyphens are used when the color is not taken into account for the calculation of the total $\sigma$. The acronym ``ns=no stars'' is adopted when no models are in the range of J$-$Ks reported in the header of the table.}
\label{Table:results}
\begin{tabular}{l c c c c c c c c c}
\hline
Opacity set: &Rouleau & & & & & & & \\
\hline
J$-$Ks$_{\rm av}\sim1.5$ & $1.1 \la J-Ks \la1.9 $ &  $N_{\rm obs, stars}=1630$ & & & & & &  &\\
\hline
$a_{\rm amC} [\mu m]$ & $\sigma_{\rm J-Ks}$ & $\sigma_{\rm J-[3.6]}$ & $\sigma_{\rm J-[8.0]}$ & $\sigma_{\rm Ks-[3.6]}$ & $\sigma_{\rm Ks-[8.0]}$ & $\sigma_{\rm [3.6]-[8.0]}$ & $\sigma_{\rm [5.8]-[8.0]}$ & $\sigma_{\rm [3.6]-[4.5]} $ & $\sigma$  \\
\hline
0.035&       1.1 &       1.5 &       1.5  &     1.6  & 1.5 &       1.3 &  - &  - & 1.4 \\
0.06&       1.0 &       1.3 &       1.3 &       1.4 &    1.4 &       1.4 &      - &      - &    1.3\\
0.12&      0.94 &      0.91 &       1.1 &       1.1 &       1.4 &       1.5 &       - &   - & 1.1 \\
0.2&       1.1 &      0.86 &       1.2 &       1.0 &       1.5 &       1.7 &       -  &       - &       1.2 \\
0.3&       1.2 &       1.1 &       1.3 &       1.1 &       1.5 &       1.7 &       - &      - &       1.3\\
0.5&       1.5 &       1.7 &       1.5 &       1.5 & 1.3 &       1.4 &       - &       - & 1.5 \\
0.7&       1.4 &       3.9 &       3.8 &       5.4 & 4.3 &       2.9 &       - &       - & 3.6\\
\hline
J~$-$~Ks$_{\rm av}\sim2.2$ & $1.9\la J-Ks\la2.7$  &    $N_{\rm obs, stars}=212$ & & & & & &  &\\
\hline
$a_{\rm amC} [\mu m]$ & $\sigma_{\rm J-Ks}$ & $\sigma_{\rm J-[3.6]}$ & $\sigma_{\rm J-[8.0]}$ & $\sigma_{\rm Ks-[3.6]}$ & $\sigma_{\rm Ks-[8.0]}$ & $\sigma_{\rm [3.6]-[8.0]}$ & $\sigma_{\rm [5.8]-[8.0]}$ & $\sigma_{\rm [3.6]-[4.5]} $ & $\sigma$\\
\hline
0.035&      0.88 &      0.87 &      0.72 &      0.86 & 0.66 &      0.49 &      - &      - & 0.75 \\
0.06&      0.87 &      0.90 &      0.74 &      0.86 & 0.67 &      0.50 &      - &      - &  0.76 \\
0.12&      0.87 &      0.75 &      0.64 &      0.62 & 0.55 &      0.50 &      - &      - &  0.65 \\
0.2&      0.80 &      0.71 &      0.84 &      0.66 & 0.89 &      0.93 &       - &      - & 0.81 \\
0.3&      0.86 &      0.75 &      0.89 5&      0.66 &  0.93 &  1.0 &       - &    - &      0.85\\
0.5&       1.0 &       2.5 &       1.9 &       3.2 & 2.0 &       1.1 &       - &       - & 1.9 \\
0.7&       1.1 &       3.9 &       3.8 &       5.4 & 4.4 &       3.0 &      - &       -  & 3.6 \\
\hline
J$-$Ks$_{\rm av}\sim3.0$ &  $2.7\la J-Ks\la3.4 $ &   $N_{\rm obs, stars}=117$ & & & & & &  \\
\hline
$a_{\rm amC} [\mu m]$ & $\sigma_{\rm J-Ks}$ & $\sigma_{\rm J-[3.6]}$ & $\sigma_{\rm J-[8.0]}$ & $\sigma_{\rm Ks-[3.6]}$ & $\sigma_{\rm Ks-[8.0]}$ & $\sigma_{\rm [3.6]-[8.0]}$ & $\sigma_{\rm [5.8]-[8.0]}$ & $\sigma_{\rm [3.6]-[4.5]} $ & $\sigma$  \\
\hline
0.035&       1.3 &       1.0 &       1.3 &      0.83 & 1.2 &      0.62 &      0.61 &      0.69 & 0.94 \\
0.06&       1.4 &       1.4 &       1.8 &       1.1 &  1.7 &      0.88 &     0.98 &      0.93 &  1.3 \\
0.12&       1.6 &      0.90 &      0.97 &      0.46 &   0.67&      0.44 &    0.51 &      0.33 &   0.73 \\
0.2&       1.2 &      0.96 &       1.9 &      0.95 &  2.2 &       1.7 &      1.1 &       1.2 &   1.4 \\
0.3&       1.2 &      0.98 &       2.1 &      0.96 &  2.4 &       1.9 &       1.3 &       1.3 &       1.5\\
0.5&       1.3 &       2.5 &       2.7 &       3.2 & 3.4 &      0.94 &      0.63 &       1.5 &  2.0 \\
0.7&       1.1 &       3.2 &       5.2 &       4.2 & 6.8 &       3.5 &       1.2 &       4.6 &   3.7 \\
\hline
J~$-$~Ks$_{\rm av}\sim3.7$ & $3.4\la J-Ks \la4.2$ &   $N_{\rm obs, stars}=43$ & & & & & &  &\\
\hline
$a_{\rm amC} [\mu m]$ & $\sigma_{\rm J-Ks}$ & $\sigma_{\rm J-[3.6]}$ & $\sigma_{\rm J-[8.0]}$ & $\sigma_{\rm Ks-[3.6]}$ & $\sigma_{\rm Ks-[8.0]}$ & $\sigma_{\rm [3.6]-[8.0]}$ & $\sigma_{\rm [5.8]-[8.0]}$ & $\sigma_{\rm [3.6]-[4.5]} $ & $\sigma$   \\
\hline
0.035 &       1.3 &       1.2  &       1.7 &       1.0 & 1.6 &       1.4 &       1.2 &       1.0 & 1.3 \\
0.06 &       1.7  &       1.1  &       1.5 &      0.70 & 1.2 &       1.1 &      0.99 &      0.88 & 1.1 \\
0.12 &       1.2  &      0.70 &       1.1 &      0.51  & 1.2 &       1.3 &      0.76 &      0.74 & 0.93 \\
0.2 &      0.98  &       1.0  &       2.0 &       1.2 & 2.5 &       2.5 &       1.4 &       1.7 &  1.7 \\
0.3&       1.0 &      0.87 &       1.8 &      0.93 &       2.3 &       2.6 &       1.5 &       1.7 & 1.6 \\
0.5 &      0.97  &       2.8  &       3.1 &       3.4 &  3.8 &       1.4 &      0.50 &      2.0 & 2.2 \\
0.7 &       1.0  &       1.7 &       3.1 &       2.5 & 4.3 &       3.7 &       1.3 &       3.4 & 2.6 \\
\hline
J~$-$~Ks$_{\rm av}\sim4.5$ &  $4.2\la J-Ks \la5.0 $ &   $N_{\rm obs, stars}=10$ & & & & & & & \\
\hline
$a_{\rm amC} [\mu m]$ & $\sigma_{\rm J-Ks}$ & $\sigma_{\rm J-[3.6]}$ & $\sigma_{\rm J-[8.0]}$ & $\sigma_{\rm Ks-[3.6]}$ & $\sigma_{\rm Ks-[8.0]}$ & $\sigma_{\rm [3.6]-[8.0]}$ & $\sigma_{\rm [5.8]-[8.0]}$ & $\sigma_{\rm [3.6]-[4.5]} $ & $\sigma$ \\
\hline
0.035&       1.3 &      0.70 &       1.4 &       1.1 &     1.7 &      0.79 &      0.45 &      0.83 &    1.0\\
0.06&       1.1 &      0.29 &      0.47 &      0.56 &    0.64 &      0.46 &      0.27 &      0.45 &   0.53 \\
0.12&      0.84 &      0.84 &       2.1 &      0.59 &    1.8 &       1.5 &      0.30 &      0.48 &  1.0\\
0.2&       1.4 &       1.6 &       4.5 &       1.5 & 4.1 &       3.2 &      0.71 &       1.7 & 2.3 \\
0.3&       1.2 &      0.88 &       2.4 &      0.53 &       2.3 &       2.8 &      0.71 &       1.2 &       1.5\\
0.5&       1.1 &       3.9 &       4.6 &       3.5  &    4.0 &      0.68 &      0.34 &       1.6 &      2.5\\
0.7&      ns &       &        &       &       &       &      &   &   \\
\hline
\end{tabular}
\end{center}
\end{table*}

\begin{table*}
\begin{center}
\caption{The same as in Table~\ref{Table:results} but for the Jaeger400 data set.}
\label{Table:results2}
\begin{tabular}{l c c c c c c c c c}
\hline
Opacity set:  & Jaeger400 & & & & & & & \\
\hline
J~$-$~Ks$_{\rm av}\sim1.5$ & $1.1\la J-Ks \la1.9$ &  $N_{\rm obs, stars}=1630$ & & & & & &  &\\
\hline
$a_{\rm amC} [\mu m]$ & $\sigma_{\rm J-Ks}$ & $\sigma_{\rm J-[3.6]}$ & $\sigma_{\rm J-[8.0]}$ & $\sigma_{\rm Ks-[3.6]}$ & $\sigma_{\rm Ks-[8.0]}$ & $\sigma_{\rm [3.6]-[8.0]}$ & $\sigma_{\rm [5.8]-[8.0]}$ & $\sigma_{\rm [3.6]-[4.5]} $ & $\sigma$  \\
\hline
0.035&       1.0 &       1.0 &       1.6&       1.5 &       1.9 &       2.0 &      - &      - &       1.5\\
0.06&       1.0 &      0.82&      0.94&      0.75 &       1.1 &       1.4 &       - &       - &      1.0\\
0.12&       1.1 &      0.78 &      0.89 &      0.79 &       1.2&       1.4 &       - &       - &       1.0 \\
0.2&       1.2 &      0.81 &       1.1 &      0.91 &       1.4 &       1.7 &       - &       - &       1.2\\
0.3&       1.2 &      0.96 &       1.2 &       1.0 &       1.5 &       1.7 &       - &       - &       1.3 \\
0.5&       1.4 &       1.4 &       1.6 &       1.3 &       1.5 &       1.6 &       -&       - &       1.5\\
0.7&       1.6 &       1.7 &       1.6 &       1.5 &       1.4 &       1.3&       - &       - &       1.5 \\
\hline
J-Ks$_{\rm av}\sim2.2$ & $1.9\la J-Ks \la2.7$  &    $N_{\rm obs, stars}=212$ & & & & & &  &\\
\hline
$a_{\rm amC} [\mu m]$ & $\sigma_{\rm J-Ks}$ & $\sigma_{\rm J-[3.6]}$ & $\sigma_{\rm J-[8.0]}$ & $\sigma_{\rm Ks-[3.6]}$ & $\sigma_{\rm Ks-[8.0]}$ & $\sigma_{\rm [3.6]-[8.0]}$ & $\sigma_{\rm [5.8]-[8.0]}$ & $\sigma_{\rm [3.6]-[4.5]} $ & $\sigma$  \\
\hline
0.035&      0.90 &      0.86 &       1.0 &      0.83 &       1.1 &       1.2 &     - &      - &      0.99 \\
0.06&      0.87 &      0.69 &      0.59 &      0.50 &      0.51 &      0.48 &      - &      - &      0.61 \\
0.12&      0.88&      0.72 &      0.68 &      0.54 &      0.62 &      0.61 &      - &      - &      0.67 \\
0.2&      0.82 &      0.74 &      0.82 &      0.68 &      0.84 &      0.84 &       - &      - &      0.79 \\
0.3&      0.85 &      0.71 &      0.80 &      0.64 &      0.84 &      0.88 &       -  &     - &      0.79\\
0.5&      0.93 &      0.75 &      0.77 &      0.54 &      0.72 &      0.85 &       - &     - &      0.76\\
0.7&      0.88 &      0.82 &      0.64 &      0.86 &      0.59 &      0.62 &       - &      - &      0.73\\
\hline
J~$-$~Ks$_{\rm av}\sim3.0$ &  $2.7\la J-Ks \la3.4$ &   $N_{\rm obs, stars}=117$ & & & & & &  \\                                                                         \hline
$a_{\rm amC} [\mu m]$ & $\sigma_{\rm J-Ks}$ & $\sigma_{\rm J-[3.6]}$ & $\sigma_{\rm J-[8.0]}$ & $\sigma_{\rm Ks-[3.6]}$ & $\sigma_{\rm Ks-[8.0]}$ & $\sigma_{\rm [3.6]-[8.0]}$ & $\sigma_{\rm [5.8]-[8.0]}$ & $\sigma_{\rm [3.6]-[4.5]} $ & $\sigma$ \\
\hline
0.035&       1.2 &      0.63 &       1.2&      0.42 &       1.3 &       1.3 &       1.5 &      0.99 &       1.1 \\
0.06&       1.2&      0.65 &      1.0 &      0.47&      0.99 &      0.72 &      0.97 &      0.68 &      0.83\\
0.12&       1.1&      0.67&       1.1 &      0.62 &       1.3 &      0.93 &      0.54&      0.88 &      0.90 \\
0.2&       1.1 &      0.93 &       1.8 &      0.97 &      2.1 &       1.5 &      0.87 &       1.4 &       1.3 \\
0.3&       1.2 &      0.97 &       1.9 &      0.98 &       2.3 &       1.7 &       1.0 &       1.5 &        1.4\\
0.5&       1.3 &      0.87 &       1.3&      0.62 &       1.4&       1.3 &      0.88 &       1.1&       1.1 \\
0.7&       1.2 &       2.3 &       2.8 &       2.4 &       3.1 &       1.1 &      0.59 &       1.5 &    1.9 \\
\hline
J~$-$~Ks$_{\rm av}\sim3.7$ & $3.4\la J-Ks\la4.2$ &   $N_{\rm obs, stars}=43$ & & & & & &  &\\
\hline
$a_{\rm amC} [\mu m]$ & $\sigma_{\rm J-Ks}$ & $\sigma_{\rm J-[3.6]}$ & $\sigma_{\rm J-[8.0]}$ & $\sigma_{\rm Ks-[3.6]}$ & $\sigma_{\rm Ks-[8.0]}$ & $\sigma_{\rm [3.6]-[8.0]}$ & $\sigma_{\rm [5.8]-[8.0]}$ & $\sigma_{\rm [3.6]-[4.5]} $ & $\sigma$  \\
\hline
0.035&       1.2 &      0.44 &      0.86 &      0.35 &      0.75 &       1.5 &      2.0 &      0.63 &      0.96\\
0.06&      0.75 &      0.46 &      0.64 &      0.53 &      0.81 &      0.66 &      0.56 &      0.79 &      0.65 \\
0.12&       1.4 &      0.45 &      0.77 &      0.72 &       1.2 &       1.1 &      0.49 &       1.1 &      0.90 \\
0.2&       1.1 &      0.89 &       1.7 &       1.2 &       2.3 &       2.2 &       1.1 &       1.8 &       1.5 \\
0.3&       1.0 &      0.86 &       1.7 &       1.0 &       2.2 &       2.4 &       1.2 &       1.9 &       1.5\\
0.5&       1.1 &       1.2 &      1.0 &       1.0 &        0.98 &       1.0 &      0.77 &      0.67 &      0.97 \\
0.7&       1.5 &       2.1 &       2.4 &       2.1 &       2.6 &       1.3 &      0.69 &       1.2 &       1.7\\
\hline
J~$-$~Ks$_{\rm av}\sim4.5$ &  $4.2\la J-Ks \la 5.0$ &   $N_{\rm obs, stars}=10$ & & & & & & & \\
\hline
$a_{\rm amC}[\mu m]$ & $\sigma_{\rm J-Ks}$ & $\sigma_{\rm J-[3.6]}$ & $\sigma_{\rm J-[8.0]}$ & $\sigma_{\rm Ks-[3.6]}$ & $\sigma_{\rm Ks-[8.0]}$ & $\sigma_{\rm [3.6]-[8.0]}$ & $\sigma_{\rm [5.8]-[8.0]}$ & $\sigma_{\rm [3.6]-[4.5]} $ & $\sigma$  \\
\hline
0.035&       1.9 &       1.2 &       1.5 &      0.48 &      0.83 &      0.53 &      0.59 &      0.44 &      0.94 \\
0.06&       2.1 &       1.1 &       1.7 &      0.73 &       1.4 &      0.86 &      0.52 &      0.64 &       1.1 \\
0.12&       1.3 &       1.8 &       3.9 &       1.2 &       3.0 &       2.2 &      0.40 &       1.4 &       1.9 \\
0.2&       1.5 &       1.5 &       4.0 &       1.5 &       3.8 &       2.8 &      0.55 &       1.8 &       2.2 \\
0.3&       1.5 &       1.4 &       3.6 &       1.0 &       3.2 &       2.7 &      0.54 &       1.6 &       2.0 \\
0.5&       2.1 &       2.9 &       2.6 &       1.7 &       1.4 &       1.3 &      0.34 &      0.50 &       1.6 \\
0.7&      0.94 &       2.6 &       3.1 &       2.5 &       2.9 &      0.53 &      0.19 &      0.94 &       1.7 \\
\hline
\end{tabular}
\end{center}
\end{table*}

\begin{table*}
\begin{center}
\caption{The same as in Table~\ref{Table:results} but for the Jaeger600 data set. }
\label{Table:results3}
\begin{tabular}{l c c c c c c c c c}
\hline
Opacity set:  &Jaeger600 & & & & & & & & \\
\hline
J~$-$~Ks$_{\rm av}\sim1.5$ & $1.1\la$J-Ks$\la1.9$ &  $N_{\rm obs, stars}=1630$ & & & & & &  &\\
\hline
$a_{\rm amC} [\mu m]$ & $\sigma_{\rm J-Ks}$ & $\sigma_{\rm J-[3.6]}$ & $\sigma_{\rm J-[8.0]}$ & $\sigma_{\rm Ks-[3.6]}$ & $\sigma_{\rm Ks-[8.0]}$ & $\sigma_{\rm [3.6]-[8.0]}$ & $\sigma_{\rm [5.8]-[8.0]}$ & $\sigma_{\rm [3.6]-[4.5]} $ & $\sigma$  \\
\hline
0.035&      0.98 &      0.97 &       1.5 &       1.5 &       1.9 8&       1.9 &      - &       - &       1.5 \\
0.06&       1.1 &      0.59 &       1.0 &      0.58 &       1.4 &       1.9 &       - &       - &       1.1 \\
0.12&       1.1 &      0.60 &      0.97 &      0.58 &       1.4 &       1.9 &       -  &       - &  1.1 \\
0.2&       1.2 &      0.75 &       1.1 &      0.71 &       1.5 &       2.0 &       - &       - &       1.2\\
0.3&       1.4 &       1.1 &       1.4 &      0.85 &       1.5 &       1.9 &       - &     - &       1.4\\
0.5&       1.6 &       1.5 &       1.3 &       1.2 &       1.2 &       1.5 &       - &      - &       1.4\\
0.7&       1.5 &       2.2 &       1.5 &       2.4 &       1.3 &       1.3 &       - &       - &       1.7\\
\hline
J~$-$~Ks$_{\rm av}\sim2.2$ & $1.9\la J-Ks \la 2.7$  &    $N_{\rm obs, stars}=212$ & & & & & &  &\\
\hline
$a_{\rm amC} [\mu m]$ & $\sigma_{\rm J-Ks}$ & $\sigma_{\rm J-[3.6]}$ & $\sigma_{\rm J-[8.0]}$ & $\sigma_{\rm Ks-[3.6]}$ & $\sigma_{\rm Ks-[8.0]}$ & $\sigma_{\rm [3.6]-[8.0]}$ & $\sigma_{\rm [5.8]-[8.0]}$ & $\sigma_{\rm [3.6]-[4.5]} $ & $\sigma$  \\
\hline
0.035&       1.0 &      0.79 &      0.86 &      0.75 &      0.86 &       1.0 &      - &       - &      0.89 \\
0.06&       1.0 &      0.83 &      1.0 &      0.84 &       1.1 &       1.3 &       - &      - &       1.0 \\
0.12&      0.99 &      0.96 &       1.2 &       1.0 &       1.4 &       1.4 &       - &      - &       1.2 \\
0.2&      0.78 &      0.86 &       1.2 &       1.1 &       1.5 &       1.5 &       - &      - &       1.2 \\
0.3&      0.87 &      0.91 &       1.2 &      0.96 &       1.4 &       1.5 &       - &      - &       1.1\\
0.5&      0.84 &      0.90 &      0.84 &      0.90 &      0.83 &       1.2 &       - &      - &      0.91 \\
0.7&      0.84 &       2.9 &       2.3 &       4.0 &       2.7 &       1.5 &       - &       - &       2.4 \\
\hline
J~$-$~Ks$_{\rm av}\sim3.0$ &  $2.7\la J-Ks\la 3.4$ &   $N_{\rm obs, stars}=117$ & & & & & &  \\
\hline
$a_{\rm amC} [\mu m]$ & $\sigma_{\rm J-Ks}$ & $\sigma_{\rm J-[3.6]}$ & $\sigma_{\rm J-[8.0]}$ & $\sigma_{\rm Ks-[3.6]}$ & $\sigma_{\rm Ks-[8.0]}$ & $\sigma_{\rm [3.6]-[8.0]}$ & $\sigma_{\rm [5.8]-[8.0]}$ & $\sigma_{\rm [3.6]-[4.5]} $ & $\sigma$  \\
\hline
0.035&       1.3 &      0.97 &       1.0 &      0.93 &      0.96 &      0.80 &      1.1 &      0.60 &      0.95 \\
0.06&       1.2 &       1.1 &       2.4 &       1.2 &       2.9 &       2.2 &      1.8 &       1.6 &    1.8 \\
0.12&       1.2 &       1.2 &       2.6 &       1.4 &       3.3 &       2.4 &      1.9 &       1.7 &        2.0\\
0.2&       1.2 &       1.5 &       3.2 &       1.6 &       3.9 &       3.0 &          2.3 &       2.2 &    2.4 \\
0.3&       1.2 &       1.3 &       3.0 &       1.3 &       3.5 &       2.9 &       2.3 &       2.0 &       2.2 \\
0.5&       1.1&       3.1 &       3.0 &       3.6 &       3.6 &       1.1 &       1.4 &       1.5 &     2.3 \\
0.7&       1.1 &       4.0 &       5.3 &       4.7 &       6.5 &       2.6 &     0.82&       4.0 &     3.6 \\
\hline
J~$-$~Ks$_{\rm av}\sim3.7$ & $3.4\la J-Ks\la 4.2$ &   $N_{\rm obs, stars}=43$ & & & & & &  &\\
\hline
$a_{\rm amC} [\mu m]$ & $\sigma_{\rm J-Ks}$ & $\sigma_{\rm J-[3.6]}$ & $\sigma_{\rm J-[8.0]}$ & $\sigma_{\rm Ks-[3.6]}$ & $\sigma_{\rm Ks-[8.0]}$ & $\sigma_{\rm [3.6]-[8.0]}$ & $\sigma_{\rm [5.8]-[8.0]}$ & $\sigma_{\rm [3.6]-[4.5]} $ & $\sigma$  \\
\hline
0.035&       1.1 &      0.84 &      0.80 &       1.0 &      0.94 &      0.75 &       1.1 &      0.46 &      0.88 \\
0.06&       1.2 &      0.97 &       2.1 &       1.2 &       2.8 &       3.0 &       2.0 &       1.9 &  1.9 \\
0.12&       1.2 &       1.1 &       2.4 &       1.5 &       3.2 &       3.5 &       2.3 &       2.1 &       2.2\\
0.2&       1.2 &       1.2 &       2.7 &       1.6 &         3.6 &       3.9 &       2.6 &       2.4 & 2.4\\
0.3&       1.1 &      0.91 &       2.2 &       1.1 &       2.9 &       3.7 &       2.6 &       2.2 &       2.1\\
0.5&       1.4 &       3.2 &       3.1 &       3.4 &         3.4 &       1.2 &       1.3 &       1.5 &   2.3 \\
0.7&      0.85 &       2.2 &       2.9 &       2.8&           3.8 &       2.3 &      0.48&       2.9& 2.3 \\
\hline
J~$-$~Ks$_{\rm av}\sim4.5$ &  $4.2\la J-Ks\la 5.0$ &   $N_{\rm obs, stars}=10$ & & & & & & & \\
\hline
$a_{\rm amC} [\mu m]$ & $\sigma_{\rm J-Ks}$ & $\sigma_{\rm J-[3.6]}$ & $\sigma_{\rm J-[8.0]}$ & $\sigma_{\rm Ks-[3.6]}$ & $\sigma_{\rm Ks-[8.0]}$ & $\sigma_{\rm [3.6]-[8.0]}$ & $\sigma_{\rm [5.8]-[8.0]}$ & $\sigma_{\rm [3.6]-[4.5]} $ & $\sigma$  \\
\hline
0.035&       1.4 &       1.3 &       3.2 &       1.2 & 3.0 &       2.5 &      0.83 &       1.2 &  1.8 \\
0.06&       1.8 &       1.4 &       4.5 &       1.4 & 4.3 &       3.6 &      0.97 &       1.8 &  2.4\\
0.12&       1.3 &       2.1 &       6.1 &       1.9 & 5.5 &       4.3 &       1.2 &       2.2 &   3.1 \\
0.2&       1.2 &       1.9 &       6.4 &       1.9 &       5.7 &       4.8 &       1.4 &       2.4 &       3.2 \\
0.3&       2.1 &       2.5 &       3.0 &       1.4 &       2.6 &       3.6 &       1.1 &       1.5 &       2.2 \\
0.5&      0.97 &       4.8 &       4.8 &       4.3 &       4.2 &      0.80 &      0.75 &       1.4 &    2.7 \\
0.7&       1.5 &       2.8 &       5.0 &       3.2 &       5.1 &       2.1 &      0.11 &       2.6 &       2.8 \\
\hline
\end{tabular}
\end{center}
\end{table*}

\begin{table*}
\begin{center}
\caption{The same as in Table~\ref{Table:results} but for the Jaeger1000 data set.}
\label{Table:results4}
\begin{tabular}{l c c c c c c c c c}
\hline
Opacity set:  & Jaeger1000 & & & & & & & \\
\hline
J~$-$~Ks$_{\rm av}\sim1.5$ & $1.1\la J-Ks\la 1.9$ &  $N_{\rm obs, stars}=1630$ & & & & & &  &\\
\hline
$a_{\rm amC} [\mu m]$ & $\sigma_{\rm J-Ks}$ & $\sigma_{\rm J-[3.6]}$ & $\sigma_{\rm J-[8.0]}$ & $\sigma_{\rm Ks-[3.6]}$ & $\sigma_{\rm Ks-[8.0]}$ & $\sigma_{\rm [3.6]-[8.0]}$ & $\sigma_{\rm [5.8]-[8.0]}$ & $\sigma_{\rm [3.6]-[4.5]} $ & $\sigma$  \\
\hline
0.035&       1.0 &       1.4 &       2.0 &       1.7 &       2.2 &       2.2 &     - &       - &       1.7 \\
0.06&       1.1 &       1.2 &       1.3 &       1.2 &  1.4 &       1.5 &       - &       - & 1.3 \\
0.12&       1.1 &       1.2 &       1.2 &       1.2& 1.3 &       1.5 &       - &       - & 1.3 \\
0.2&       1.4 &       1.5 &       1.5 &       1.5 &       1.6 &       1.6 &       - &      - &       1.5 \\
0.3&       1.5&       2.0 &       1.7 &       2.1 &       1.6 &       1.4&        - &       - &       1.7\\
0.5&       1.1 &       4.3 &       4.8 &       6.5 &   5.7 &       4.3 &       - &       - &   4.5\\
0.7&      0.96 &       5.0 &       6.9 &       8.0 &  8.4 &       7.4 &       - &       - &  6.1 \\
\hline
J~$-$~Ks$_{\rm av}\sim2.2$ & $1.9\la J-Ks\la 2.7$  &    $N_{\rm obs, stars}=212$ & & & & & &  &\\
\hline
$a_{\rm amC} [\mu m]$ & $\sigma_{\rm J-Ks}$ & $\sigma_{\rm J-[3.6]}$ & $\sigma_{\rm J-[8.0]}$ & $\sigma_{\rm Ks-[3.6]}$ & $\sigma_{\rm Ks-[8.0]}$ & $\sigma_{\rm [3.6]-[8.0]}$ & $\sigma_{\rm [5.8]-[8.0]}$ & $\sigma_{\rm [3.6]-[4.5]} $ & $\sigma$  \\
\hline
0.035&       1.0 &      0.91 &       1.0 &      0.72 & 1.0 &       1.1 &     - &     - &      0.98 \\
0.06&      0.97 &      0.80 &      0.73 &      0.57 &  0.63 &      0.68 &       - &     - &       0.73 \\
0.12&      0.93 &      0.76 &      0.77 &      0.53 &  0.70 &      0.80 &      - &      - &   0.75\\
0.2&      0.86 &      0.69 &      0.70 &      0.51 &   0.65 &      0.81 &       - &      - &  0.70 \\
0.3&       1.0 &       2.2 &       1.8 &       2.8 &       1.9 &       1.1 &    - & - &       1.8 \\
0.5&      0.75 &       4.7 &       5.2 &       6.9 &   6.3 &       4.8 &       - &       - &   4.8 \\
0.7&      0.83 &       4.0 &       5.0 &       6.4 &   6.3 &       5.3 &       - &       - &    4.7 \\
\hline
J~$-$~Ks$_{\rm av}\sim 3.0$ &  $2.7\la J-Ks\la 3.4$ &   $N_{\rm obs, stars}=117$ & & & & & &  \\
\hline
$a_{\rm amC} [\mu m]$ & $\sigma_{\rm J-Ks}$ & $\sigma_{\rm J-[3.6]}$ & $\sigma_{\rm J-[8.0]}$ & $\sigma_{\rm Ks-[3.6]}$ & $\sigma_{\rm Ks-[8.0]}$ & $\sigma_{\rm [3.6]-[8.0]}$ & $\sigma_{\rm [5.8]-[8.0]}$ & $\sigma_{\rm [3.6]-[4.5]} $ & $\sigma$ \\
\hline
0.035&       1.2 &      0.71 &       1.1 &      0.42 &  1.1 &       1.0 &       1.4 &      0.75 &  0.96 \\
0.06&       1.2 &      0.71 &       1.0 &      0.38 &  0.91 &      0.83 &     0.82 &      0.41 &   0.79 \\
0.12&       1.2 &      0.69 &       1.2 &      0.42 &   1.3 &       1.2 &      1.3 &      0.54 &   0.99 \\
0.2&       1.3 &      0.87 &       1.4 &      0.63 &   1.4 &       1.3 &      1.4 &      0.65 &       1.1 \\
0.3&       1.3 &       3.9 &       4.9 &       4.8 &       6.1 &       2.0 &       1.0 &       2.7 &       3.3 \\
0.5&       1.1 &       3.5 &       6.4 &       4.7 &       8.3 &       4.7 &  2.5 &       5.1 &         4.5 \\
0.7&       ns &       &        &       &   &       &       &       &      \\
\hline
J~$-$~Ks$_{\rm av}\sim3.7$ & $3.4\la J-Ks\la 4.2$ &   $N_{\rm obs, stars}=43$ & & & & & &  &\\
\hline
$a_{\rm amC} [\mu m]$ & $\sigma_{\rm J-Ks}$ & $\sigma_{\rm J-[3.6]}$ & $\sigma_{\rm J-[8.0]}$ & $\sigma_{\rm Ks-[3.6]}$ & $\sigma_{\rm Ks-[8.0]}$ & $\sigma_{\rm [3.6]-[8.0]}$ & $\sigma_{\rm [5.8]-[8.0]}$ & $\sigma_{\rm [3.6]-[4.5]} $ & $\sigma$  \\
\hline
0.035&       1.2 &      0.63 &      0.89 &      0.29 &      0.73 &      0.95 &       1.6 &      0.31 &      0.82 \\
0.06&       1.1 &      0.55 &      0.68 &      0.25 &      0.67 &      0.93 &      0.87 &      0.40 &      0.69 \\
0.12&       1.1 &      0.49 &      0.92 &      0.40 &       1.2 &       1.5&       1.3 &      0.70 &      0.94 \\
0.2&       1.1 &      0.84 &       1.1 &      0.74 &       1.2 &       1.6 &       1.5 &      0.76 &       1.1 \\
0.3&       1.0 &       3.6 &       4.2 &       4.2 &       5.2 &       2.4 &      0.92 &       2.5 &       3.0 \\
0.5&      ns &       &       &       &       &     &      &   &       \\
0.7&       ns & & & &       &  &  & &  \\
\hline
J~$-$~Ks$_{\rm av}\sim4.5$ &  $4.2\la J-Ks\la 5.0$ &   $N_{\rm obs, stars}=10$ & & & & & & & \\
\hline
$a_{\rm amC}[\mu m]$ & $\sigma_{\rm J-Ks}$ & $\sigma_{\rm J-[3.6]}$ & $\sigma_{\rm J-[8.0]}$ & $\sigma_{\rm Ks-[3.6]}$ & $\sigma_{\rm Ks-[8.0]}$ & $\sigma_{\rm [3.6]-[8.0]}$ & $\sigma_{\rm [5.8]-[8.0]}$ & $\sigma_{\rm [3.6]-[4.5]} $ & $\sigma$  \\
\hline
0.035&       1.4&       1.4 &       2.3 &      0.76 &       1.5 &      0.91 &      0.86 &      0.37 &       1.2\\
0.06&       1.9 &       1.8 &       2.1 &      0.74 &      0.96 &      0.89 &      0.30 &      0.31 &       1.1 \\
0.12&       1.4 &       1.0 &       2.2 &      0.36 &       1.8 &       1.9 &      0.63 &      0.642&       1.3 \\
0.2&       1.4 &       3.3 &       3.3 &       2.5 &       2.3 &      0.92 &      0.41 &      0.59 &       1.8 \\
0.3&       1.1 &       5.6 &       7.3 &       5.0 &       6.4 &       1.4 &      0.20 &       2.0 &       3.6 \\
0.5&       ns &       &  & &       &       &       &       &       \\
0.7&    ns &       &    & &       &       &       &       & \\
\hline
\end{tabular}
\end{center}
\end{table*}

\begin{table*}
\begin{center}
\caption{The same as in Table~\ref{Table:results} but for the Hanner data set. }
\label{Table:results5}
\begin{tabular}{l c c c c c c c c c}
\hline
Opacity set:    &Hanner & & & & & & & & \\
\hline
J~$-$~Ks$_{\rm av}\sim1.5$ & $1.1\la J-Ks\la 1.9$ &  $N_{\rm obs, stars}=1630$ & & & & & &  &\\
\hline
$a_{\rm amC} [\mu m]$ & $\sigma_{\rm J-Ks}$ & $\sigma_{\rm J-[3.6]}$ & $\sigma_{\rm J-[8.0]}$ & $\sigma_{\rm Ks-[3.6]}$ & $\sigma_{\rm Ks-[8.0]}$ & $\sigma_{\rm [3.6]-[8.0]}$ & $\sigma_{\rm [5.8]-[8.0]}$ & $\sigma_{\rm [3.6]-[4.5]} $ & $\sigma$  \\
\hline
0.035&       1.1 &       1.1 &       1.3 &       1.1 &       1.4 &       1.5 &     -&       - &       1.3\\
0.06&       1.1 &       1.1 &       1.2 &       1.0 &  1.3 &       1.4 &      - &      - &       1.2 \\
0.12&       1.1 &      0.97 &       1.1 &      0.95 &       1.3 &       1.5 &       -&      -& 1.2\\
0.2&       1.2 &       1.2 &       1.4 &       1.1 & 1.5 &       1.7 &      - &       - &1.3 \\
0.3&       1.4 &       1.7 &       1.5 &       1.7 &       1.5 &       1.4 &       - &       - &       1.6\\
0.5   &    1.1 &       4.3 &       4.9 &       6.6 & 5.8 &       4.4 &       -&       -& 4.5 \\
0.7   &    0.95 &       4.9 &       6.9 &       7.9 & 8.4 &       7.5 &       - &     - & 6.1\\
\hline
J~$-$~Ks$_{\rm av}\sim 2.2$ & $1.9\la J-Ks\la 2.7$  &    $N_{\rm obs, stars}=212$ & & & & & &  &\\
\hline
$a_{\rm amC} [\mu m]$ & $\sigma_{\rm J-Ks}$ & $\sigma_{\rm J-[3.6]}$ & $\sigma_{\rm J-[8.0]}$ & $\sigma_{\rm Ks-[3.6]}$ & $\sigma_{\rm Ks-[8.0]}$ & $\sigma_{\rm [3.6]-[8.0]}$ & $\sigma_{\rm [5.8]-[8.0]}$ & $\sigma_{\rm [3.6]-[4.5]} $ & $\sigma$  \\
\hline
0.035&      0.97 &      0.77 &      0.67 &      0.53 &      0.55 &      0.53 &      - &      - &      0.67\\
0.06&      0.99 &      0.80&      0.72 &      0.55 &      0.62 &      0.62 &       - &      - &      0.72 \\
0.12&      0.93 &      0.78 &      0.78 &      0.60&      0.74 &      0.76 &       - &     -&      0.76 \\
0.2&      0.84 &      0.72 &      0.84 &      0.61 &      0.85 &      0.93 &       - &      - &      0.80 \\
0.3&       1.0 &       2.1 &       1.7 &       2.7 &       1.8 &       1.1 &       - &    - &       1.7 \\
0.5&      0.77 &       4.6 &       5.3 &       6.9 &       6.4 &       5.1 &      - &      - &       4.9 \\
0.7&      0.85 &       4.0 &       5.2 &       6.4 &       6.5&       5.5 &      - &       -&       4.7\\
\hline
J~$-$~Ks$_{\rm av}\sim 3.0$ &  $2.7\la J-Ks\la3.4$ &   $N_{\rm obs, stars}=117$ & & & & & &  \\
\hline
$a_{\rm amC} [\mu m]$ & $\sigma_{\rm J-Ks}$ & $\sigma_{\rm J-[3.6]}$ & $\sigma_{\rm J-[8.0]}$ & $\sigma_{\rm Ks-[3.6]}$ & $\sigma_{\rm Ks-[8.0]}$ & $\sigma_{\rm [3.6]-[8.0]}$ & $\sigma_{\rm [5.8]-[8.0]}$ & $\sigma_{\rm [3.6]-[4.5]} $ & $\sigma$  \\
\hline
0.035&       1.3 &      0.73 &       1.1 &      0.41 &      0.95 &      0.76 &   0.76 &      0.72 &    0.79 \\
0.06&       1.3 &      0.73 &       1.2 &      0.44 &       1.1 &      0.94 &   0.94 &      0.94 &      0.88\\
0.12&       1.2 &      0.76 &       1.4 &      0.62&       1.5 &       1.2&      1.2 &       1.1&    1.0\\
0.2&       1.3 &      0.90 &       1.8 &      0.80 &       2.1&       1.7 &     1.7 &       1.5 &    1.4 \\
0.3&       1.1 &       3.5 &       4.1 &       4.4 &       5.1 &       1.4 &      0.80 &       2.0 &       2.8 \\
0.5&       1.1 &       3.7&       6.9 &       4.8 &       8.9 &       5.3 &     5.3 &       2.8&       4.9 \\
0.7&       ns &      &   & &       &       &       &       & \\
\hline
J~$-$~Ks$_{\rm av}\sim 3.7$ & $3.4\la J-Ks\la 4.2$ &   $N_{\rm obs, stars}=43$ & & & & & &  &\\
\hline
$a_{\rm amC} [\mu m]$ & $\sigma_{\rm J-Ks}$ & $\sigma_{\rm J-[3.6]}$ & $\sigma_{\rm J-[8.0]}$ & $\sigma_{\rm Ks-[3.6]}$ & $\sigma_{\rm Ks-[8.0]}$ & $\sigma_{\rm [3.6]-[8.0]}$ & $\sigma_{\rm [5.8]-[8.0]}$ & $\sigma_{\rm [3.6]-[4.5]} $ & $\sigma$  \\                                                                           \hline
0.035&       1.2 &      0.56 &      0.66 &      0.29 &
      0.59 &      0.65 &      0.53 &      0.38 &
      0.61 \\
0.06&       1.2 &      0.55 &      0.77 &      0.37 &
      0.83 &      0.92 &      0.73 &      0.48&
      0.73\\
0.12&       1.1 &      0.63 &       1.2 &      0.71 &
       1.5 &       1.6 &       1.2 &      0.92&
       1.1 \\
0.2&       1.1 &      0.60 &       1.2 &      0.63 &
       1.6 &       2.1 &       1.5 &       1.2 &
       1.2 \\
0.3&       1.1 &       4.0 &       4.6 &       4.7 &       5.6 &       2.3 &      1.0 &       2.4 &       3.2 \\
0.5&      ns &        &    & &
       &       &       &       &
       \\
0.7&       ns &   &   &    &
       &       &       &   & \\
\hline
J~$-$~Ks$_{\rm av}\sim 4.5$ &  $4.2\la J-Ks\la5.0$ &   $N_{\rm obs, stars}=10$ & & & & & & & \\
\hline
$a_{\rm amC} [\mu m]$ & $\sigma_{\rm J-Ks}$ & $\sigma_{\rm J-[3.6]}$ & $\sigma_{\rm J-[8.0]}$ & $\sigma_{\rm Ks-[3.6]}$ & $\sigma_{\rm Ks-[8.0]}$ & $\sigma_{\rm [3.6]-[8.0]}$ & $\sigma_{\rm [5.8]-[8.0]}$ & $\sigma_{\rm [3.6]-[4.5]} $ & $\sigma$  \\                                                                           \hline
0.035&       1.2 &      0.99 &       1.3 &      0.39 &
      0.69 &      0.86 &      0.24 &      0.24 &
      0.73 \\
0.06&       1.9 &       1.5 &       1.5 &      0.49 &
      0.65 1&      0.76 &      0.25 &      0.22 &
      0.91 \\
0.12&       1.2 &       1.1 &       2.8 &      0.79 &
       2.5 &       2.1 &      0.57 &      0.91 &
       1.5\\
0.2&       1.9 &       2.8 &       2.8&       1.8 &
       1.9 &       1.8 &      0.55 &      0.66 &
       1.8 \\
0.3&       1.2 &       6.3 &       7.8 &       5.6 &       6.7 &       1.1 &      0.14 &       1.8 &       3.8 \\
0.5&       ns &   &   &   &
       &  &   &  &
       \\
0.7&   ns &   &    &       &
       &       &   &       &
       \\
       \hline
\end{tabular}
\end{center}
\end{table*}

\end{appendix}
\label{lastpage}
\end{document}